\documentclass[11pt,a4paper,fleqn]{article}
\usepackage[utf8]{inputenc}
\usepackage{authblk}
\usepackage{geometry}
\usepackage{xcolor}
\usepackage{natbib}
\usepackage{graphicx}
\usepackage{pdflscape}
\usepackage{appendix}
\usepackage{setspace}

\usepackage{colortbl}

\usepackage{threeparttable}
\usepackage{booktabs}
\usepackage{makecell}

\usepackage{amsmath}
\usepackage{amssymb}
\usepackage{bm}
\usepackage{xfrac}

\usepackage{enumitem}
\newlist{steps}{enumerate}{1}
\setlist[steps, 1]{label = Step \arabic*:}

\usepackage{dcolumn}
\newcolumntype{d}[1]{D{.}{.}{#1}}

\makeatletter
\renewenvironment{thebibliography}[1]
         {\section*{\refname}%
          \@mkboth{\MakeUppercase\refname}{\MakeUppercase\refname}%
          \list{\@biblabel{\@arabic\c@enumiv}}%
               {\settowidth\labelwidth{\@biblabel{#1}}%
                \leftmargin\labelwidth
                \advance\leftmargin\labelsep
                \@openbib@code
                \usecounter{enumiv}%
                \let\p@enumiv\@empty
                \itemsep=0pt
                \parsep=0pt
                \leftmargin=\parindent
                \itemindent=-\parindent
                \renewcommand\theenumiv{\@arabic\c@enumiv}}%
          \sloppy
          \clubpenalty4000
          \@clubpenalty \clubpenalty
          \widowpenalty4000%
          \sfcode`\.\@m}
         {\def\@noitemerr
           {\@latex@warning{Empty `thebibliography' environment}}%
          \endlist}
\makeatother

\usepackage{caption}
\usepackage{subcaption}
\definecolor{nblue}{HTML}{000660}
\usepackage[colorlinks=true,urlcolor=nblue,linkcolor=nblue,citecolor=nblue]{hyperref}

\title{\LARGE \textbf{Non-linear dimension reduction in factor-augmented vector autoregressions}}
\author{\MakeUppercase{Karin Klieber}\thanks{Oesterreichische Nationalbank. \textit{Address}: Otto-Wagner-Platz 3, 1090 Vienna, Austria. \textit{Email}: \href{mailto:karin.klieber@soenb.at}{karin.klieber@oenb.at}.  I thank Florian Huber, Niko Hauzenberger, Michael Pfarrhofer, Anna Stelzer, Andreas Tsopanakis as well as the participants of the 2nd Workshop on High-Dimensional Data Analysis at the Universidad Carlos III de Madrid and the ICMAIF 2023 conference for valuable comments and suggestions. The views expressed in this paper do not necessarily reflect those of the Oesterreichische Nationalbank or the Eurosystem.}}
\affil{\textit{Oesterreichische Nationalbank}}

\date{\today}

\begin{document}

\maketitle\thispagestyle{empty}\normalsize\vspace*{-2em}\small

\begin{center}
\begin{minipage}{0.8\textwidth}
\noindent\small This paper introduces non-linear dimension reduction in factor-augmented vector autoregressions to analyze the effects of different economic shocks. I argue that controlling for non-linearities between a large-dimensional dataset and the latent factors is particularly useful during turbulent times of the business cycle. In simulations, I show that non-linear dimension reduction techniques yield good forecasting performance, especially when data is highly volatile. In an empirical application, I identify a monetary policy as well as an uncertainty shock excluding and including observations of the COVID-19 pandemic. Those two applications suggest that the non-linear FAVAR approaches are capable of dealing with the large outliers caused by the COVID-19 pandemic and yield reliable results in both scenarios. 
\end{minipage}
\end{center}

\onehalfspacing\normalsize\renewcommand{\thepage}{\arabic{page}}

\newpage
\section{Introduction}\label{sec:intro}


The COVID-19 pandemic belongs to the severest health, economic and social crises in recent decades and poses the greatest challenge to the world economy since World War II. The virus has spread around the globe and paralyzed entire economic sectors and activities. 
For economic modeling, the COVID-19 pandemic entails dealing with huge, unprecedented outliers in datasets which adversely affect the reliability of established, mostly linear, economic models. To the detriment of those commonly used models, economic indicators and variables are prone to unanticipated movements and  do not respond in the way they are supposed to. Large shifts in the level of certain variables and strong deviations from their usual paths clearly aggravate the challenge of handling large outliers within existing econometric models. 
As a solution, very recent studies suggest either discarding these outliers \citep[e.g., ][]{schorfheide2020covid} or tayloring workhorse models to include COVID-19 information via priors \citep[e.g., ][]{lenza2020covid,carriero2022addressing,cascaldi2022pandemic}. \cite{primiceri2020macroeconomic} and \cite{ng2021modeling} take a structural perspective and interpret the COVID-19 pandemic as an exclusive shock to the economy. Another strategy is to incorporate highly non-linear techniques into existing models \citep[e.g., ][]{huber2020nowcasting,hauzenberger2022enhanced}, which I also pursue in this paper.

The model introduced in this paper extends the factor-augmented vector autoregression (FAVAR) model as proposed by \cite{Bernanke2005FAVAR} to a more general framework. This approach allows to flexibly model the relationship between a large number of regressors and the factors. Similar to non-linear dynamic factor models \citep[see, e.g., ][]{feng2018deep,Polson2018,dixon2019deep,wang2019deep,andreini2020deep} I assume that a small number of unobserved factors can explain the underlying dynamics of many economic and financial variables without restricting it to be linear. While the existing literature on non-linear FAVAR models mainly focuses on time variation or state dependencies in the coefficients and/or in the error variances \citep[see, e.g., ][]{korobilis2013tvp,mumtaz2014tv_favar,eickmeier2015favar,hf2018msfavar}, this paper accounts for potential non-linear relationships between the high-dimensional dataset and its lower-dimensional factor representation.
For the non-linear FAVAR, I apply non-linear dimension reduction techniques borrowed from the machine learning literature \citep{roweis2000lle,heaton2008introduction,Goodfellow2016} for constructing the latent factors. Recent approaches for dimension reduction non-linearly compress the information in a dataset, thereby allowing for uncovering more complex patterns in the underlying panel of economic or financial time series \citep[see, e.g., ][]{gallant_white1992,Chakraborty2017ML,heaton2017,Mullainathan2017ML,Polson2018,Kelly2018AE,Coulombe2019ML,coulombe2020forest}.

The proposed approach can be seen as a general form of the commonly used FAVAR model enriched with non-linearities in the relation between the covariates and the factors. In particular, it not only captures a high-dimensional dataset in a reduced form but nests various functional forms in the factor structure making it a powerful tool for economic analysis. In two different empirical application, I assess how controlling for non-linearities in the factor structures affects dynamic responses to economic shocks. I develop a \textit{`Locally Embedded FAVAR'} and a \textit{`Deep Dynamic FAVAR'} and compare them to the commonly used linear FAVAR. While the former uses locally linear embedding (LLE) from the manifold learning literature \citep{roweis2000lle}, the latter can be interpreted as a modification of the deep dynamic factor model as proposed in \cite{dixon2019deep} and \cite{andreini2020deep} and is based on an autoencoder.

Before I present the performance of the proposed approaches in two empirical applications, I investigate the properties of the models using artificial data. The analysis reveals that the non-linear techniques yield superior forecasting performance and controlling for non-linearities yields stable and reliable results when datasets involve large outliers similar to the ones observed during the COVID-19 pandemic. 

As a next step, I consider two empirical applications based on US data. First, I identify an expansionary monetary policy shock and compare the impulse responses generated by the linear and non-linear FAVAR approaches when the sample ends in 2019 and when the pandemic observations are included. To recover the structural shocks of the models, I rely on the well established identification schemes of imposing short-run restrictions on variables that are assumed to respond with a certain delay to a cut in interest rates \citep[see, e.g.,][]{Bernanke2005FAVAR, christiano2005zerores, stock2005restrictions,Boivin2009FAVAR}. 

The second application imposes an uncertainty shock on the macroeconomic and financial variables of the US economy. This involves constructing an uncertainty index which I do by following \cite{JLN2015uncertainty}. I identify the structural vector autoregression (SVAR) by adding the uncertainty proxy and ordering it first \citep[see, e.g.,][]{bloom2009uncertainty, koop2014uncertainty,carriero2015uncertainty, baker2016uncertainty, carriero2018uncertainty, carriero2021uncertainty}. Again, this analysis is carried out for a sample that ends in 2019 and one that includes the pandemic.

Model results show that the FAVAR approaches with either linear or non-linear compression techniques yield similar impulse responses to monetary policy as well as uncertainty shocks in tranquil times. However, when I include the pandemic observations, non-linear techniques yield tighter confidence bands and provide responses in line with economic theory whereas linear models suffer from large uncertainty bands and ambiguous responses. Overall, for both shocks considered, results suggest that the non-linear FAVAR models reliably measure responses to economic shocks, especially, in turbulent times such as the COVID-19 pandemic.


The remainder of this paper is structured as follows. Section \ref{sec:favar} presents the proposed general FAVAR model which nests a broad range of dimension reduction techniques. The identification strategies chosen to recover the structural shocks are discussed in Section \ref{sec:ident}. Section \ref{sec:simulation} applies the proposed FAVAR approaches to synthetic data. Section \ref{sec:results} describes the dataset, analyzes the properties of the latent factor in great detail and provides the results of the impulse response analysis for a monetary policy and an uncertainty shock before and during the COVID-19 crisis. The last section summarizes and concludes the paper.

\section{A general FAVAR}\label{sec:favar}

Let $\bm D_t$ denote an $N \times 1$ vector of macroeconomic and financial variables observed at time $t=1, \dots, T$. The number of variables is large compared to the number of observations (i.e., $T \ll N$). I assume that the economy is driven by the dynamics of the variables in $\bm D_t$, which may feature highly non-linear dependencies, and that those dynamics can be captured in a small, $Q$-dimensional set of latent factors $\bm F_t$. In the following, the observation equation of the model is given by:
\begin{equation}\label{eq:obseq}
\bm D_t = g(\bm F_t, \bm v_t), \quad \bm v_t \sim \mathcal{N}(\bm 0,\bm \Sigma_{v}).
\end{equation}
In this very general framework, the functional form of the observation equation is unknown and potentially highly non-linear. To model the relationship between the observed variables and the latent factors I approximate function $g$ via dimension reduction techniques. That is, we learn the latent factors and obtain its estimates $\hat{\bm F_t}$ via linearly and non-linearly compressing its dimension with methods discussed in Section \ref{sec:factors}.

A suitable econometric model which combines unobserved and observed variables in a vector autoregression is the factor-augmented vector autoregression model. Introduced by \cite{Bernanke2005FAVAR}, the FAVAR model is capable of achieving parsimony and at the same time including a broad range of information necessary to capture the dynamics in a large dataset. This is accomplished by assuming that the $K$-dimensional vector of endogenous variables $\bm y_t$ is comprised of the set of latent factors $\bm F_t$ and small number of $R$ observed macroeconomic variables $\bm Z_t$ (such as, e.g., the policy instrument), i.e.,  $\bm y_t = [\bm F'_t, \bm Z'_t]'$ (with $K=R+Q$) and follows a VAR model with $p$ lags given by
\begin{equation}
\bm y_t = \bm c + \bm A_1 \bm y_{t-1} + \dots + \bm A_p \bm y_{t-p} + \bm \epsilon_t, \label{eq:redVAR} 
\end{equation}
where $\bm c$ denotes the $K$-dimensional vector of constants and the $K \times K$ matrices $\bm A_1, \dots, \bm A_p$ contain the reduced form coefficients for each lag and $\bm \epsilon_t$ is the normally distributed error term with zero mean and a $K \times K$ variance-covariance matrix $\bm \Sigma_{\bm \epsilon}$.

I apply the two-step approach as in, e.g., \cite{Bernanke2005FAVAR}, \cite{Boivin2009FAVAR} and \cite{korobilis2013tvp} and start by estimating the latent factors in Eq. \ref{eq:obseq}. Section \ref{sec:factors} provides deeper insights into the first step of the procedure. Next, the dynamics of the factors are estimated in a Bayesian VAR model. I apply the standard Minnesota prior on the VAR coefficients to shrink unimportant coefficients towards zero \citep{doan1984forecasting, sims1998bayesian,giannone2015prior}. Details on the prior specifications can be found in Appendix \ref{sec:App Tech}. Since a structural analysis of the system (e.g., impulse response analysis) needs some kind of effect size measures, I use a linear approximation which captures the functional relationship between the variables similar to regression coefficients. This way I make sure that the process is computationally tractable even if a closed-form inverse is not available, what is the case in many highly non-linear models. A detailed discussion is provided in Section \ref{subsec:approx}.

\subsection{Dimension reduction techniques}\label{sec:factors}

For extracting estimates of the latent factors $\hat{\bm F_t}$ I implement three different dimension reduction techniques. First, the construction of principal components (PCs) allows to implement the standard approach, referred to as the linear FAVAR. Second, I use locally linear embedding (LLE) from the manifold learning literature, labeled \textit{Locally Embedded FAVAR}. Third, I construct the latent factors by applying an autoencoder and refer to it as the \textit{Deep Dynamic FAVAR}. 
With this modeling choices the aim is to elaborate on the impact of different degrees of non-linearities. Manifold learning techniques can be seen as a generalization of PCA, which are able to preserve the global structure of the data even if it does not lie in a linear subspace \citep{roweis2000lle,bengio2013replearning}. The autoencoder, on the other hand, is based on neural networks and, as such, is able to learn any functional form under relatively few assumptions \citep{hornik1989neuralnet,bank2023autoencoders}. It is the most flexible approach I test in this setup in order to learn the structure of the data.
As emphasized in Section \ref{sec:favar}, the general FAVAR model is not limited to those dimension reduction techniques but is capable of incorporating various functional forms to obtain the latent factors.\\

\noindent \textbf{Linear FAVAR.} The most popular and commonly used dimension reduction technique to obtain latent factors is principal component analysis (PCA). The principal components of $\bm D$ are obtained by performing a truncated singular value decomposition (SVD) of the sample covariance matrix of $\bm D$. The resulting factor matrix $\hat{\bm F}$ is of dimension $T \times Q$ and, for an appropriate $Q$, summarizes the main information in the data \citep{stock2002macroeconomic}. Formally, the latent factors are defined $\hat{\bm F}$ as
\begin{equation}
\hat{\bm F} = \bm D \Lambda(\bm D' \bm D),\label{eq:PCA}
\end{equation}
with $\Lambda$ being the truncated eigenvector matrix of $\bm D' \bm D$ with dimension $N \times Q$.\\

\noindent \textbf{Locally Embedded FAVAR.} Originating from the field of image recognition, methods from the manifold learning literature are increasingly used in various research areas that deal with high-dimensional datasets. One of these non-linear methods for dimensionality reduction is locally linear embedding (LLE) introduced by \cite{roweis2000lle}. The algorithm aims to infer a lower-dimensional representation of the dataset while trying to preserve its geometric features.

This is done by finding the $k$-nearest neighbors of each column $\bm d_{\bullet i}$ ($i=1,\dots,N)$ of $\bm D$ and approximating the vector by weighted linear combinations of its $k$-nearest neighbors. We determine the $k$-nearest neighbors in terms of the Euclidean distance and select the optimal number of $k$ by applying the algorithm of \cite{kayo2006}. The algorithm preselects a set of potential candidates for $k$ and then runs through the steps in the LLE algorithm to find its optimal value. For the approximation of the data vectors, the weight matrix $\bm \Omega$ for the linear combinations of the $k$-nearest neighbors is obtained by minimizing the following cost function:
\begin{equation}
C(\bm \Omega) = \sum_{i}( \bm d_{\bullet i} - \sum_{j} \omega_{ij} \bm d_{\bullet j})^2,\label{eq:LLE weights}
\end{equation}
where $\omega_{ij}$ denotes the $(i,j)$th element of $\bm \Omega$. This minimization problem is subject to two constraints. First, matrix $\bm \Omega$ must be row-stochastic, i.e. each row of the matrix sums to one. Second, the reconstruction of each $\bm d_{\bullet i}$ is only considering its neighbors, implying non-zero weights only if $\bm d_{\bullet j}$ is a neighbor of $\bm d_{\bullet i}$.

Given the optimal weights, the algorithm requires the minimization of the cost for $\hat{\bm F}$ being the new data points given by
\begin{equation}
\Phi (\hat{\bm F}) = \sum_i | \mathfrak{\bm f}_{\bullet i} - \sum_j \Omega_{ij} \mathfrak{\bm f}_{\bullet j}|^2,\label{eq:LLE newdata}
\end{equation}
with $\mathfrak{\bm f}_{\bullet i}$ denoting the $i$th column of $\hat{\bm F}$. To obtain a well-behaved problem $\mathfrak{\bm f}_{\bullet i}$ is constrained to have zero mean and unit variance. 
The factors are then extracted by solving
\begin{equation}
\bm M = (\bm I_t + \bm \Omega)' (\bm I_t + \bm \Omega)\label{eq:LLE factors}
\end{equation}
and finding the $Q+1$ eigenvectors  of $\bm M$ corresponding to the $Q+1$ smallest eigenvalues. Discarding the bottom eigenvector gives the $Q$ factors, which represented the dataset in a low-dimensional and neighborhood-preserving manner.\\

\noindent \textbf{Deep Dynamic FAVAR.} For the Deep Dynamic FAVAR model, the latent factors $\hat{\bm F}$ are obtained by implementing an autoencoder and making use of the non-linear, lower-dimensional representation of the dataset. Belonging to the family of deep learning algorithms, autoencoders non-linearly convert a high-dimensional input to a transformed representation by first encoding the input to a lower-dimensional internal representation (the latent factors) and then decoding those latent factors back to the original dimension.

Autoencoders enjoy increasing attention in econometric analysis. First applications in economic forecasting and economic modeling can be found in, e.g., \cite{heaton2017,farrell2018deep, Polson2018,Kelly2018AE, cabanilla2019forecasting, dixon2019deep, andreini2020deep, hkkl2020real}. The proposed Deep Dynamic FAVAR approach is closely related to the recent literature dealing with factor models in combination with autoencoders or deep neural networks \citep{cabanilla2019forecasting, dixon2019deep, andreini2020deep}. Extending these concepts for structural analysis, I incorporate the highly non-linear factors to a VAR setting. 

In contrast to the dimension reduction techniques discussed so far, deep learning algorithms generate the latent factors by learning the functional form of the observation equation in Eq. \ref{eq:obseq}. This is done by introducing a set of parameters, which are optimized to find a good representation of the dataset \citep{Goodfellow2016}. In particular, the algorithm involves applying a number of $l \in \{1, \dots ,L\}$ non-linear transformations, i.e., activation functions, to $\bm D$. I repeat this process in $L$ hidden layers. The number of neurons which are input to the transformation process in each hidden layer is denoted by $m_l$. That is, in each layer a univariate activation function denoted by $h_1, \dots, h_L$ is applied to the neurons of the previous layer collected in matrix $\hat{\bm D}^{(l)}$. Note that for the first layer this corresponds to the original dataset ($\hat{\bm D}^{(1)} = \bm D$). Formally, this boils down to
\begin{equation}
h_l^{W^{(l)},b_l} = h_l \left( \sum_{i=1}^{m_l} \bm W^{(l)}_{\bullet i} \hat{\bm d}^{(l)}_{\bullet i} + b_l \right), \quad 1 \leq l \leq L, \label{eq:Autoencoder1}
\end{equation}
with $\hat{\bm d}^{(l)}$ denoting the $i$th column of matrix $\hat{\bm D}^{(l)}$. The parameters of the activation function to be determined are $\bm W^{(l)}$, which represents a weighting matrix and $b_l$, which denotes a bias term associated with layer $l$. $\bm W^{(l)}_{\bullet i}$ corresponds to the $i$th column of the weighting matrix. By minimizing a loss function of choice, the optimal values for the weight matrix and the bias term for each layer are determined. The estimation of these parameters is crucial. Provided that the algorithm adopted learns the correct parameters, a feed-forward network such as the autoencoder can basically approximate any functional form. This implies that, given the optimal parameters, the autoencoder can model the relationships in an underlying dataset regardless of their complexity and non-linearity \citep{hornik1989neuralnet,hornik1991approximation,Goodfellow2016}.

Finally, the deep dynamic factors $\hat{\bm F}$ are extracted after applying $L$ layers to the dataset:
\begin{equation}
\hat{\bm F} = (h_1^{W^{(1)}, b_1} \circ \dots \circ h_L^{W^{(L)}, b_L} ) (\bm D)\label{eq:Autoencoder2}.
\end{equation}

\begin{table}[htb!]
\caption{Summary of the model features for the Deep Dynamic FAVAR}\label{tab:ddf}
{\scriptsize
\begin{center}
\begin{tabular}{l lll}
\toprule
\bfseries  & \bfseries Hyperparameter & \bfseries Sets in the cross validation &  \bfseries Final choice \tabularnewline
\midrule
\bfseries Model structure & number of latent factors & \{2,3,4,5,6\} & 5 \tabularnewline
\bfseries & number of hidden layers & \{1,2,3,4,5\} & 3 \tabularnewline
\bfseries  & number of neurons for each layer & evenly downsizing the original dimension & \{126, 86, 46\} \tabularnewline  
\bfseries & penalisation & none & none \tabularnewline
\bfseries & dropout layers and rates & none & none \tabularnewline
\bfseries & batch norm layers & none & none \tabularnewline
\bfseries & activation function & tanh and ReLU & ReLU \tabularnewline
\midrule
\bfseries Optimization & size of mini batches & 24 & 24 \tabularnewline
\bfseries & number of epochs & 100 & 100 \tabularnewline
\bfseries & optimization algorithm & Adam with default parameters & Adam \tabularnewline
\bfseries & loss function & mean squared error  & mean squared error \tabularnewline

\bottomrule
\end{tabular}
    \begin{minipage}{\textwidth}
     \scriptsize
    \emph{Note:} The table gives details on the sets of parameters used in the cross validation and on the final choice of parameters used in the algorithm.
      \end{minipage}
\end{center}}
\end{table}

The deep dynamic FAVAR depends on a large set of hyperparameters. In the empirical application, I choose the hyperparameters according to the results of a cross validation exercise. The final model is comprised of five latent factors, three hidden layers with 126 neurons in the first hidden layer, 86 neurons in the second and 46 neurons in the third hidden layer. The activation function suggested by the cross validation exercise is ReLU. \cite{glorot2011relu}, for example, show that ReLU is often preferred to other activation functions because rectifying neurons are capable of creating a sparse representation of the dataset. As the most common choices for the loss function and optimization algorithm, I use a mean squared error loss function and adaptive moment estimation (Adam).
For the optimization of the algorithm, I use $24$ minibatches corresponding to the average duration of a business cycle in the US, i.e., six years.\footnote{The National Bureau of Economic Research publishes data on the duration of US business cycle expansions and contractions what allows for a determination of the average duration of a business cycle in the US. Details on the data can be found in Appendix \ref{sec:App Data}.} 
The optimization algorithm is repeated in $100$ epochs. Table \ref{tab:ddf} provides an overview of all choices on the deep learning algorithm implemented for the empirical application.\\

\subsection{Linear approximation for measuring effect sizes of highly non-linear models}\label{subsec:approx}

I wish to study how the economy reacts to structural shocks. This is achieved by comparing impulse response functions.
For the FAVAR approach, this analysis requires a mapping between the latent factors $\bm F_t$ and the underlying macroeconomic and financial variables in $\bm D_t$. Especially, when dealing with non-linear latent factors there is the need for an effect size measure similar to a regression coefficient that enables the estimation of the relationship between the factors and the dataset even in highly non-linear models. As many of these models are not computationally tractable, there is the need for approximations in order to measure the effects of the latent factors on the observed variables. I suggest using a linear approximation, which originates from the machine learning literature and the attempt of estimating effect sizes in black-box models \citep{crawford2018approx, crawford2019approx}. \cite{huber2020nowcasting}, for example, use such approximations to cast highly non-linear models in a Gaussian state space form.

In particular, I apply the Moore-Penrose pseudoinverse, which allows for a linkage between the non-linear factors and the economic and financial variables in the dataset even if the inverse of the unobserved factor matrix $\bm F$ does not exist \citep{theodoridis2015machine}. I define $\bm F$  as the full data matrix of the unobserved factors, i.e, $\bm F = (\bm F_1, \dots, \bm F_T)'$ and $\bm F^{+}$ as the Moore-Penrose pseudoinverse of $\bm F$. To find a set of linearized coefficients $\hat{\bm \theta}$ I solve
\begin{equation*}
\hat{\bm \theta} = \text{Proj}(\bm F, \bm D) = \bm F^{+} \bm D,\label{eq:MPI}
\end{equation*}
with $\bm D = (\bm D_1, \dots, \bm D_T)'$. Note that if $\bm F$ has full rank $\hat{\bm \theta}$ produces the least squares estimate of $\bm \theta$. If  $\bm F$ has less than full rank, the application of the Moore-Penrose pseudoinverse makes sure that the correct fitted values for $\bm D$ can be found.

\section{Identification strategy}\label{sec:ident}

When it comes to analyzing the effect of economic shocks on a set of macroeconomic variables the reduced form residuals in the model described above are not the ones of main interest. Instead we would like to recover the structural shocks of the VAR model. These, however, are only identified with further restrictions. Proposals in the literature involve restrictions on short- or long-run responses of variables \citep[see, e.g., ][]{king1991stochastic,Bernanke2005FAVAR,christiano2005zerores,stock2005restrictions,pagan2008econometric}, or on the sign of impulse responses \citep[see, e.g., ][]{uhlig2005sign,baumeister2015sign,  uhlig2015favar,antolin2018narrative}. Other studies suggest using dynamics in volatilities of the residuals \citep[see, e.g., ][]{rigobon2003identification,lanne2008identifying, lanne2010structural, lutkepohl2020identification} or introducing proxy variables \citep[see, e.g., ][]{bloom2009uncertainty,mertens2013dynamic,carriero2015uncertainty,gertler2015monetary,angelini2019proxy}. 
In the empirical section of the paper, I implement two applications with different identification strategies. While the first one relies on short-run restrictions justified by plausible economic reasoning, the second application involves estimating a proxy variable for uncertainty.

The monetary policy shock is modelled with identification by short-run restrictions. This strategy is based on economic reasoning and restricts contemporaneous effects of certain variables to be zero \citep[see, e.g.,][]{Bernanke2005FAVAR, christiano2005zerores, stock2005restrictions, Boivin2009FAVAR}. The implementation of this idea involves orthogonalizing the reduced form errors, i.e., making the errors mutually uncorrelated. This is obtained by Cholesky decomposition and results in a recursive structural model, which requires an ordering based on economic justifications \citep{kilian2017structural}. 
I divide the dataset into fast- and slow-moving variables based on theoretical considerations and economic intuition as suggested by \cite{Bernanke2005FAVAR}. Fast-moving variables include financial variables, prices and monetary aggregates and are assumed to respond immediately to unanticipated shocks. Slow-moving variables, such as wages or consumption, do not show contemporaneous effects after a monetary policy shock by assumption. A detailed description of the dataset including the classification of the variables are presented in Appendix \ref{sec:App Data}.

The second empirical application involves identifying an uncertainty shock, which is achieved by estimating the structural vector autoregression with an external instrument (or Proxy-SVAR). I follow \cite{JLN2015uncertainty} and construct an uncertainty index, which functions as the instrument (or proxy) variable. I define the measure of uncertainty as the volatility of expected forecast errors. In particular, I construct an uncertainty estimate of each variable in the dataset by applying the stochastic volatility approach of \cite{Kastner2014stochvol} and extract one common factor by using the first principal component of the covariance matrix of all uncertainty estimates. For further details on the approach I refer to \cite{JLN2015uncertainty}. Having obtained the proxy for uncertainty, I include it in the set of endogenous variables and order it first in the VAR \citep[see, e.g.,][]{bloom2009uncertainty, koop2014uncertainty,carriero2015uncertainty, JLN2015uncertainty,baker2016uncertainty,  carriero2018uncertainty, carriero2021uncertainty}.

\section{Simulation study}\label{sec:simulation}

In this section I apply the proposed FAVAR approaches to synthetic data and investigate the properties of the non-linear models in more detail. I do so by conducting a fully-fledged out-of-sample forecasting exercise. To mimic the complex dynamics observed in macroeconomic and financial variables during severe crises such as the COVID-19 pandemic, I introduce non-linearities in the data generating process (DGP). This is achieved by modeling a non-linear VAR structure, which generates outliers in the data. Moreover, I assume that the covariation in the synthetic series can be captured by a few latent factors and that the relationship between the main dataset and the latent factors is non-linear.

In particular, the relationship between the main dataset $\bm D_t$ and the latent factors $\bm y_t$ is given by
\begin{equation*}
\bm D_t = \bm \lambda (\bm y_{t-1} y_{2,t-2}) + \bm v_t, \quad \bm v_t  \sim N (0, \bm \Sigma_{v}).
\end{equation*}
I assume that $\bm y_t$ is a $3$-dimensional vector of latent factors. $\bm D_t$ denotes the observed dataset including $20$ variables which are observed for $350$ periods ($T=350$). $\bm \lambda$ denotes the matrix capturing the factor loadings which is of dimension $20 \times 3$ and sampled from a $\mathcal{N}(0,0.1^2)$ distribution. The variance-covariance matrix $\bm \Sigma_{v}$ is the identity matrix with dimension $20 \times 20$.

For the non-linear VAR, I let $\bm A_1$ and $\bm A_2$ denote a $3 \times 3$ coefficient matrix with off-diagonal elements sampled from a normal distrubtion of the form $\mathcal{N}(0,0.1^2)$. Centering all coefficients on $0$ ensures stationarity of the data. Moreover, I define $\bm C$ as a lower triangular matrix with off-diagonal elements drawn from a normal distribution given by $\mathcal{N}(0,0.1^2)$ and diagonal elements set to $1$. The latent factors are then modelled as:
\begin{equation*}
\bm y_t = \bm A_1 (\bm y_{t-1} y_{2,t-2}^{-1}) + \bm A_2 (\bm y_{t-1} y_{3,t-1}) + \bm C \bm u_t, \quad \bm u_t  \sim N (\bm 0, \bm I).
\end{equation*}

I take $20$ random samples from the data generating process (DGP) and conduct an out-of-sample forecasting exercise with the three competing FAVAR approaches discussed in Section \ref{sec:favar}. The hold-out is comprised of the last $200$ periods. I compute the predictive densities on an expanding window basis, i.e., I forecast the first period in the hold-out only with the data up to this point and repeatedly add the subsequent observation until I end up at the end of the sample.

\begin{figure}[htb!]
\caption{Simulated dataset. \label{fig:simdata}}

\begin{minipage}{0.49\textwidth}
\centering
 \textit{Observed dataset ($\bm D$)}
\end{minipage}
\begin{minipage}{0.49\textwidth}
\centering
 \textit{Latent factors ($\bm y$)}
\end{minipage}
\begin{minipage}{0.49\textwidth}
\centering
\includegraphics[scale=.38]{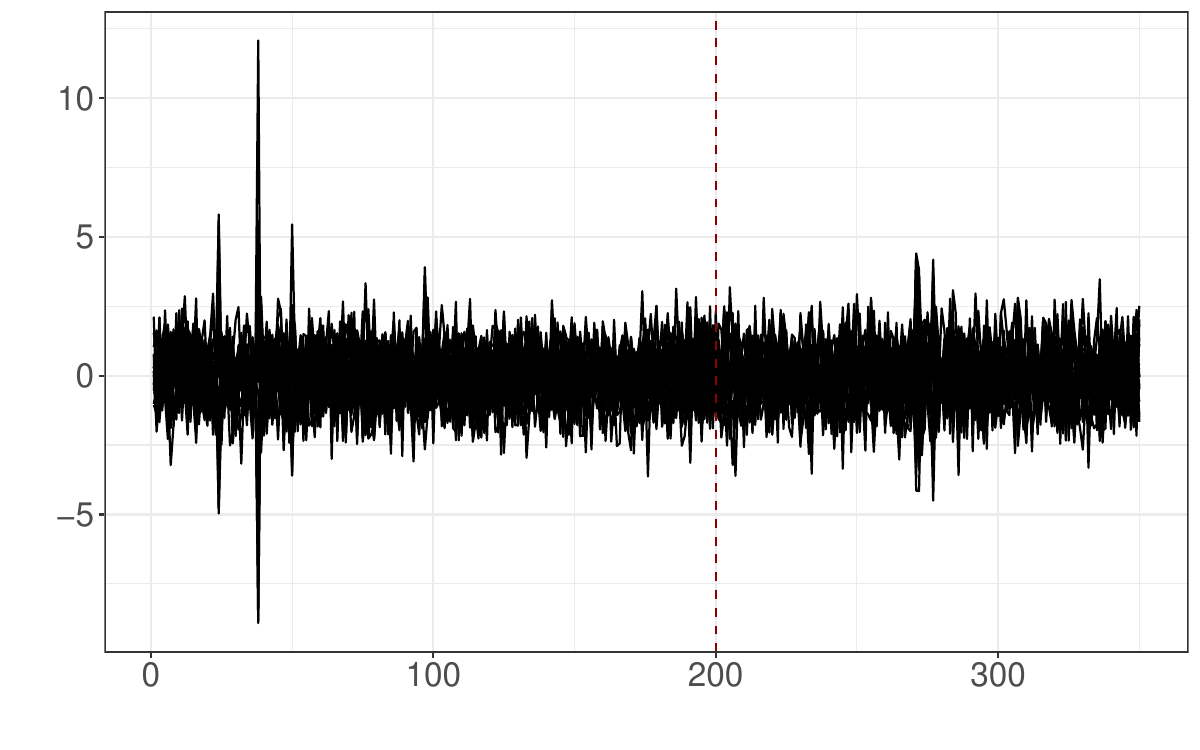}
\end{minipage}
\begin{minipage}{0.49\textwidth}
\centering
\includegraphics[scale=.38]{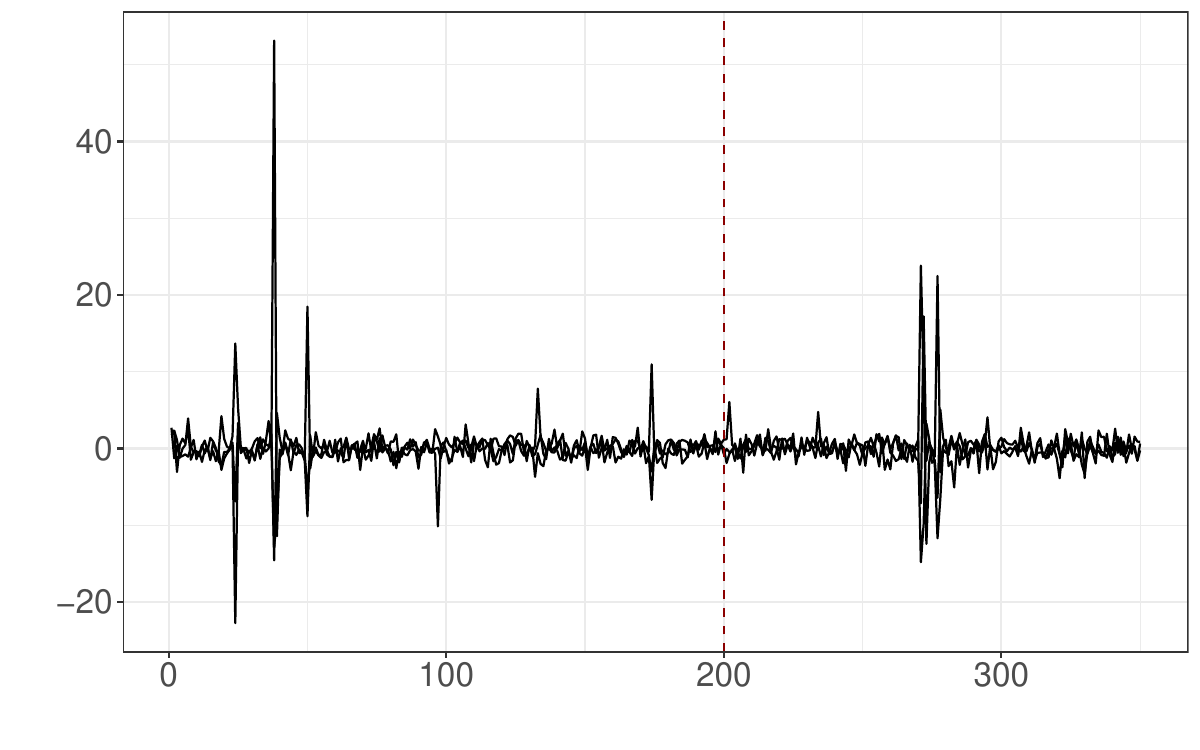}
\end{minipage}

\begin{minipage}{\textwidth}
\footnotesize
    \emph{Note:} The graph shows all observed variables (left panel) and the latent factors (right panel) from a randomly selected sample of the non-linear DGP. 
\end{minipage}

\end{figure}

Figure \ref{fig:simdata} presents one randomly selected realization from the DGP described above with the left panel plotting all variables in the dataset and the right panel plotting the latent factors. Mimicking ups and downs of the business cycle as well as severe crises such as the COVID-19 pandemic, the DGP includes substantial movements and severe outliers next to more quiet periods over the sample.

To evaluate the performance of the different models, Table \ref{tab:sim_performance} reports root mean squared errors (RMSEs) for point forecasts and continuous ranked probability scores \citep[CRPS, ][]{gneiting2007strictly} for density forecasts. All values are relative to the linear model. To gain deeper insights on the benefits of the non-linear approaches and when they are most pronounced I separately evaluate the forecasting performance during highly volatile and tranquil periods of the DGP. This is accomplished by labeling periods in which the latent factors exceed the interquantile range by a factor of seven as ``Crisis Times''. Moreover, I allocate variables to different groups depending on how severely they are affected by highly volatile periods. This allocation is carried out by clustering variables whose values exceed the interquantile range by certain levels. Variables classified as ``heavily affected” are those exhibiting outliers that exceed the interquantile range by a factor of seven. ``Affected” defines variables with outliers exceeding two times the interquantile range. All other variables, where no outliers are detected, are included in ``not affected”.

\begin{table}[!tbp]
{\small
\caption{Relative point and density forecasting performance for simulated data \label{tab:sim_performance}} 
\begin{center}
\begin{tabular}{llcccccc}
\toprule
\multicolumn{1}{l}{\textbf{Model}}&\multicolumn{1}{l}{\textbf{Variables}}&\multicolumn{1}{c}{}&\multicolumn{2}{c}{\textbf{Crisis Times}}&\multicolumn{1}{c}{}&\multicolumn{2}{c}{\textbf{Tranquil Times}}\tabularnewline
\addlinespace[2pt] 
 \cline{4-5} \cline{7-8}
\addlinespace[5pt] 
\multicolumn{1}{c}{}&\multicolumn{1}{c}{}&\multicolumn{1}{c}{}&\multicolumn{1}{c}{{RMSE}}&\multicolumn{1}{c}{{CRPS}}&\multicolumn{1}{c}{}&\multicolumn{1}{c}{{RMSE}}&\multicolumn{1}{c}{{CRPS}}\tabularnewline
\midrule
\addlinespace[5pt] 
\rowcolor{gray!15} 
\textbf{Deep Dynamic FAVAR} & Overall& &  \textbf{0.92} &      \textbf{0.92}& &   1.01&   1.01\tabularnewline
\addlinespace[2pt] 
  & Heavily affected& &  \textbf{0.87}&      \textbf{0.86}&&   1.03&   1.01\tabularnewline
  & Affected&  & \textbf{0.91}&      \textbf{0.91}&&   1.00&   1.00\tabularnewline
  & Not affected& &  \textbf{0.87}&     \textbf{0.83}& &  \textbf{0.98}&   \textbf{0.99}\tabularnewline
   \midrule
   \addlinespace[5pt] 
\rowcolor{gray!15} 
\textbf{Locally Embedded FAVAR} &Overall&  & \textbf{0.99}&      \textbf{0.99}& &  1.00&   1.00\tabularnewline
\addlinespace[2pt] 
 &  Heavily affected&&   \textbf{0.98}   &   \textbf{0.98}& &  1.00&   1.01\tabularnewline
 &  Affected&  & \textbf{0.99}&      \textbf{0.99}&  & 1.00  & 1.00\tabularnewline
 &  Not affected& &  \textbf{0.94}&      \textbf{0.90}& &  1.00&   1.00\tabularnewline
\bottomrule
\end{tabular}
\begin{minipage}{0.92\textwidth}
\scriptsize 
    \textit{Note}: The table shows point forecasting performance in terms of root mean squared errors (RMSE) as well as density forecasting performance in terms of continuous ranked probability scores (CRPS). All metrics are relative to the linear FAVAR. Values below one (bold numbers) show that the non-linear model outperforms the linear one. Variables are separated depending on how strongly they are affected by crises. ``Heavily affected" defines variables exhibiting outliers that exceed the interquantile range by a factor of seven. ``Affected" variables are those with outliers exceeding two times the interquantile range. Variables without any outliers are classified as ``not affected".
\end{minipage}
\end{center}}
\end{table}
 
Overall, I find that major gains from using non-linear models are obtained during highly volatile times. The Deep Dynamic FAVAR, being the most flexible model, outperforms the other two approaches by significant margins. This holds for point as well as for density forecasting performance. Improvements of the Locally Embedded FAVAR against the linear approach are rather small when averaging across all variables. 

Zooming into the different groups of variables reveals that the Deep Dynamic FAVAR gives the highest forecasting accuracy for heavily affected variables, closely followed by variables being not affected. Those gains are high in terms of RMSEs and even higher when considering CRPS. This implies that the deep learning model flexibly adapts to times of high uncertainty and is able to spread its strengths to many variables in the system. This can also be seen in  Table \ref{tab:sim_performance_details} in Appendix \ref{sec:App Res}, which shows the forecasting performance of each variable. Evidently, the improvement of the Deep Dynamic FAVAR upon the linear model is not restricted to a few specific variables but can be found across most of them.
The Locally Embedded FAVAR shows some gains for heavily affected models which are, however, rather small. Highest gains can be found for variables without outliers. From this I conclude that this model is also able to handle certain degrees of non-linearities but is not flexible enough to accurately model the highly affected and affected variables. 

For tranquil times all models yield a very similar performance. This suggests that highly non-linear techniques benefit the most against linear models when data is characterized by high volatility and complex relationships between variables. Nonetheless, I do not see any adverse effects of the high flexibility of the deep learning model or the locally linear embedding approach when variables show unobtrusive behavior.

\section{Empirical application}\label{sec:results}

In this section, I first introduce the dataset for the empirical application and provide an in-depth analysis of the latent factors. The discussion of the latent factors obtained from the different approaches in a profound manner helps to achieve a better understanding of the role of potential non-linearities and to give basic economic interpretation. I proceed with presenting the results of the impulse response analysis of a monetary policy shock as well as an uncertainty shock before and during the COVID-19 pandemic. I compare the responses of selected macroeconomic and financial quantities between linear and non-linear FAVAR models to investigate whether controlling for non-linearities alter the results.

\subsection{Data}\label{subsec:data}

I use 166 quarterly variables from the US database discussed in \cite{mccracken2020fred}. The data runs from 1965Q1 to 2020Q4. I assess and compare the properties of the different FAVAR approaches for the period before the COVID-19 outbreak (i.e., 1965Q1 - 2019Q4) and for the period including the COVID-19 pandemic (i.e., 1965Q1 - 2020Q4). For the data transformation, I choose a mixed approach where most variables are included with year-on-year growth rates and some enter the analysis in levels.
A detailed description of the data transformation can be found in Appendix \ref{sec:App Data}. Each time series is standardized to get series with zero mean and unit sample variance. 

Similar to \cite{Bernanke2005FAVAR}, the observed variables included in the VAR are industrial production, the unemployment rate, inflation and the policy instrument. Since the sample includes the prolonged period at the zero lower bound, I choose the shadow federal funds rate as the policy measure \citep[see, e.g., ][]{damjanovic2016shadow,potjagailo2017spillover,lombardi2018shadow}. In particular, I use the shadow rate suggested by \cite{wu2016shadow}, who show that their measure allows to study the reaction of macroeconomic variables to monetary policy even when hitting the zero lower bound.  

The latent factors are obtained by reducing the dimension of all other variables ($N=162$). The number of factors to be included in the model is chosen according to the cross validation exercise for the autoencoder which suggests a number of five factors (i.e., $Q=5$). Moreover, a closer examination of the principal components obtained from the linear approach shows that the first five factors explain close to 70 \% of the variation in the input dataset what seems to be reasonably high. The lag order is set equal to four ($p=4$).

\subsection{Structure of the latent factors}\label{subsec:latent}

In this section, I discuss the properties of the factors obtained from the different dimension reduction techniques presented in Section \ref{sec:factors}. 
Since the FAVAR model is based on the assumption that the main dynamics of the economy can be captured in a lower dimensional representation, the latent factors play a key role. Depending on the method used to extract them, they may significantly differ in processing the signals they receive from variables in the underlying dataset. I focus on two aspects: First, I compare the shape of each factor obtained from the three dimension reduction techniques. This way, I can shed light on the effect of potential non-linearities. Second, I identify the 15 most important variables characterizing the factors, which allows to give them an economic meaning. For each factor I provide a figure comprised of four plots including the respective time series in the upper panel and the variable importance measures in the lower panel. Given that the dataset is extensive and includes a large number of variables, I also summarize variable importance for each factor across groups \citep[as specified in ][]{mccracken2020fred} to gain a digestable overview.


\begin{figure}[htb!]
\caption{Importance of different groups to latent factors. \label{fig:vi_groups}}

\begin{minipage}{0.49\textwidth}
\centering
\hspace{5em} \textit{2019Q4}
\end{minipage}
\begin{minipage}{0.49\textwidth}
\centering
\hspace{5em} \textit{2020Q4}
\end{minipage}

\begin{minipage}{\textwidth}
\centering
\hspace{5em} (1) \textit{Linear FAVAR}
\end{minipage}

\begin{minipage}{0.49\textwidth}
\centering
\includegraphics[scale=0.44]{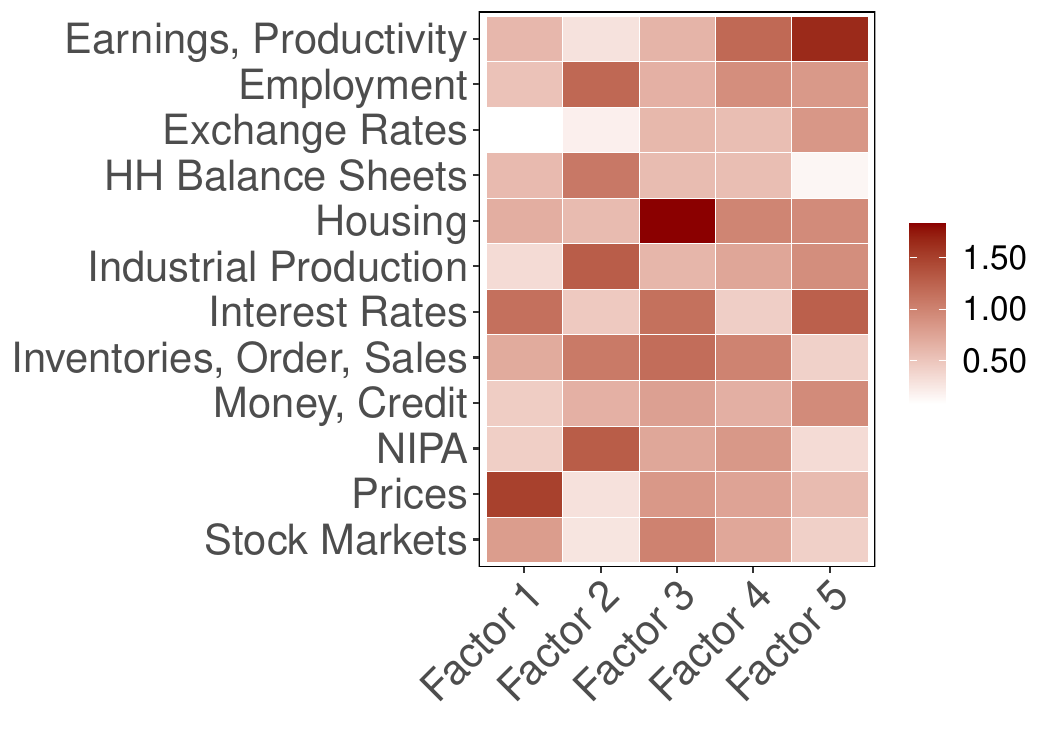}
\end{minipage}
\begin{minipage}{0.49\textwidth}
\centering
\includegraphics[scale=0.44]{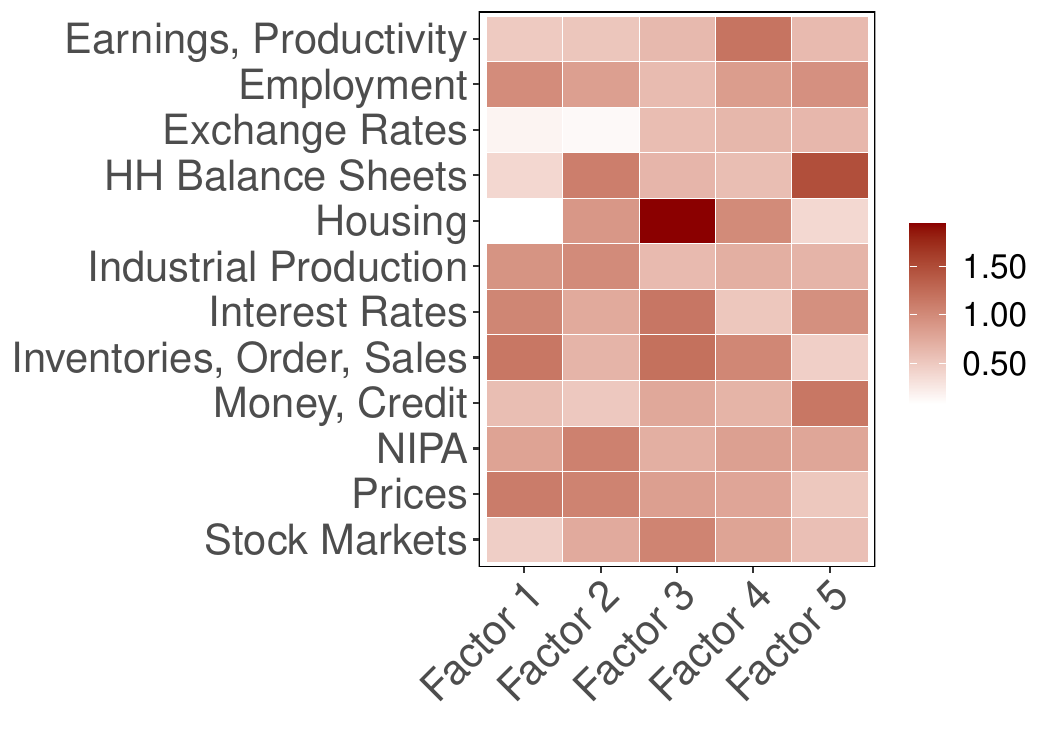}
\end{minipage}

\begin{minipage}{\textwidth}
\centering
\hspace{5em} (2) \textit{Locally Embedded FAVAR}
\end{minipage}

\begin{minipage}{0.49\textwidth}
\centering
\includegraphics[scale=0.44]{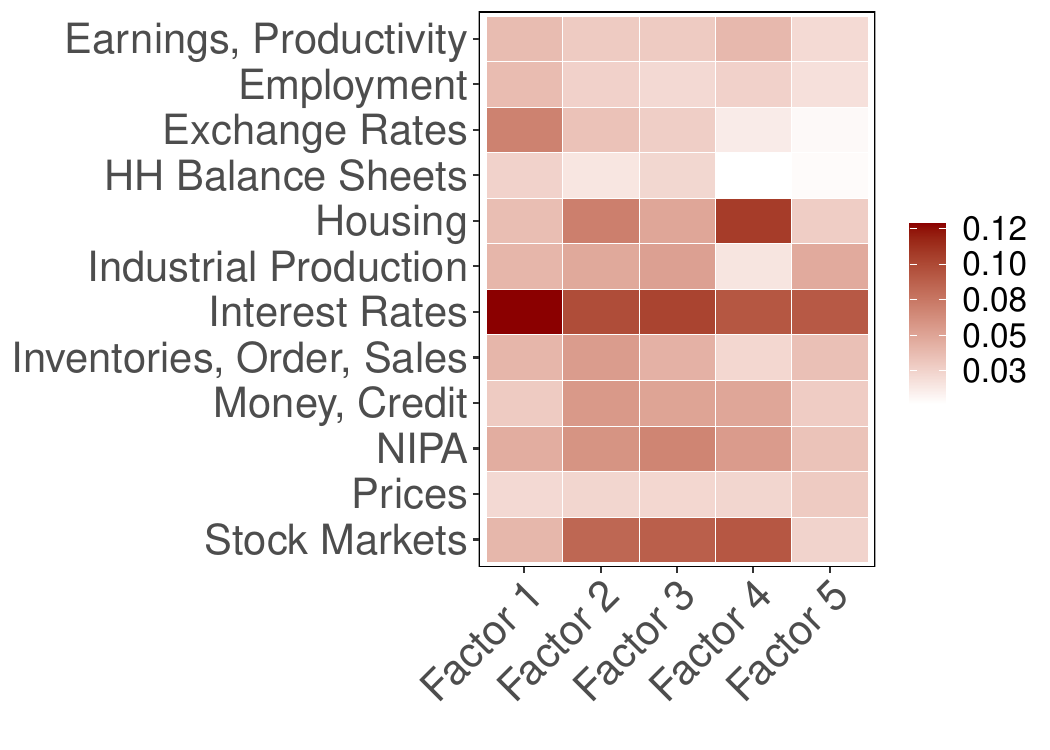}
\end{minipage}
\begin{minipage}{0.49\textwidth}
\centering
\includegraphics[scale=0.44]{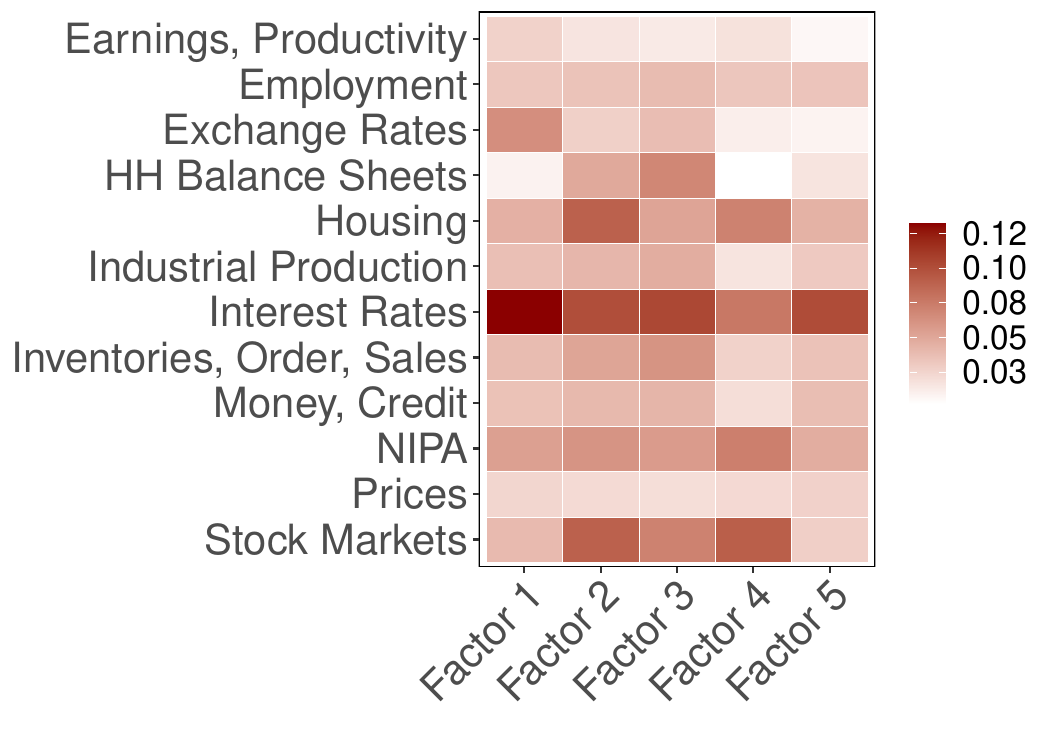}
\end{minipage}

\begin{minipage}{\textwidth}
\centering
\hspace{5em} (3) \textit{Deep Dynamic FAVAR}
\end{minipage}

\begin{minipage}{0.49\textwidth}
\centering
\includegraphics[scale=0.44]{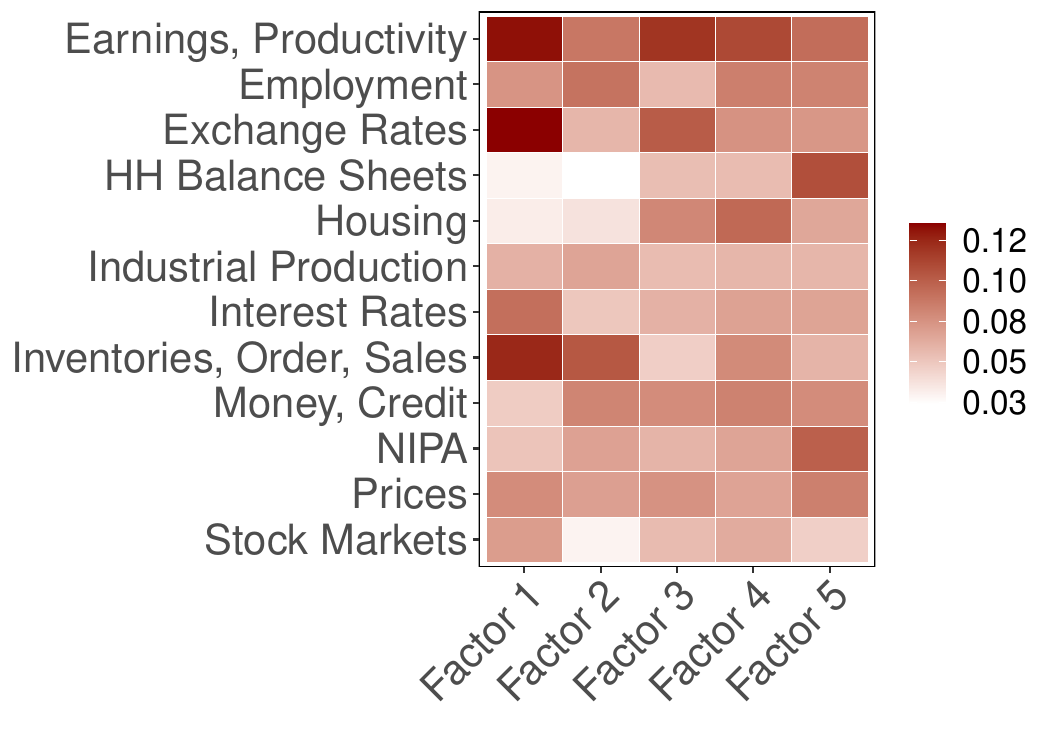}
\end{minipage}
\begin{minipage}{0.49\textwidth}
\centering
\includegraphics[scale=0.44]{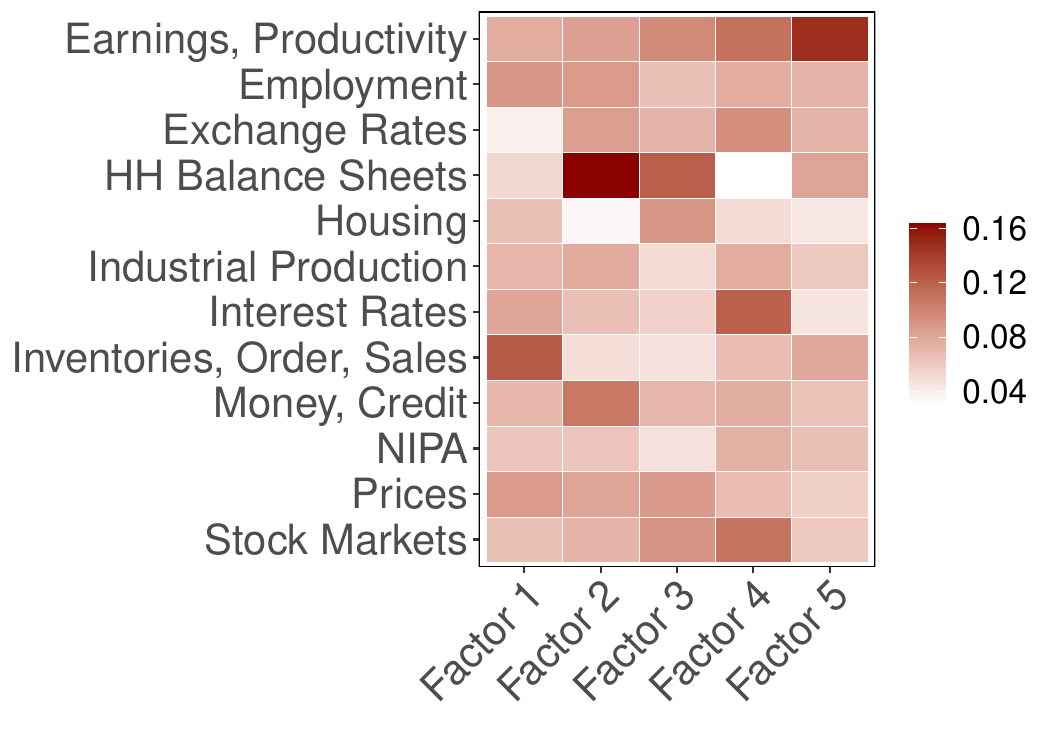}
\end{minipage}

\begin{minipage}{\textwidth}
\footnotesize
    \emph{Note:} The graph shows variable importance according to factor loadings for the linear FAVAR and the Locally Embedded FAVAR and Shapley values for the Deep Dynamic FAVAR. Values are averaged across variables belonging to a specific group defined as in \cite{mccracken2020fred}. \textit{HH} stands for households and \textit{NIPA} is the National Income and Product Accounts.
\end{minipage}

\end{figure}

Since I use linear and non-linear techniques to reduce the dimension of the dataset, the variable importance measure needs to be adapted accordingly. The variable importance for the linear FAVAR as well as the Locally Embedded FAVAR is given by absolute factor loadings. 
Due to its non-linear structure the Locally Embeddded FAVAR needs minor modifications, which I borrow from the Neighborhood Preserving Embedding algorithm proposed by \cite{he2005neighborhood}. The main modification boils down to a linear approximation of the last step (i.e., Eq. \ref{eq:LLE factors}) in the LLE algorithm. The resulting standard eigenvalue problem allows to interpret the factor loadings in a similar manner to those of PCA. For the Deep Dynamic FAVAR I compute Shapley values for each factor to get the contributions of individual variables \citep{shapley1953value,lundberg2017unified}. Details for both measures are given in Appendix \ref{subsec:VI}.
To achieve maximal comparability, I map factors according to their highest absolute correlation with each factor of the linear FAVAR. That is, I identify the factor of the Locally Embedded FAVAR as well as the Deep Dynamic FAVAR showing the highest absolute correlation with the first factor of the linear FAVAR and order it first. I repeat this exercise for all factors. \\

\noindent \textbf{Variable importance by group.} I start our discussion with the importance of variables summarized by groups (i.e., averaging importance measures across variables for each group and factor). Figure \ref{fig:vi_groups} presents the groups according to their importance for the factors of each FAVAR model for 2019Q4 (left panels) and 2020Q4 (right panels). Darker colors indicate higher weights across the variables forming a specific group. 

In general, I see that the factors cover variables from all groups. Some factors can be assigned to a specific group, others are influenced by variables stemming from various groups. For the linear FAVAR estimated with data ending before the COVID-19 pandemic, I can identify a few main drivers for each factor. The first factor is driven by price data, the second one by real activity growth. Housing variables explain the third factor whereas the fourth and fifth factors are influenced by earnings and productivity. When including the year 2020 the ranking of the main groups per factor is less clear, except for the third factor which is still mainly driven by developments in the housing sector. For the first factor, for example, I get similar weights on real activity measures, such as employment and industrial production but also on interest rates and prices. The last factor is mainly driven by the households balance sheets and monetary variables. 

For the non-linear cases the following picture arises. The Locally Embedded FAVAR puts high weight on interest rates and stock market variables for all factors. This holds for the periods ending before and after the COVID-19 crisis. The factors estimated within the Deep Dynamic FAVAR up to 2019Q4 show high weights on variables from earnings and productivity (Factor 1, 3 and 4) as well as exchange rates and order positions (Factor 1). Factor 2 is also influenced by order positions and inventories whereas the fifth factor is driven by balance sheet measures. When including 2020 I can identify Factor 1 being driven by inventories, orders and sales, Factor 2 and 3 representing the financial situation of households via balance sheet measures and Factor 4 being influenced by interest rates and stock market variables. Factor 5 is shaped by developments in earnings and productivity. \\

\noindent \textbf{Variable importance and time series behavior.} In the proposed general FAVAR setup, which includes linear and non-linear factor structures, it is not only important to identify the main drivers of the factors but also how the models extract the signals provided by variables with high weights. Hence, in the following I study closely the shape of the factors (time series plot in the upper panel) along with the main drivers (barplots in the lower panel). Figures for the sample including the COVID-19 period are relegated to Appendix \ref{sec:App Res}.

\begin{figure}[htb!]
\caption{First latent factor arising from linear and non-linear dimension reduction techniques and corresponding variable importance for 2019Q4. \label{fig:vi_factor1_2019}}

\begin{minipage}{\textwidth}
\centering
\includegraphics[scale=.42]{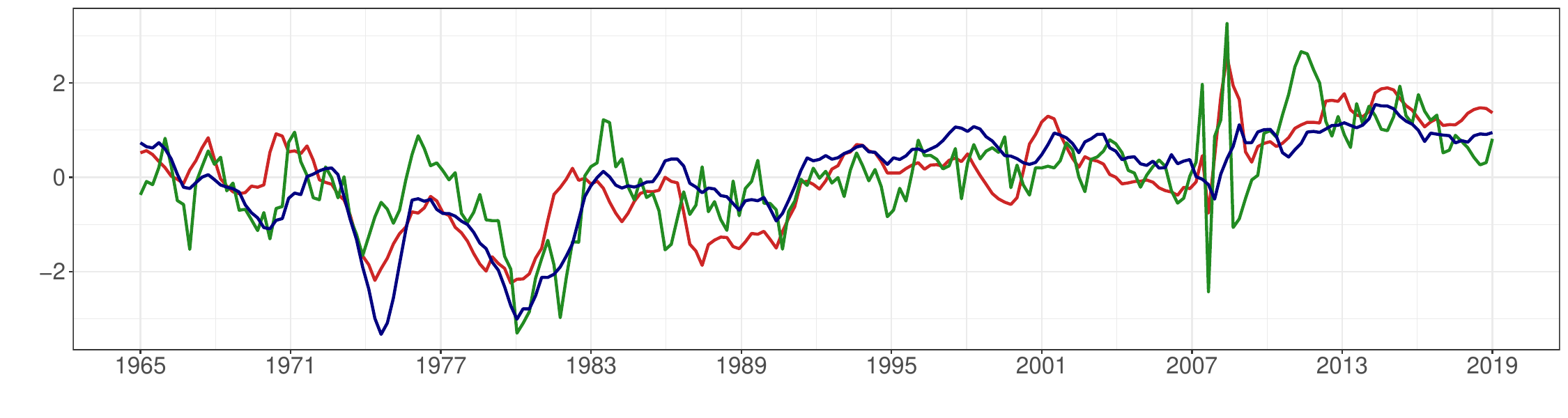}
\end{minipage}
\begin{minipage}{\textwidth}
\vspace{-1.6cm}
\centering
\hspace{1.5cm} \includegraphics[scale=.55]{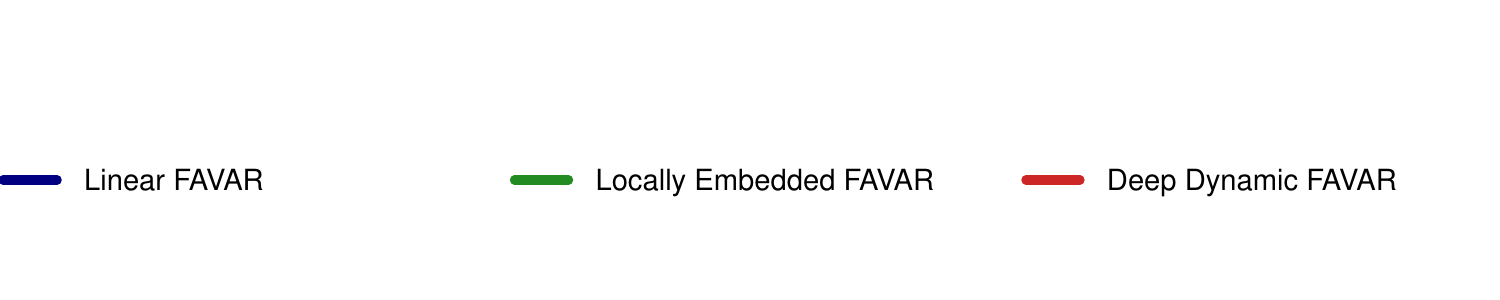}
\vspace{-0.7cm}
\end{minipage}


\begin{minipage}{0.33\textwidth}
\centering
\includegraphics[scale=0.42]{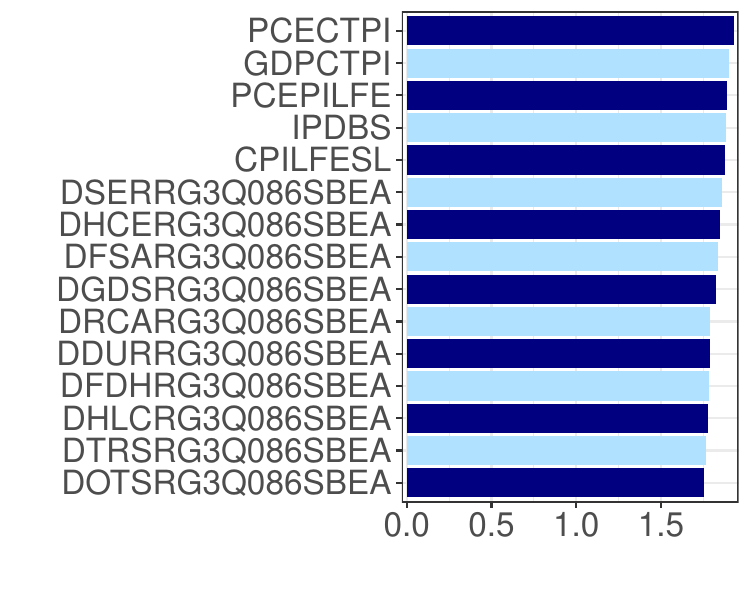}
\end{minipage}
\begin{minipage}{0.33\textwidth}
\centering
\includegraphics[scale=0.42]{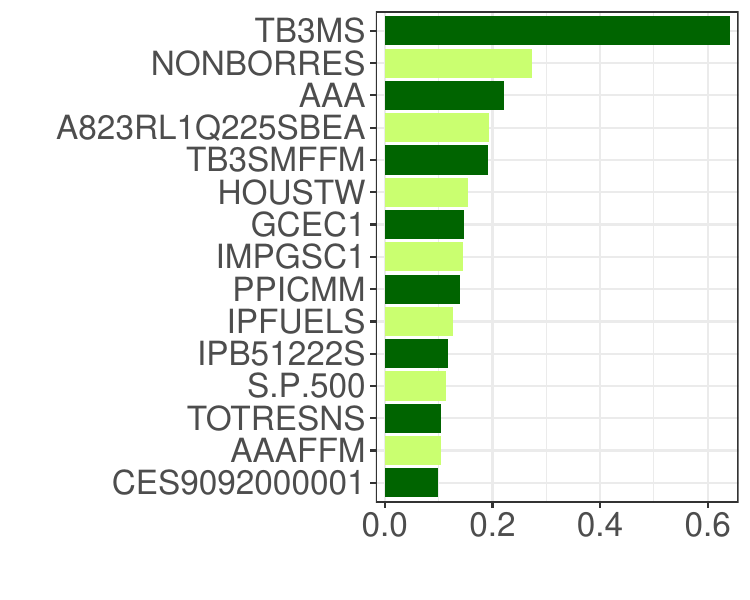}
\end{minipage}
\begin{minipage}{0.33\textwidth}
\centering
\includegraphics[scale=0.42]{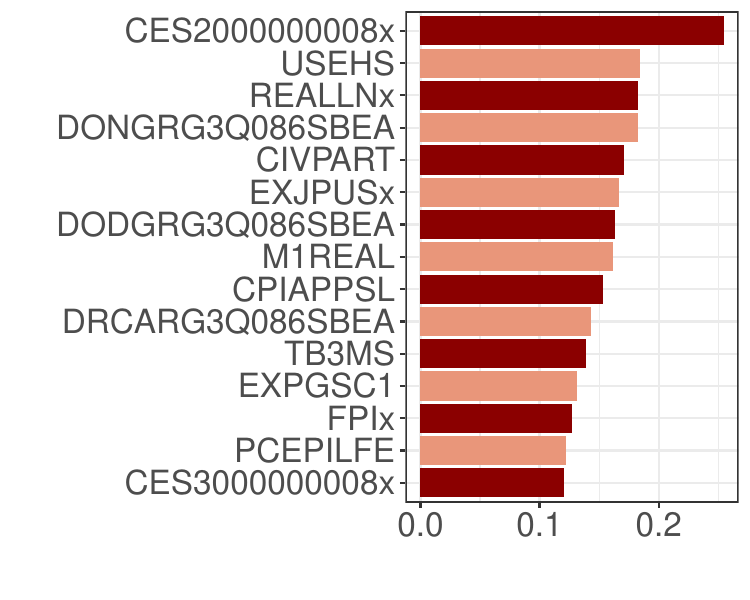}
\end{minipage}

\begin{minipage}{\textwidth}
\footnotesize
    \emph{Note:} The upper panel depicts normalized factors of the three dimension reduction techniques (mapped according to the highest correlation) with mean zero and variance one obtained from the main dataset ($N=162$) ranging from $1965$Q1 to $2019$Q4. The barplots show the 15 most important variables for each factor measured by the factor loadings for the linear FAVAR and the Locally Embedded FAVAR and Shapley values for the Deep Dynamic FAVAR. Mnemonics are those of \cite{mccracken2020fred} and can be found in Appendix \ref{sec:App Data}.
\end{minipage}

\end{figure}

For the first factor up to 2019Q4, Figure \ref{fig:vi_factor1_2019} shows that all three approaches estimate strongest movements between the mid-1970s and the mid-1980s as well as during the Global Financial Crisis (GFC). This implies that the models successfully detect times of high uncertainty in the large-dimensional dataset. The linear factor attaches a lower degree of severity to the GFC than the other two models, which can be explained by the fact that it is mainly driven by inflation series. Top-15 variables for the first linear factor include consumer price indices as well as personal consumption expenditure price indices. On the contrary, the locally embedded factor shows high volatility during the GFC. This can be traced back to interest rates, money stock variables and bond yields, mainly influencing its shape. Given the composition and prevailing conditions of the GFC, monetary and financial variables were exposed to higher fluctuations than inflation. Among the most important variables shaping the deep dynamic factor I find real hourly earnings, employment and exchange rates. This explains the rather strong reaction to the GFC but to a lesser extent than the factor of the Locally Embedded FAVAR. When including the COVID-19 observations, there is little difference between the three methods (see Figure \ref{fig:vi_factor1_2020} in the appendix), especially, between the linear and the deep learning case. Both factors follow a very similar course over time and also show similarities with respect to variable importance. Main drivers are employment, business inventories and price series. The Locally Embedded FAVAR, on the other hand, is mainly driven by interest rates and variables from the national accounts.

\begin{figure}[htb!]
\caption{Second latent factor arising from linear and non-linear dimension reduction techniques and corresponding variable importance for 2019Q4. \label{fig:vi_factor2_2019}}

\begin{minipage}{\textwidth}
\centering
\includegraphics[scale=.42]{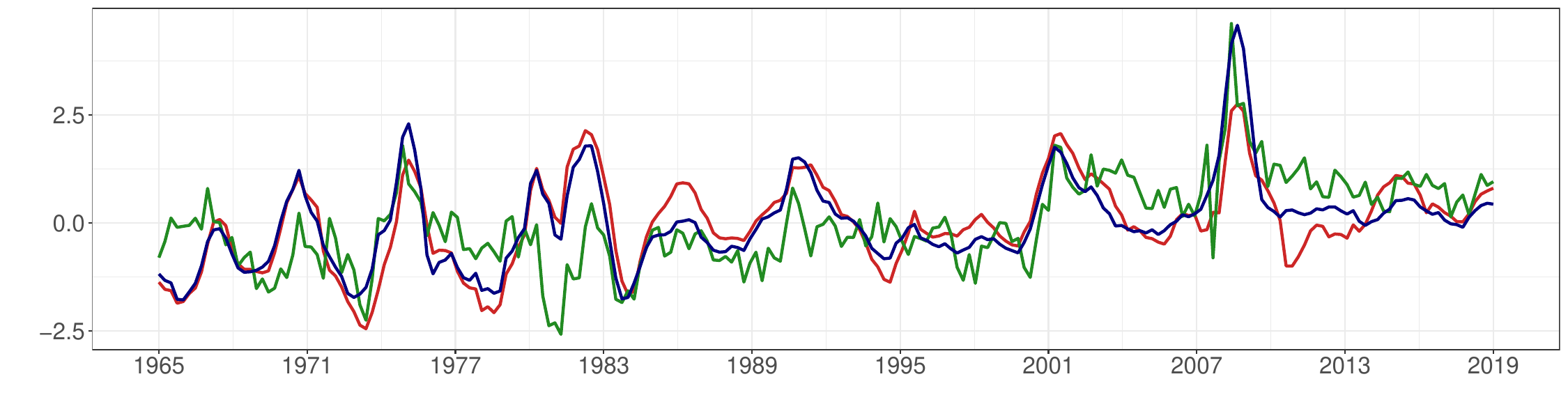}
\end{minipage}
\begin{minipage}{\textwidth}
\vspace{-1.6cm}
\centering
\hspace{1.5cm} \includegraphics[scale=.55]{legend_col.pdf}
\vspace{-0.7cm}
\end{minipage}

\begin{minipage}{0.33\textwidth}
\centering
\includegraphics[scale=0.42]{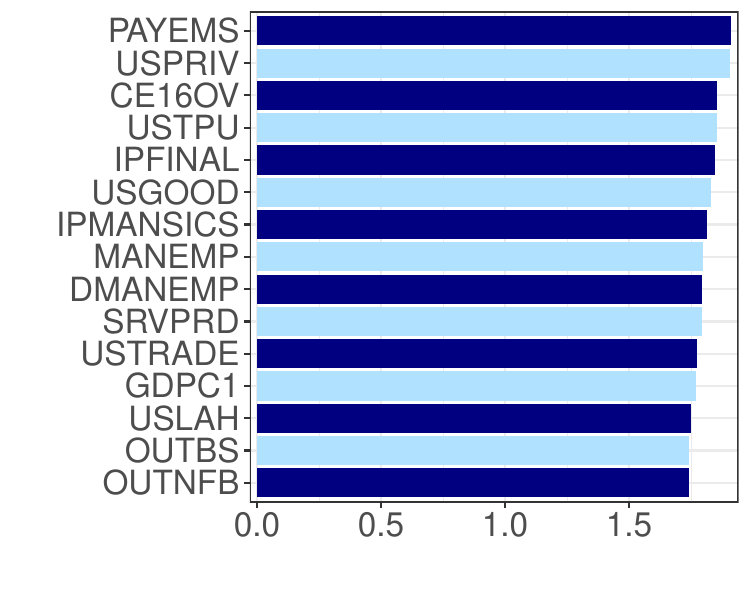}
\end{minipage}
\begin{minipage}{0.33\textwidth}
\centering
\includegraphics[scale=0.42]{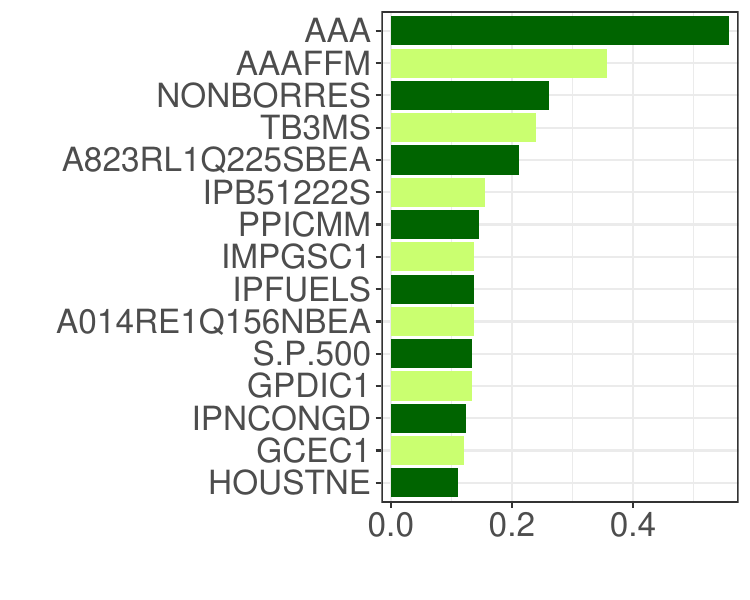}
\end{minipage}
\begin{minipage}{0.33\textwidth}
\centering
\includegraphics[scale=0.42]{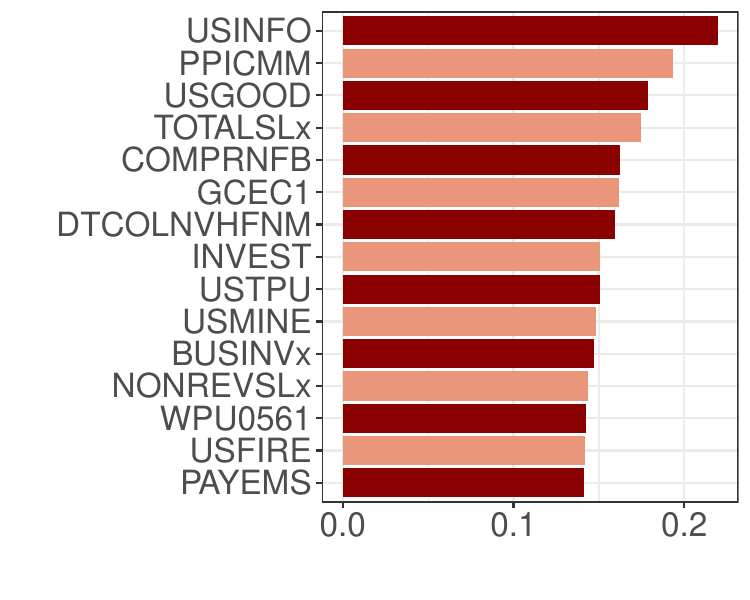}
\end{minipage}

\begin{minipage}{\textwidth}
\footnotesize
    \emph{Note:} For more details I refer to Figure \ref{fig:vi_factor1_2019}.
\end{minipage}

\end{figure}

Turning to the second factor, again, all factors differentiate between crisis and tranquil times, although attaching more weight on the dotcom bubble than in the previous case (see Figure \ref{fig:vi_factor2_2019}). Moreover, the linear factor shows the largest outlier during the GFC, since it is now mainly driven by real acitivty variables (i.e., industrial production and employment), which were prone to high uncertainty during that time. The deep dynamic factor resembles the linear one with a lower reaction to the GFC. The most important variables include variables measuring the employment situation, producer price index for commodities and outstanding credits. The factor obtained from locally linear embedding shows the highest volatility and is again influenced by interest rates and nonborrowed reserves. Including data up to 2020Q4, as presented in Figure \ref{fig:vi_factor2_2020} in the appendix, changes the top variables for the linear factor to price and employment series and for the deep dynamic factor to variables measuring the employment situation, outstanding credit and money stocks. In the case of the locally embedded factor I identify stock market and housing variables in addition to the recurring importance of interest rates.

Inspecting the third factor (Figure \ref{fig:vi_factor3_2019} and Figure \ref{fig:vi_factor3_2020} in the appendix) reveals some interesting pattern of the linear model. It clearly peaks in advance to the non-linear methods for most crises, most importantly, for the Volcker period and the GFC. In the linear FAVAR model, Factor 3 is mainly driven by housing variables, which are often considered as leading indicators in the literature \citep[see, e.g., ][]{stock1989new,marcellino2006leading}. The locally embedded factor up to 2019Q4 shows a similar behavior to its second counterpart with the most important variables being interest rates, housing starts and real disposable income. When including the COVID-19 observations I additionally observe importance of duration of unemployment. The deep dynamic factor is driven by employment variables and real money stocks for the case excluding COVID-19 periods. Considering the COVID-19 pandemic shifts highest importance to interest rates, real hourly earnings and dividend yield of the S\&P 500 stock market index.

\begin{figure}[htb!]
\caption{Third latent factor arising from linear and non-linear dimension reduction techniques and corresponding variable importance for 2019Q4. \label{fig:vi_factor3_2019}}

\begin{minipage}{\textwidth}
\centering
\includegraphics[scale=.42]{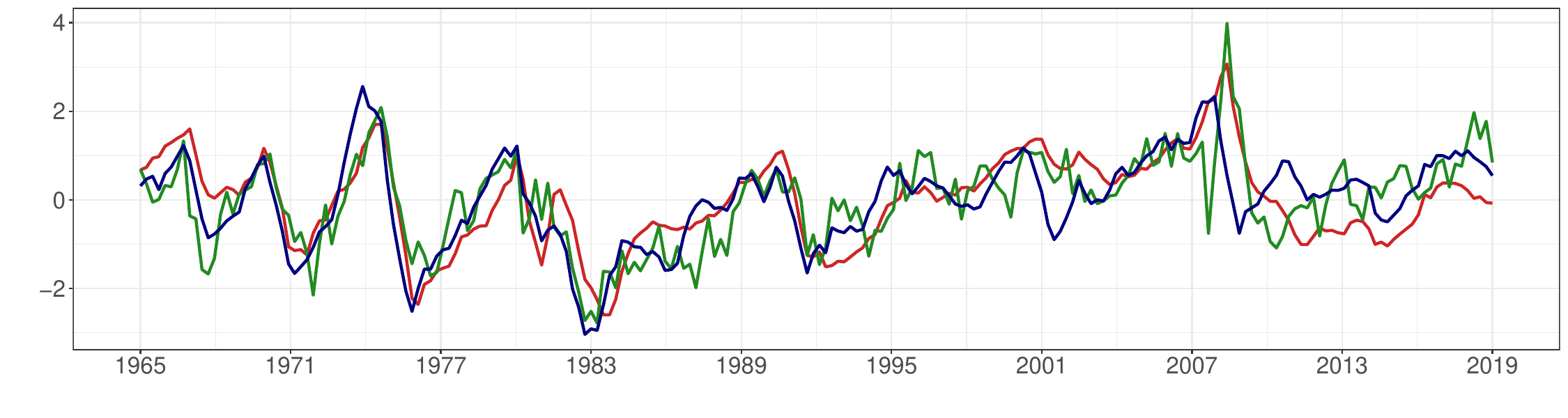}
\end{minipage}
\begin{minipage}{\textwidth}
\vspace{-1.6cm}
\centering
\hspace{1.5cm} \includegraphics[scale=.55]{legend_col.pdf}
\vspace{-0.7cm}
\end{minipage}

\begin{minipage}{0.33\textwidth}
\centering
\includegraphics[scale=0.42]{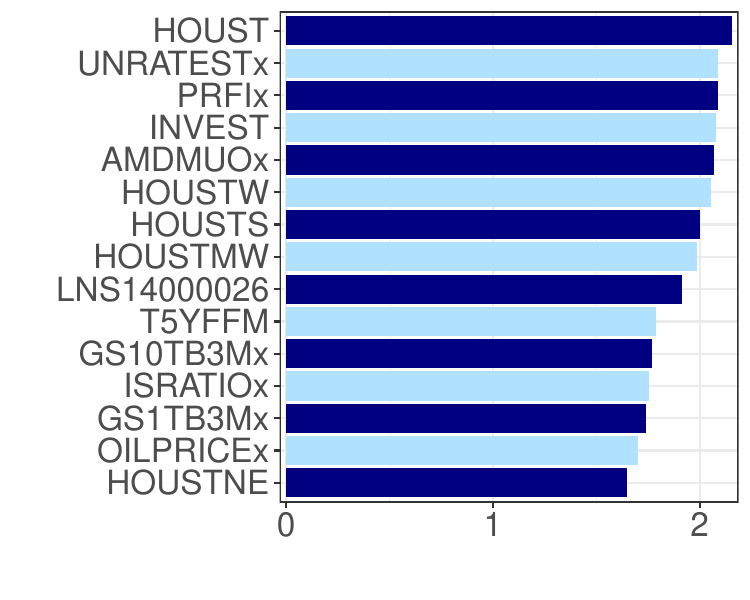}
\end{minipage}
\begin{minipage}{0.33\textwidth}
\centering
\includegraphics[scale=0.42]{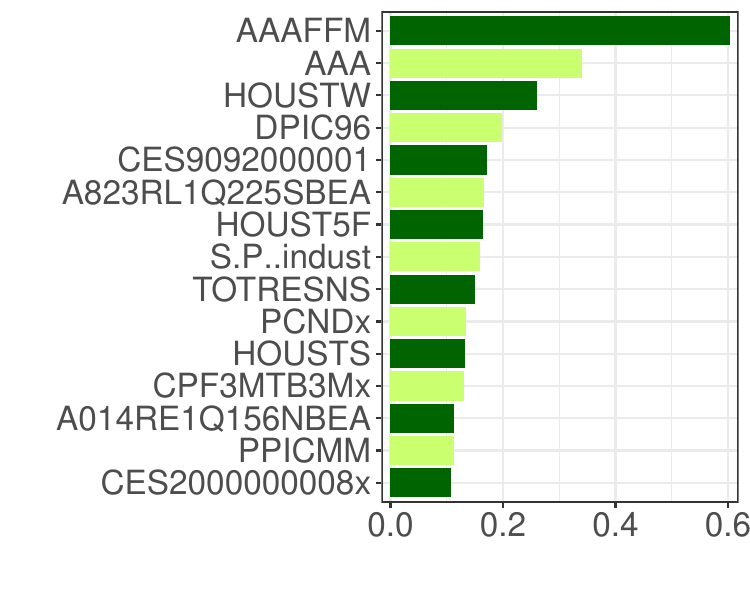}
\end{minipage}
\begin{minipage}{0.33\textwidth}
\centering
\includegraphics[scale=0.42]{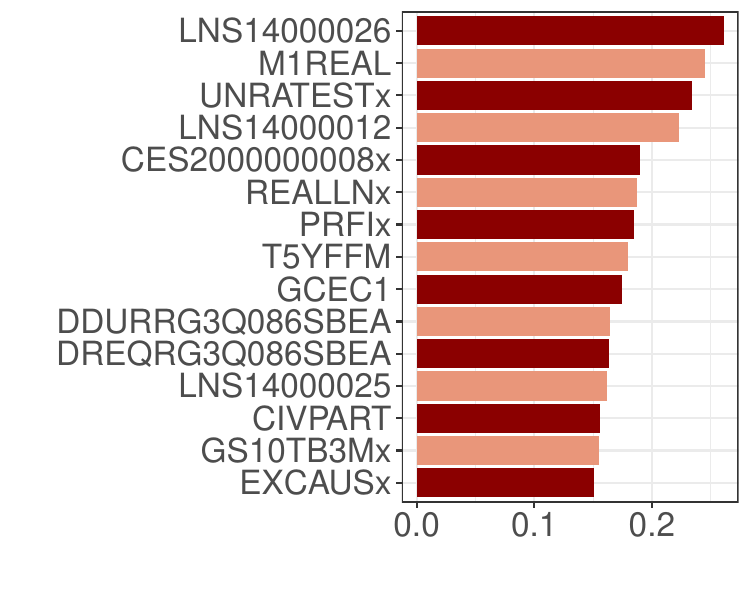}
\end{minipage}

\begin{minipage}{\textwidth}
\footnotesize
    \emph{Note:} For more details I refer to Figure \ref{fig:vi_factor1_2019}.
\end{minipage}

\end{figure}

Compared to Factors 1, 2 and 3, the correlation between Factors 4 obtained from the different dimension reduction techniques decreases significantly (see Figure \ref{fig:vi_factor4_2019} and Figure \ref{fig:vi_factor4_2020} in the appendix). The Deep Dynamic FAVAR yields a relatively smooth factor without major fluctuations or outliers. On the contrary, the locally embedded factor is highly volatile and, similar to the linear case, spikes during the GFC. This holds for both sample periods considered. The most important variables describing the linear factor for both samples include government consumption and employment/unemployment measures. As for most factors obtained from locally linear embedding, the locally embedded factor is driven by interest rates and among the first three variables I also find employment. For the Deep Dynamic FAVAR considering the sample up to 2019Q4 most important variables are overtime hours, business inventories and average hourly earnings whereas for up to 2020Q4 real hourly earnings and other statistics for employment as well as government expenditure influence the factor most.

\begin{figure}[!h]
\caption{Fourth latent factor arising from linear and non-linear dimension reduction techniques and corresponding variable importance for 2019Q4. \label{fig:vi_factor4_2019}}

\begin{minipage}{\textwidth}
\centering
\includegraphics[scale=.42]{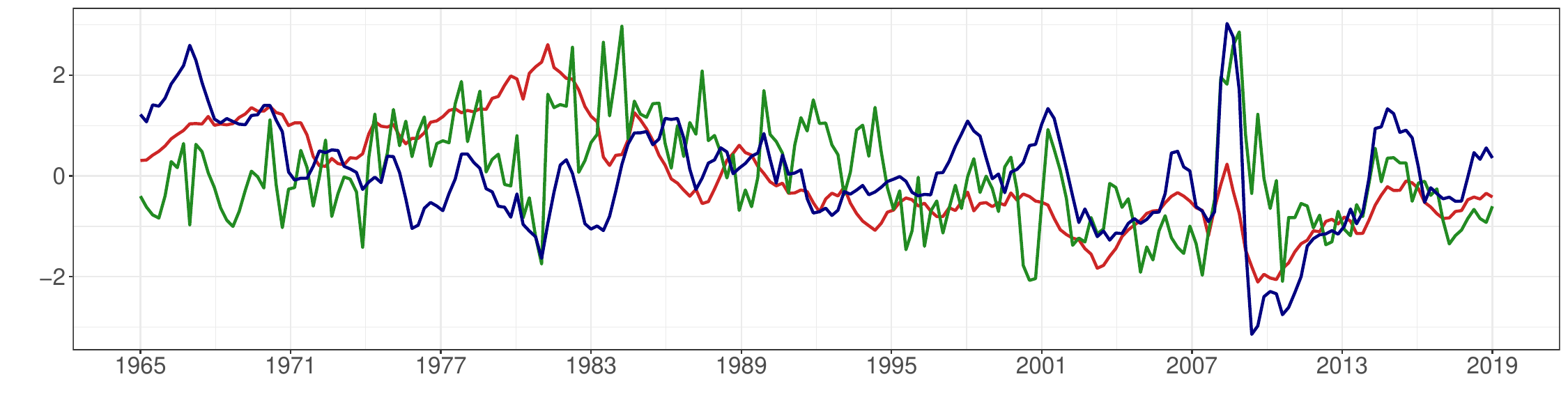}
\end{minipage}
\begin{minipage}{\textwidth}
\vspace{-1.6cm}
\centering
\hspace{1.5cm} \includegraphics[scale=.55]{legend_col.pdf}
\vspace{-0.7cm}
\end{minipage}

\begin{minipage}{0.33\textwidth}
\centering
\includegraphics[scale=0.42]{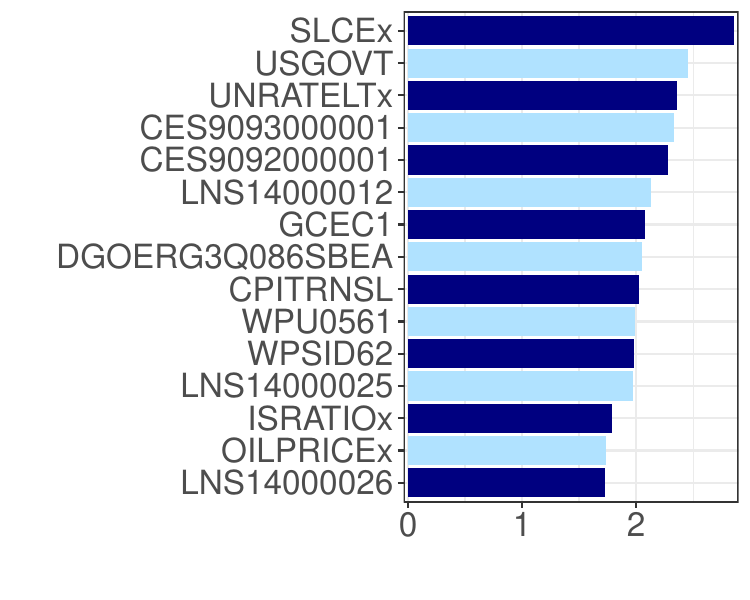}
\end{minipage}
\begin{minipage}{0.33\textwidth}
\centering
\includegraphics[scale=0.42]{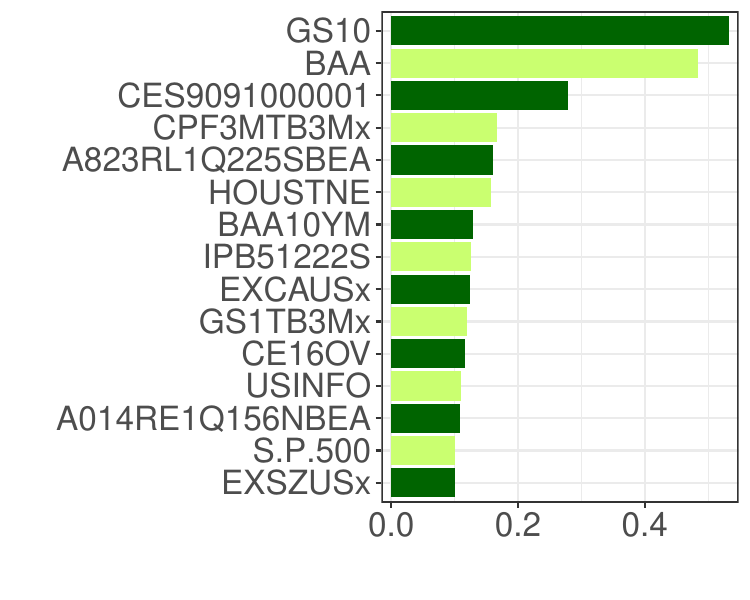}
\end{minipage}
\begin{minipage}{0.33\textwidth}
\centering
\includegraphics[scale=0.42]{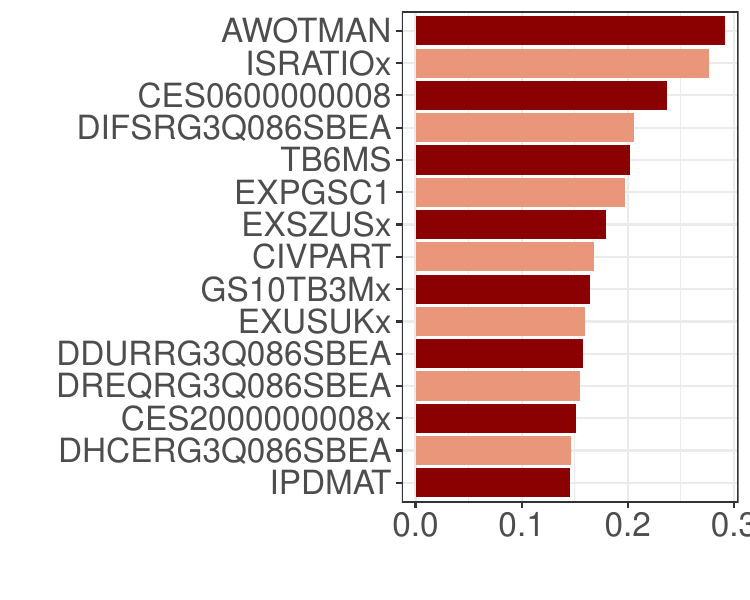}
\end{minipage}

\begin{minipage}{\textwidth}
\footnotesize
    \emph{Note:} For more details I refer to Figure \ref{fig:vi_factor1_2019}.
\end{minipage}

\end{figure}

\begin{figure}[!h]
\caption{Fifth latent factor arising from linear and non-linear dimension reduction techniques and corresponding variable importance for 2019Q4. \label{fig:vi_factor5_2019}}

\begin{minipage}{\textwidth}
\centering
\includegraphics[scale=.42]{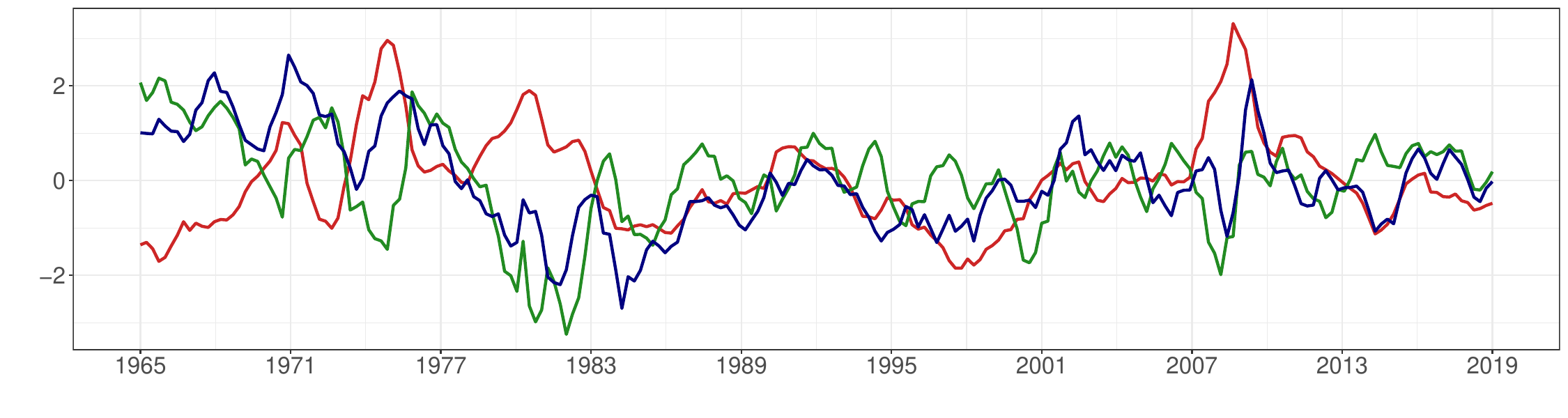}
\end{minipage}
\begin{minipage}{\textwidth}
\vspace{-1.6cm}
\centering
\hspace{1.5cm} \includegraphics[scale=.55]{legend_col.pdf}
\vspace{-0.7cm}
\end{minipage}

\begin{minipage}{0.33\textwidth}
\centering
\includegraphics[scale=0.42]{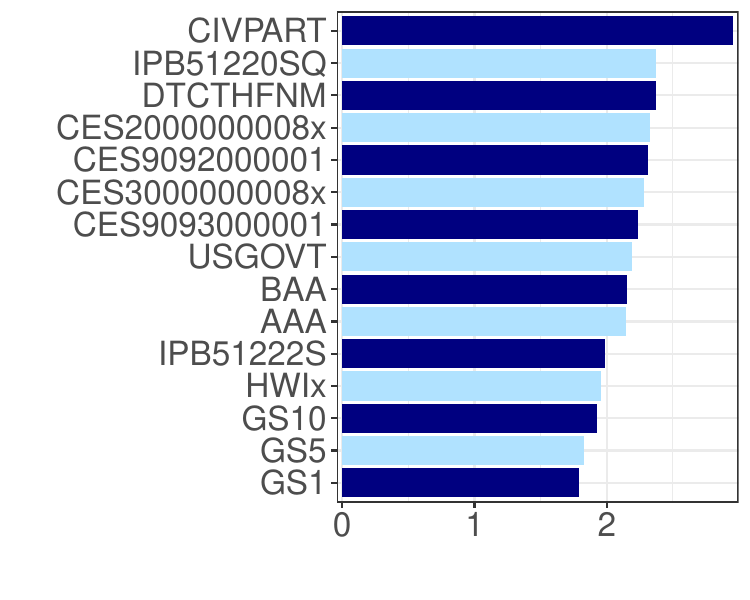}
\end{minipage}
\begin{minipage}{0.33\textwidth}
\centering
\includegraphics[scale=0.42]{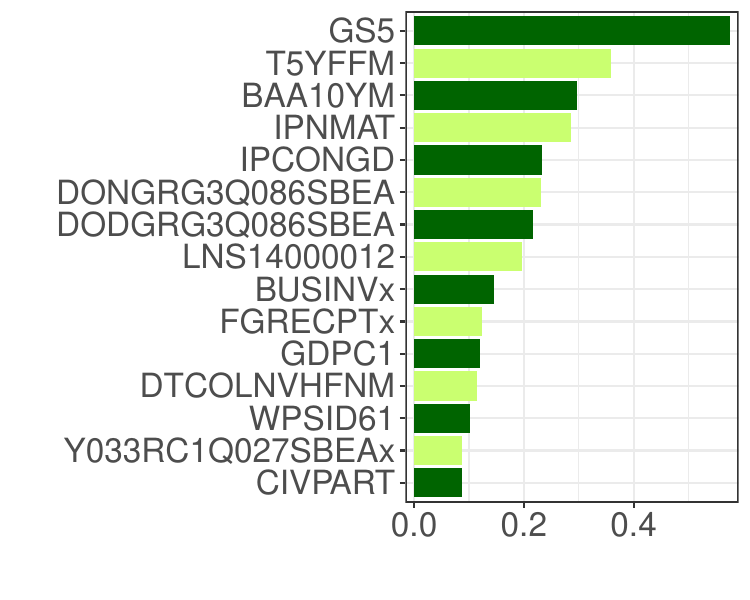}
\end{minipage}
\begin{minipage}{0.33\textwidth}
\centering
\includegraphics[scale=0.42]{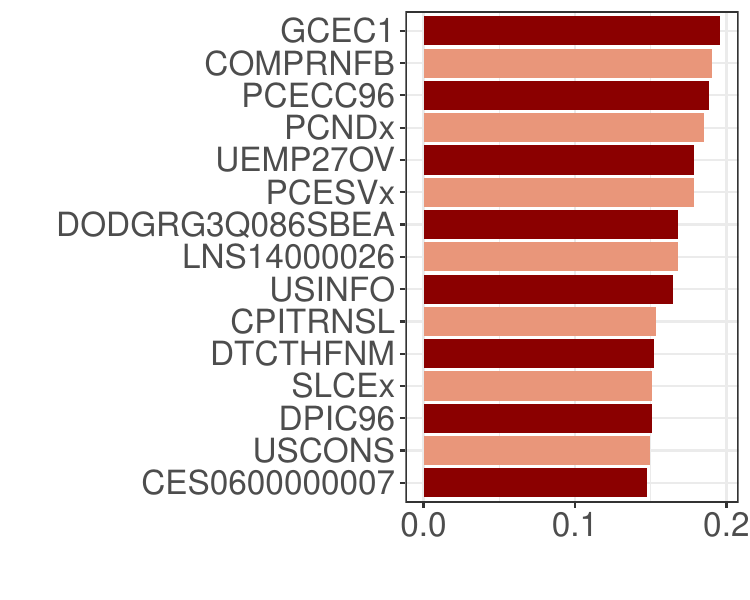}
\end{minipage}

\begin{minipage}{\textwidth}
\footnotesize
    \emph{Note:} For more details I refer to Figure \ref{fig:vi_factor1_2019}.
\end{minipage}

\end{figure}

Similarly, Figure \ref{fig:vi_factor5_2019} shows that the last factor follows its own path for each model. While the linear and the locally embedded factor follow at least the same trend, the deep dynamic factor shows a different path. Its severest peaks can be found during the Volcker period as well as during the GFC when looking at the sample up to 2019Q4, which is still true when including data up to 2020Q4 (see Figure \ref{fig:vi_factor5_2020} in the appendix). Top-3 drivers for up to 2019Q4 are government and personal consumption expenditures as well as real compensation per hour and for the full sample I get personal consumption expenditure price indices as well as the oil price. The linear FAVAR, on the other hand, reacts clearly to the COVID-19 pandemic. This can be explained by the high influence of real activity measures such as industrial production and employment as well as loans. The factor obtained from the locally embedded FAVAR estimates the highest fluctuations during the GFC. Again, this behavior may be traced to the emphasis on interest rates and monetary variables.

Summing up this discussion, I find that the linear and the deep dynamic factors behave similarly for the first three factors but differ significantly for Factor 4 and 5. Both cover various sectors of the economy and often put high focus on real activity variables, prevailing financial conditions and price developments. The locally embedded factors are characterized by higher volatility and are concentrated on monetary variables.

\subsection{Application 1 - Monetary policy shock}\label{subsec:mp}

In the first empirical application, I simulate a 100 basis points (bps) expansionary monetary policy shock and compare the impulse responses generated by the different FAVAR approaches. I trace the responses of key macroeconomic and financial variables (i.e., output growth, unemployment rate, inflation, growth of housing starts, S\&P 500 stock market index, short-term interest rate) over 16 periods (i.e., 4 years) after the shock hit the economy. 

\begin{figure}[h!]

\caption{Impulse responses of selected variables to a 100 bps expansionary monetary policy shock in 2019Q4 and 2020Q4 for the three different FAVAR approaches}\label{fig:mp}

\begin{minipage}{0.33\textwidth}
\centering
(1) \textit{Linear FAVAR}
\end{minipage}
\begin{minipage}{0.33\textwidth}
\centering
(2) \textit{Locally Embedded FAVAR}
\end{minipage}
\begin{minipage}{0.33\textwidth}
\centering
(3) \textit{Deep Dynamic FAVAR}
\end{minipage}

\begin{minipage}{0.33\textwidth}
\vspace{5pt}
\centering
\scriptsize \textit{GDPC1}
\end{minipage}
\begin{minipage}{0.33\textwidth}
\vspace{5pt}
\centering
\scriptsize \textit{GDPC1}
\end{minipage}
\begin{minipage}{0.33\textwidth}
\vspace{5pt}
\centering
\scriptsize \textit{GDPC1}
\end{minipage}

\begin{minipage}{0.33\textwidth}
\centering
\includegraphics[scale=0.38]{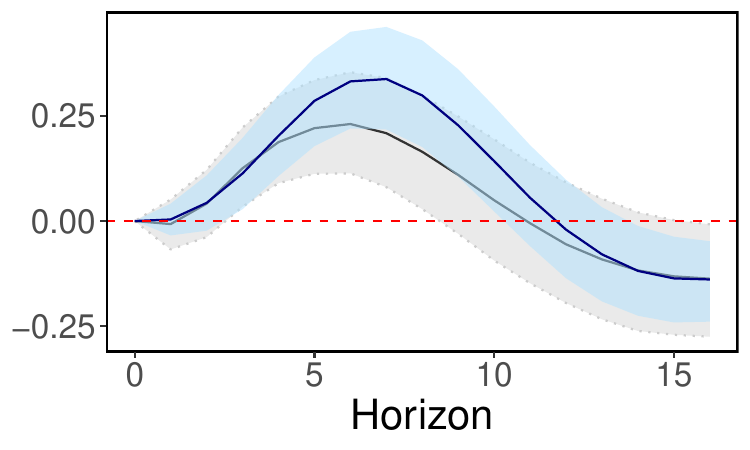}
\end{minipage}
\begin{minipage}{0.33\textwidth}
\centering
\includegraphics[scale=0.38]{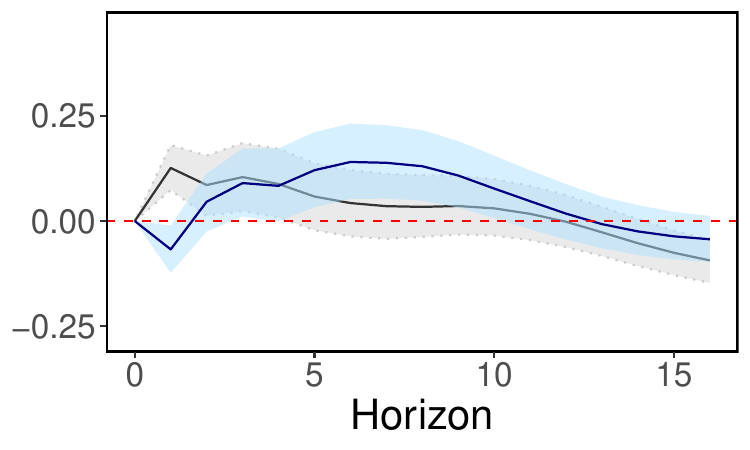}
\end{minipage}
\begin{minipage}{0.33\textwidth}
\centering
\includegraphics[scale=0.38]{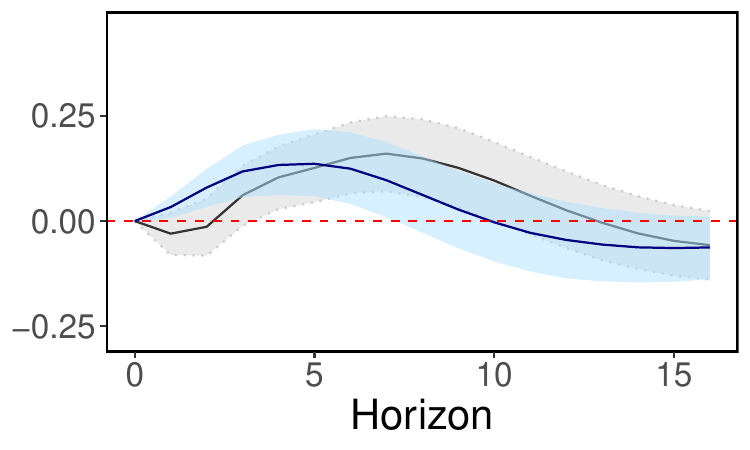}
\end{minipage}

\begin{minipage}{0.33\textwidth}
\vspace{5pt}
\centering
\scriptsize \textit{UNRATE}
\end{minipage}
\begin{minipage}{0.33\textwidth}
\vspace{5pt}
\centering
\scriptsize \textit{UNRATE}
\end{minipage}
\begin{minipage}{0.33\textwidth}
\vspace{5pt}
\centering
\scriptsize \textit{UNRATE}
\end{minipage}

\begin{minipage}{0.33\textwidth}
\centering
\includegraphics[scale=0.38]{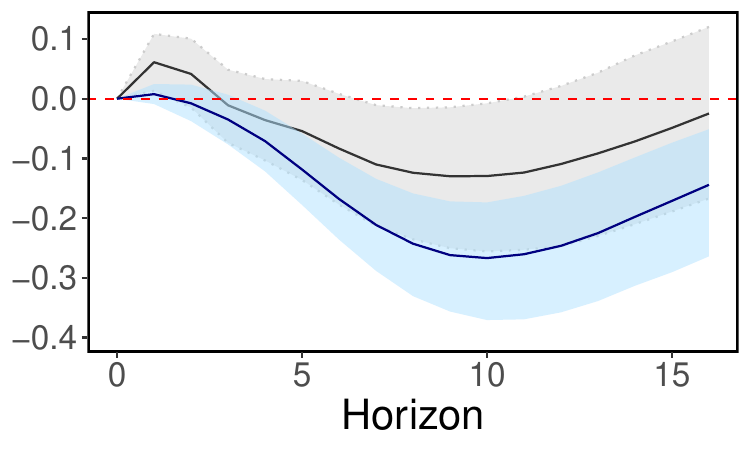}
\end{minipage}
\begin{minipage}{0.33\textwidth}
\centering
\includegraphics[scale=0.38]{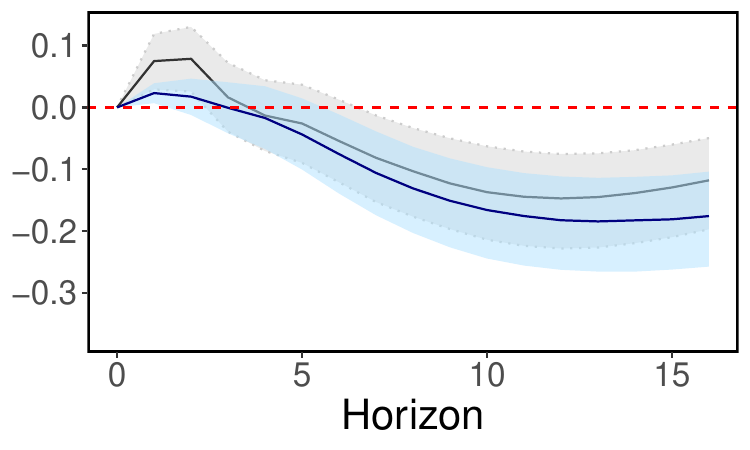}
\end{minipage}
\begin{minipage}{0.33\textwidth}
\centering
\includegraphics[scale=0.38]{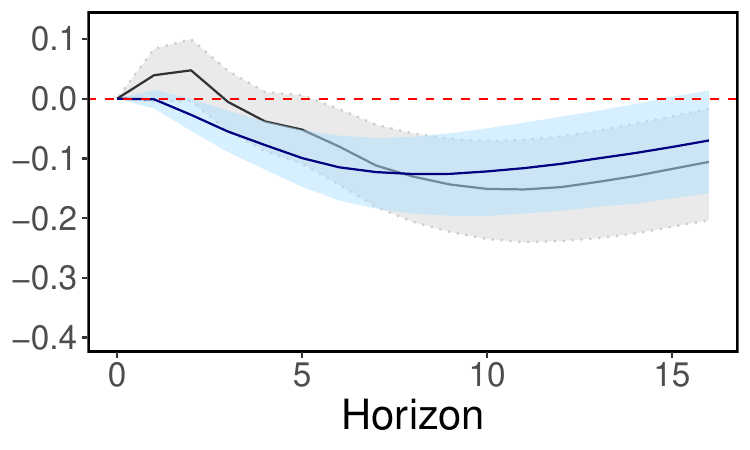}
\end{minipage}

\begin{minipage}{0.33\textwidth}
\vspace{5pt}
\centering
\scriptsize \textit{GDPCTPI}
\end{minipage}
\begin{minipage}{0.33\textwidth}
\vspace{5pt}
\centering
\scriptsize \textit{GDPCTPI}
\end{minipage}
\begin{minipage}{0.33\textwidth}
\vspace{5pt}
\centering
\scriptsize \textit{GDPCTPI}
\end{minipage}

\begin{minipage}{0.33\textwidth}
\centering
\includegraphics[scale=0.38]{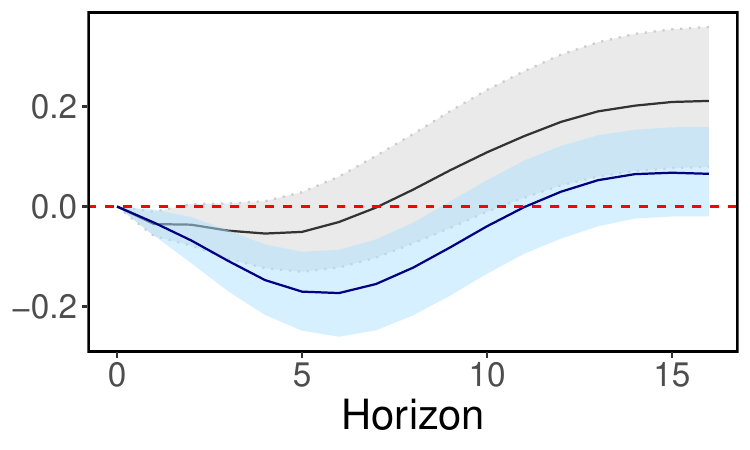}
\end{minipage}
\begin{minipage}{0.33\textwidth}
\centering
\includegraphics[scale=0.38]{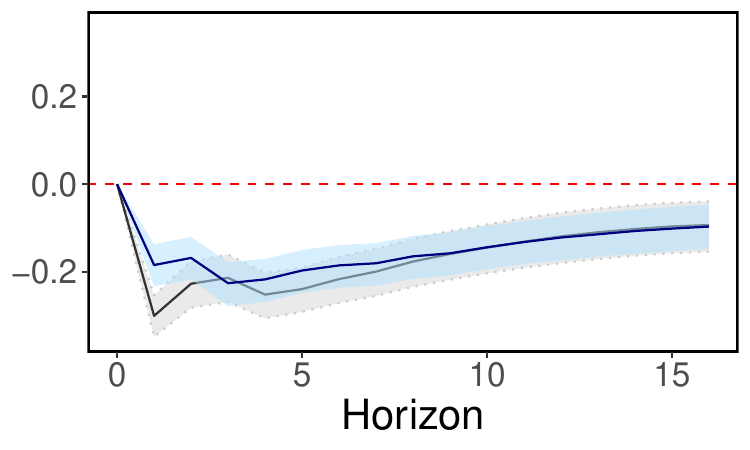}
\end{minipage}
\begin{minipage}{0.33\textwidth}
\centering
\includegraphics[scale=0.38]{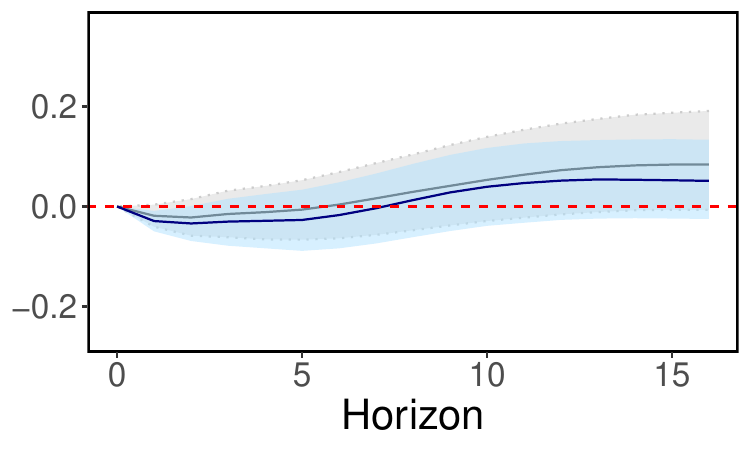}
\end{minipage}

\begin{minipage}{0.33\textwidth}
\vspace{5pt}
\centering
\scriptsize \textit{HOUST}
\end{minipage}
\begin{minipage}{0.33\textwidth}
\vspace{5pt}
\centering
\scriptsize \textit{HOUST}
\end{minipage}
\begin{minipage}{0.33\textwidth}
\vspace{5pt}
\centering
\scriptsize \textit{HOUST}
\end{minipage}

\begin{minipage}{0.33\textwidth}
\centering
\includegraphics[scale=0.37]{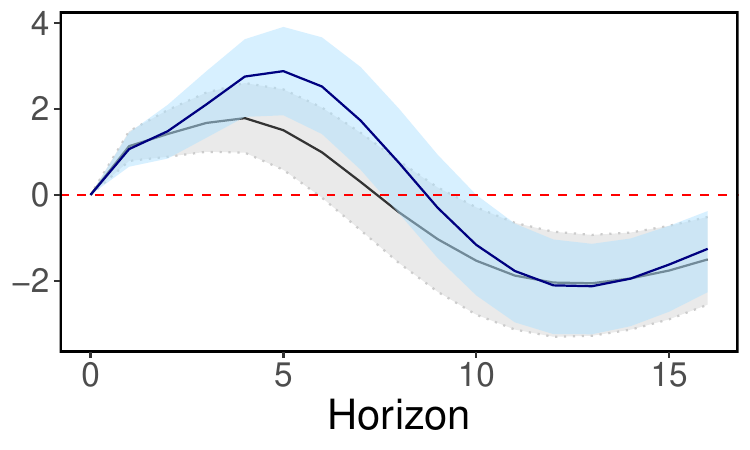}
\end{minipage}
\begin{minipage}{0.33\textwidth}
\centering
\includegraphics[scale=0.37]{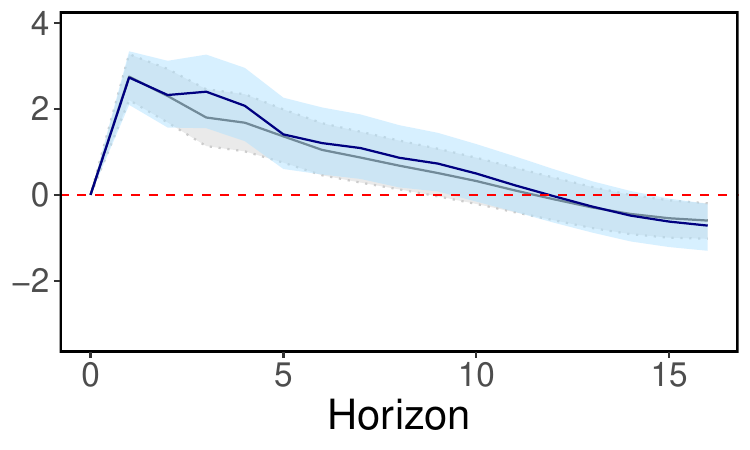}
\end{minipage}
\begin{minipage}{0.33\textwidth}
\centering
\includegraphics[scale=0.37]{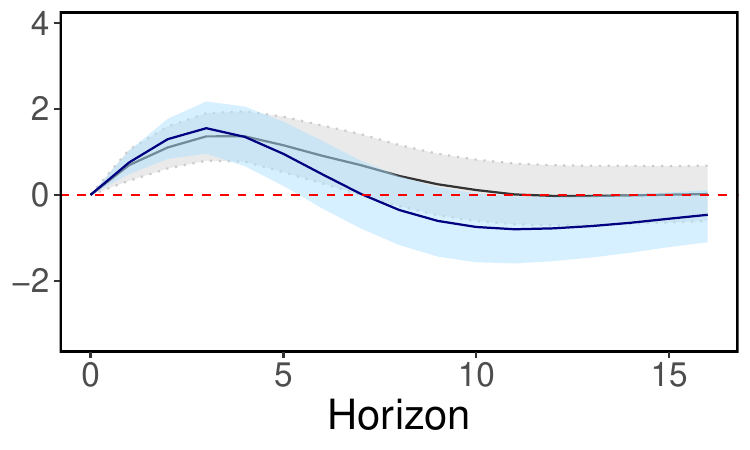}
\end{minipage}

\begin{minipage}{0.33\textwidth}
\vspace{5pt}
\centering
\scriptsize \textit{S\&P 500}
\end{minipage}
\begin{minipage}{0.33\textwidth}
\vspace{5pt}
\centering
\scriptsize \textit{S\&P 500}
\end{minipage}
\begin{minipage}{0.33\textwidth}
\vspace{5pt}
\centering
\scriptsize \textit{S\&P 500}
\end{minipage}

\begin{minipage}{0.33\textwidth}
\centering
\includegraphics[scale=0.38]{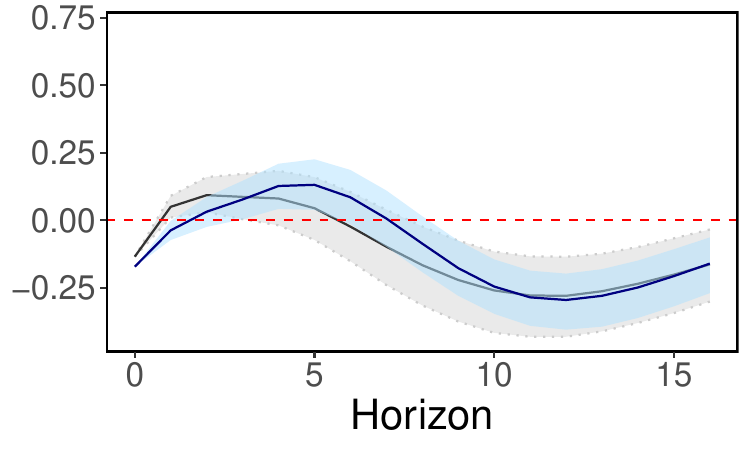}
\end{minipage}
\begin{minipage}{0.33\textwidth}
\centering
\includegraphics[scale=0.38]{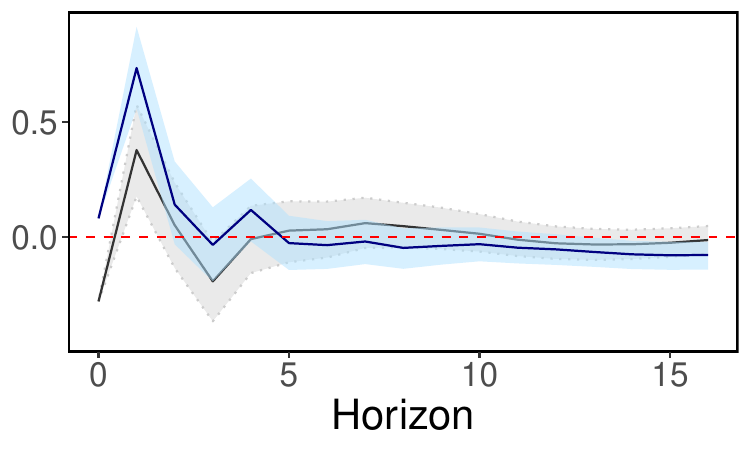}
\end{minipage}
\begin{minipage}{0.33\textwidth}
\centering
\includegraphics[scale=0.38]{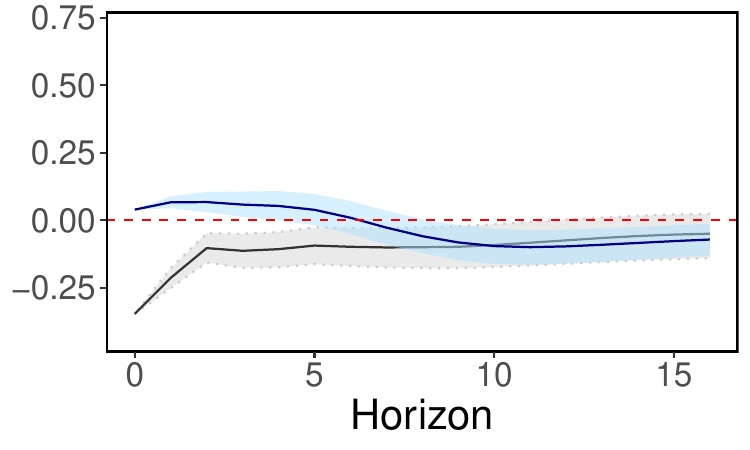}
\end{minipage}

\begin{minipage}{0.33\textwidth}
\vspace{5pt}
\centering
\scriptsize \textit{GS1}
\end{minipage}
\begin{minipage}{0.33\textwidth}
\vspace{5pt}
\centering
\scriptsize \textit{GS1}
\end{minipage}
\begin{minipage}{0.33\textwidth}
\vspace{5pt}
\centering
\scriptsize \textit{GS1}
\end{minipage}

\begin{minipage}{0.33\textwidth}
\centering
\includegraphics[scale=0.38]{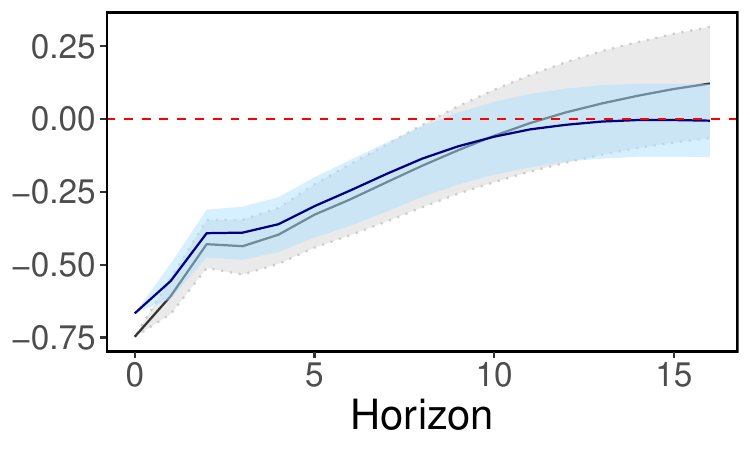}
\end{minipage}
\begin{minipage}{0.33\textwidth}
\centering
\includegraphics[scale=0.38]{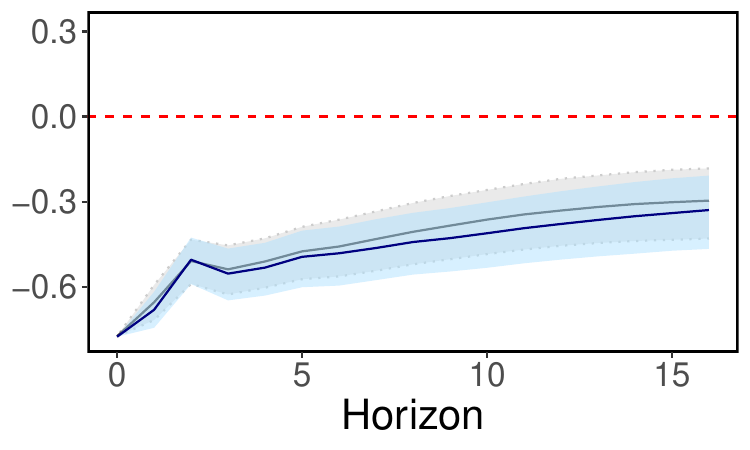}
\end{minipage}
\begin{minipage}{0.33\textwidth}
\centering
\includegraphics[scale=0.38]{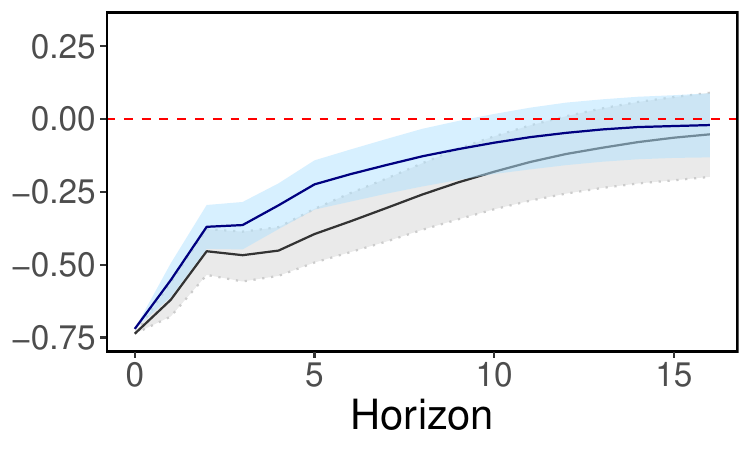}
\end{minipage}

\begin{minipage}{\textwidth}
    \footnotesize
    \emph{Note:} The blue solid lines refer to the median response and the light blue shaded area indicates the $16$th and $84$th percentiles of the posterior distribution for the data up to 2019Q4. The impulse responses including the COVID-19 observations are shown in grey. The median responses are presented in the black solid lines and the confidence bands in form of the grey shaded area. The red dashed line shows the zero line. 
\end{minipage}
\end{figure}

Figure \ref{fig:mp} depicts the impulse responses of the different variables for an expansionary monetary policy shock based on data through 2019Q4, i.e. before the outbreak of COVID-19 and the same shock based on data through 2020Q4. I compare the responses of the linear FAVAR (panel 1), the Locally Embedded FAVAR (panel 2) and the Deep Dynamic FAVAR (panel 3). The blue solid line and the light blue shaded area depict the median and the $16$th and $84$th percentiles of the posterior distribution of the pre-pandemic responses, respectively. The black solid line and the grey shaded area correspond to the median response and the $16$th and $84$th posterior percentiles of the pandemic response.

Figure \ref{fig:mp} reveals that the three different FAVAR approaches yield very similar responses of the variables of interest when modeling an expansionary monetary policy shock at the end of 2019. 
When including the pandemic observations I observe that the responses differ across the modeling techniques. The linear FAVAR yields responses that are surrounded by appreciable uncertainty bands. In contrast, the non-linear FAVAR models estimate similar reactions to the monetary policy shock in both scenarios (i.e., before and during the COVID-19 pandemic).

Comparing the responses of GDP growth (GDPC1) between the different FAVAR approaches suggests that all models yield similar results for the scenario excluding the COVID-19 pandemic. The peak is reached after six periods before approaching to zero and turning slightly negative. Considering the dataset ending in 2019, the linear FAVAR yields responses with the most pronounced impact estimate. When I include the pandemic observations, the linear FAVAR suggests a lower impact on GDP growth which is mainly insignificant. In contrast, the non-linear techniques yield significant and positive reactions similar to the case which excludes the pandemic. A similar pattern can be observed for the unemployment rate (UNRATE). Regardless of the included time periods, the non-linear approaches estimate a fall in the unemployment rate as a response to an expansionary monetary policy shock. The linear FAVAR, however, yields insignificant reactions when considering the dataset including the COVID-19 outliers.

For the inflation rate (GDPCTPI) differences between the responses of the models are more pronounced. The linear and Locally Embedded FAVAR estimated without pandemic observations yield negative reactions of inflation to the expansionary monetary policy shock during the first year after the shock hit the system. This stands in contrast with the results of the Deep Dynamic FAVAR which suggest a positive relationship between expansionary monetary policy shocks and inflation. Interestingly, when including pandemic observations the linear FAVAR model generates persistent and elevated reactions for inflation after six periods. The Locally Embedded FAVAR still yields a significantly negative reaction after the shock. The Deep Dynamic FAVAR, on the other hand, yields no significant reaction in this scenario.

Housing starts (HOUST) react positively to a cut in interest rates for all approaches when the pandemic observations are excluded. When taking the COVID-19 crisis into account, this pattern persists only for the non-linear models, i.e., Locally Embedded FAVAR and Deep Dynamic FAVAR. The linear FAVAR shows no significant reactions during the first year after the shock and even estimate negative responses for the second year after the shock.

For the S\&P 500, I observe slightly negative reactions on impact for all FAVAR approaches when considering the data until the end of 2020. Excluding the observations of the pandemic leads to a reversal of this behavior for the Deep Dynamic FAVAR and the Locally Embedded FAVAR and suggests positive reactions of stock markets to an expansionary monetary policy shock. The Locally Embedded FAVAR is the only approach which estimates positive responses for two and three quarters after the shock in both scenarios.

The application of an expansionary monetary policy shock involves reducing the policy rate measured by the shadow rate by 100 bps on impact. As a consequence, the short-term interest rate (GS1)  rate falls but slightly less than the 100 bps shock of the shadow rate. I observe this pattern for all models and both scenarios.

\subsection{Application 2 - Uncertainty shock}\label{subsec:unc}

\begin{figure}[h!]
\caption{Impulse responses of selected variables to a positive uncertainty shock (25 bps) in 2019Q4 and 2020Q4 for the three different FAVAR approaches}\label{fig:unc}

\begin{minipage}{0.33\textwidth}
\centering
(1) \textit{Linear FAVAR}
\end{minipage}
\begin{minipage}{0.33\textwidth}
\centering
(2) \textit{Locally Embedded FAVAR}
\end{minipage}
\begin{minipage}{0.33\textwidth}
\centering
(3) \textit{Deep Dynamic FAVAR}
\end{minipage}

\begin{minipage}{0.33\textwidth}
\vspace{5pt}
\centering
\scriptsize \textit{GDPC1}
\end{minipage}
\begin{minipage}{0.33\textwidth}
\vspace{5pt}
\centering
\scriptsize \textit{GDPC1}
\end{minipage}
\begin{minipage}{0.33\textwidth}
\vspace{5pt}
\centering
\scriptsize \textit{GDPC1}
\end{minipage}

\begin{minipage}{0.33\textwidth}
\centering
\includegraphics[scale=.38]{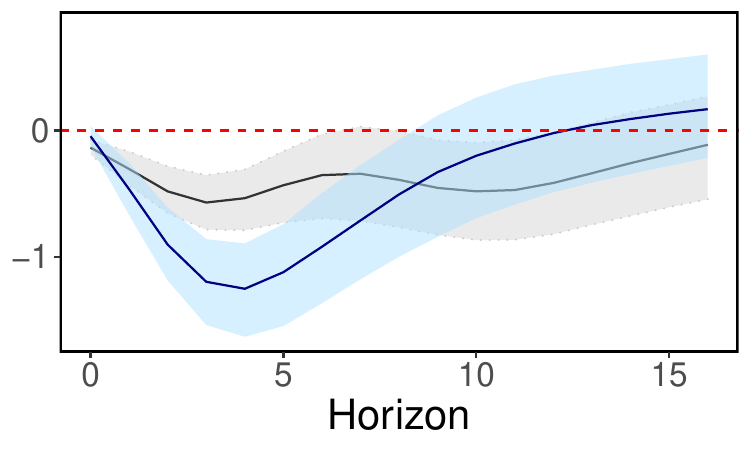}
\end{minipage}
\begin{minipage}{0.33\textwidth}
\centering
\includegraphics[scale=.38]{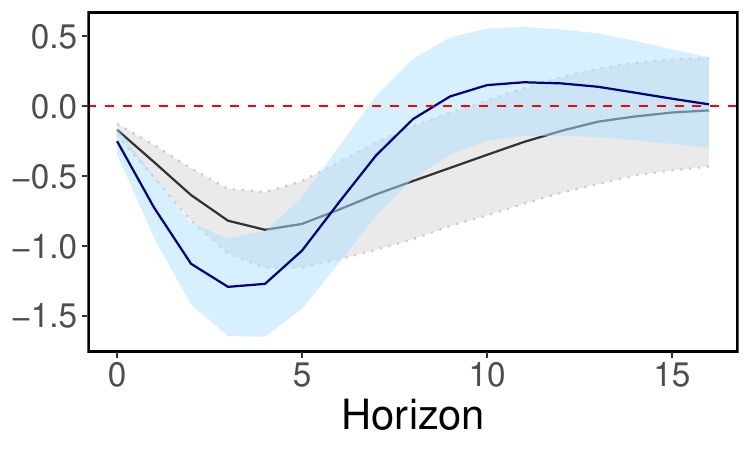}
\end{minipage}
\begin{minipage}{0.33\textwidth}
\centering
\includegraphics[scale=.38]{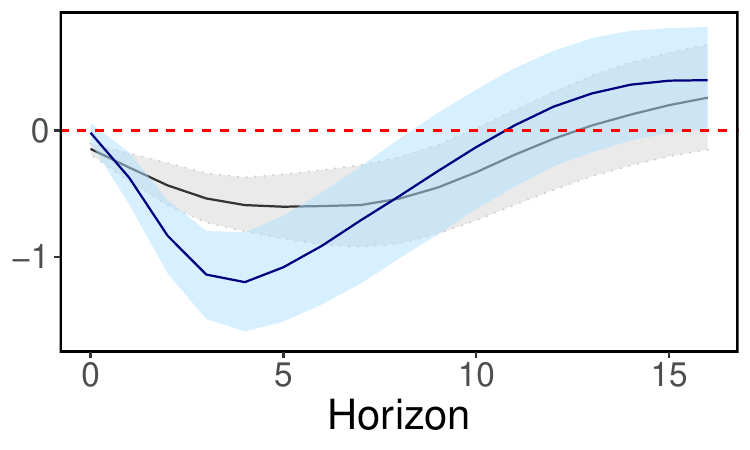}
\end{minipage}

\begin{minipage}{0.33\textwidth}
\vspace{5pt}
\centering
\scriptsize \textit{UNRATE}
\end{minipage}
\begin{minipage}{0.33\textwidth}
\vspace{5pt}
\centering
\scriptsize \textit{UNRATE}
\end{minipage}
\begin{minipage}{0.33\textwidth}
\vspace{5pt}
\centering
\scriptsize \textit{UNRATE}
\end{minipage}

\begin{minipage}{0.33\textwidth}
\centering
\includegraphics[scale=0.38]{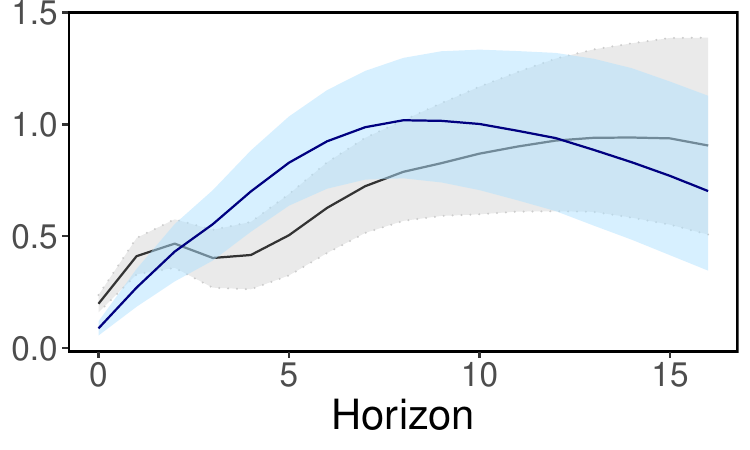}
\end{minipage}
\begin{minipage}{0.33\textwidth}
\centering
\includegraphics[scale=0.38]{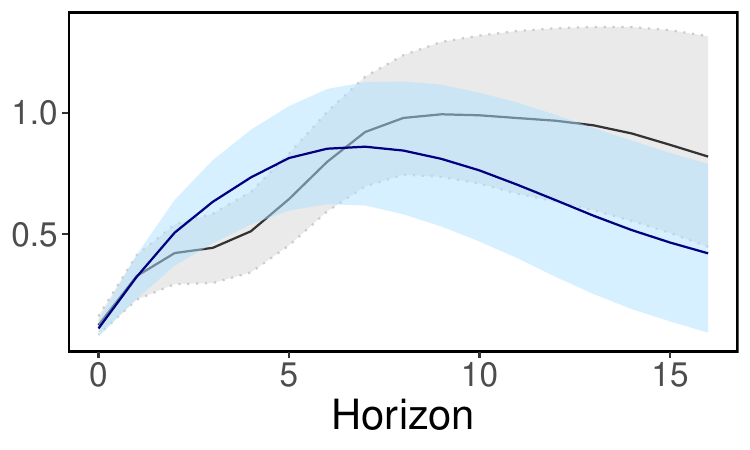}
\end{minipage}
\begin{minipage}{0.33\textwidth}
\centering
\includegraphics[scale=0.38]{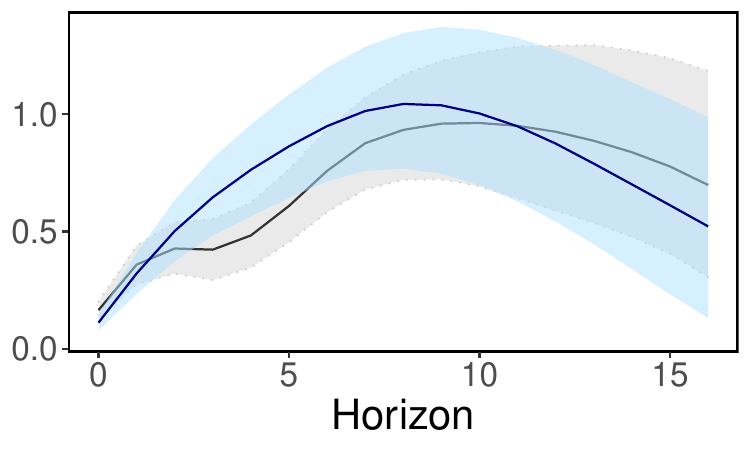}
\end{minipage}

\begin{minipage}{0.33\textwidth}
\vspace{5pt}
\centering
\scriptsize \textit{GDPCTPI}
\end{minipage}
\begin{minipage}{0.33\textwidth}
\vspace{5pt}
\centering
\scriptsize \textit{GDPCTPI}
\end{minipage}
\begin{minipage}{0.33\textwidth}
\vspace{5pt}
\centering
\scriptsize \textit{GDPCTPI}
\end{minipage}

\begin{minipage}{0.33\textwidth}
\centering
\includegraphics[scale=0.38]{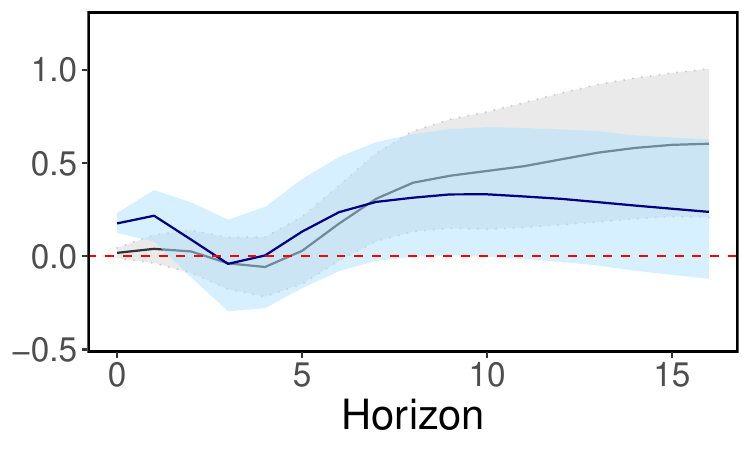}
\end{minipage}
\begin{minipage}{0.33\textwidth}
\centering
\includegraphics[scale=0.38]{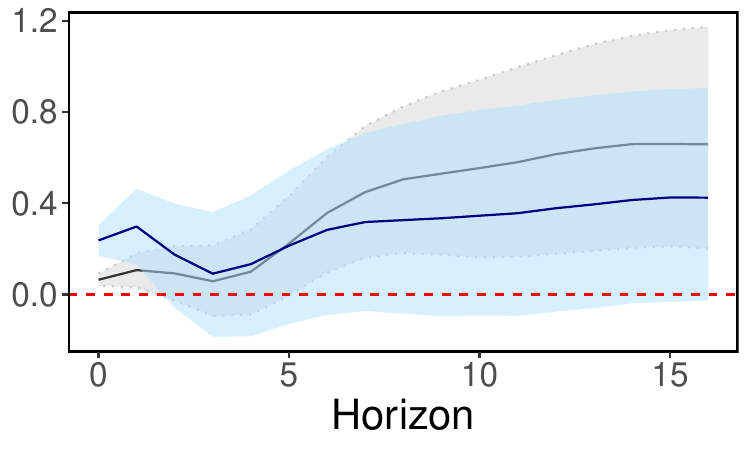}
\end{minipage}
\begin{minipage}{0.33\textwidth}
\centering
\includegraphics[scale=0.38]{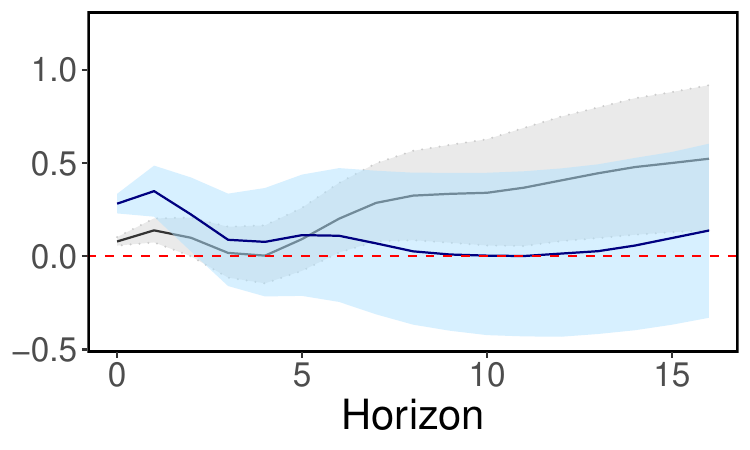}
\end{minipage}

\begin{minipage}{0.33\textwidth}
\vspace{5pt}
\centering
\scriptsize \textit{HOUST}
\end{minipage}
\begin{minipage}{0.33\textwidth}
\vspace{5pt}
\centering
\scriptsize \textit{HOUST}
\end{minipage}
\begin{minipage}{0.33\textwidth}
\vspace{5pt}
\centering
\scriptsize \textit{HOUST}
\end{minipage}

\begin{minipage}{0.33\textwidth}
\centering
\includegraphics[scale=0.38]{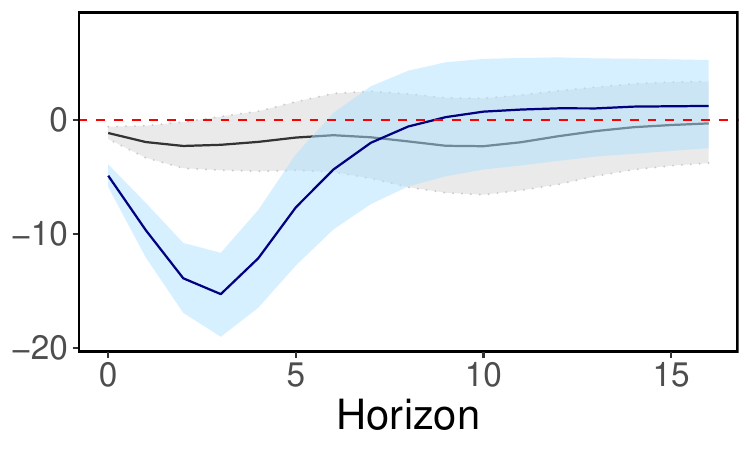}
\end{minipage}
\begin{minipage}{0.33\textwidth}
\centering
\includegraphics[scale=0.38]{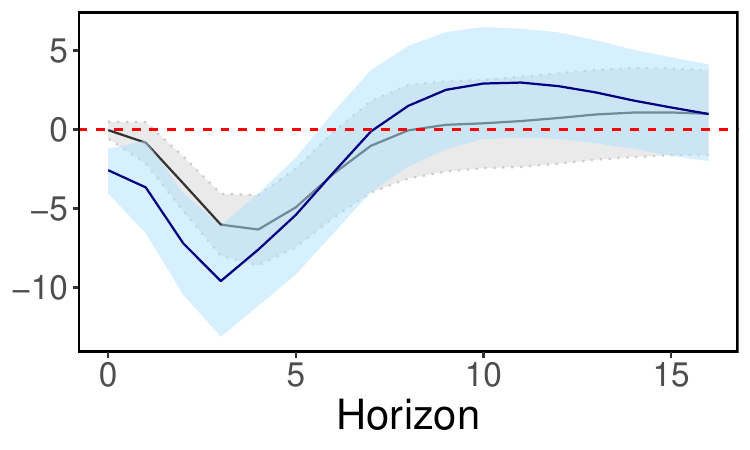}
\end{minipage}
\begin{minipage}{0.33\textwidth}
\centering
\includegraphics[scale=0.38]{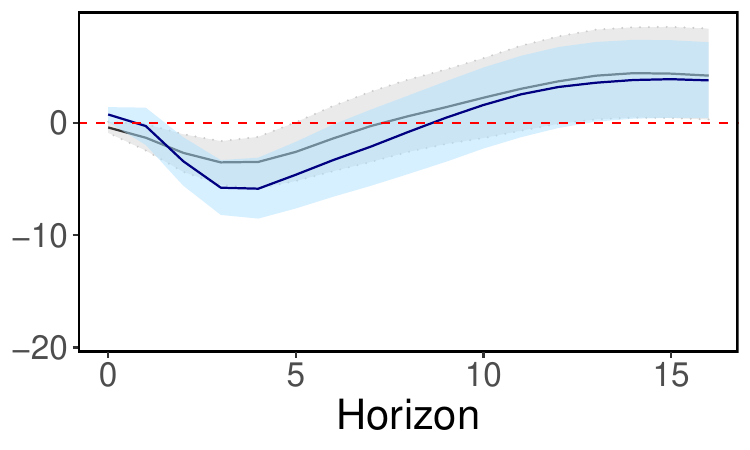}
\end{minipage}

\begin{minipage}{0.33\textwidth}
\vspace{5pt}
\centering
\scriptsize \textit{S\&P 500}
\end{minipage}
\begin{minipage}{0.33\textwidth}
\vspace{5pt}
\centering
\scriptsize \textit{S\&P 500}
\end{minipage}
\begin{minipage}{0.33\textwidth}
\vspace{5pt}
\centering
\scriptsize \textit{S\&P 500}
\end{minipage}

\begin{minipage}{0.33\textwidth}
\centering
\includegraphics[scale=0.38]{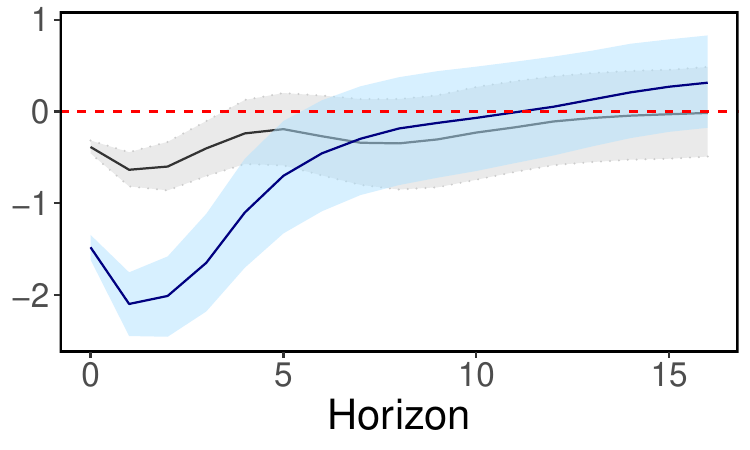}
\end{minipage}
\begin{minipage}{0.33\textwidth}
\centering
\includegraphics[scale=0.38]{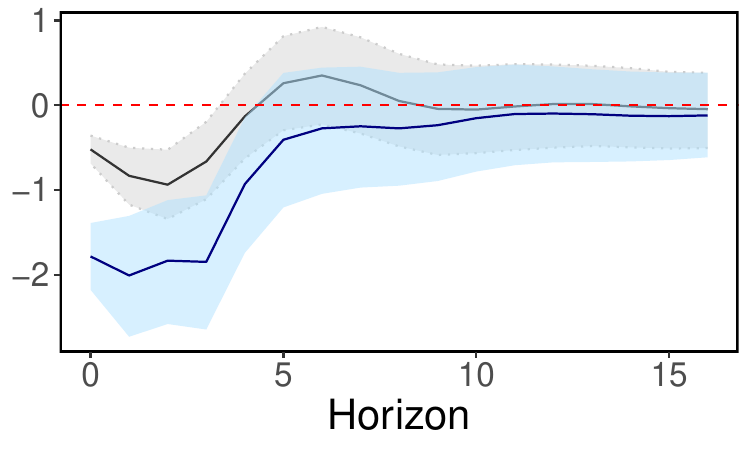}
\end{minipage}
\begin{minipage}{0.33\textwidth}
\centering
\includegraphics[scale=0.38]{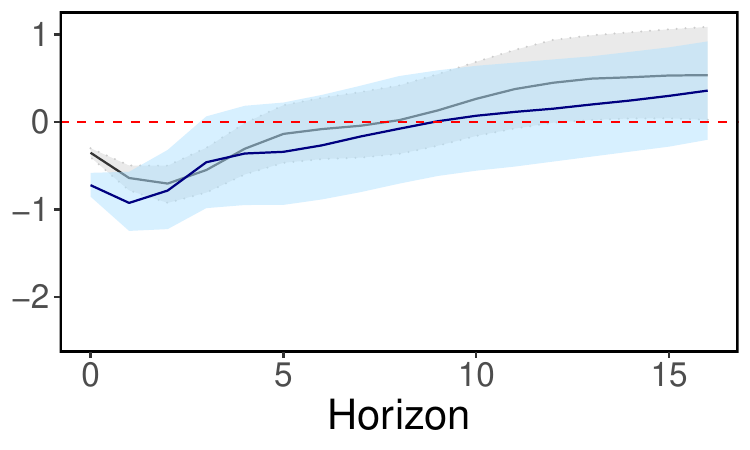}
\end{minipage}

\begin{minipage}{0.33\textwidth}
\vspace{5pt}
\centering
\scriptsize \textit{GS1}
\end{minipage}
\begin{minipage}{0.33\textwidth}
\vspace{5pt}
\centering
\scriptsize \textit{GS1}
\end{minipage}
\begin{minipage}{0.33\textwidth}
\vspace{5pt}
\centering
\scriptsize \textit{GS1}
\end{minipage}

\begin{minipage}{0.33\textwidth}
\centering
\includegraphics[scale=0.38]{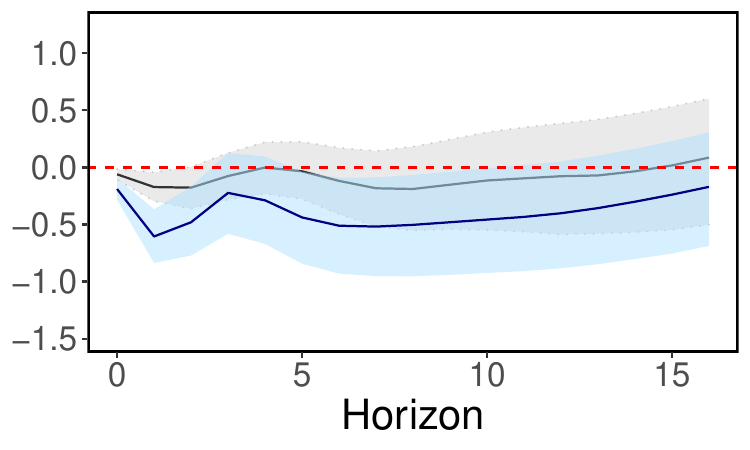}
\end{minipage}
\begin{minipage}{0.33\textwidth}
\centering
\includegraphics[scale=0.38]{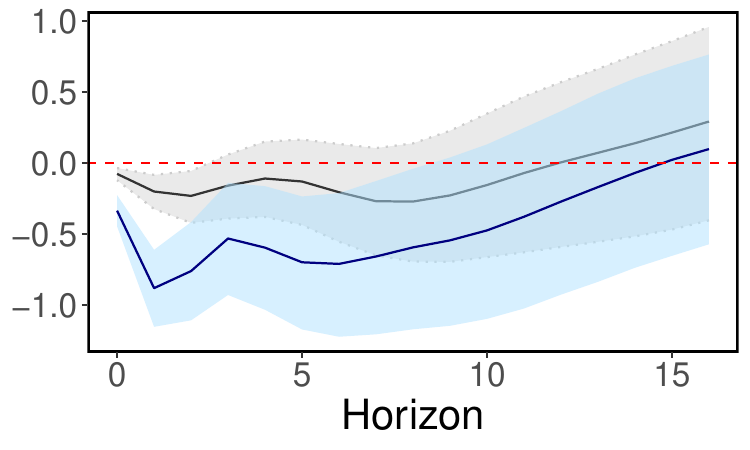}
\end{minipage}
\begin{minipage}{0.33\textwidth}
\centering
\includegraphics[scale=0.38]{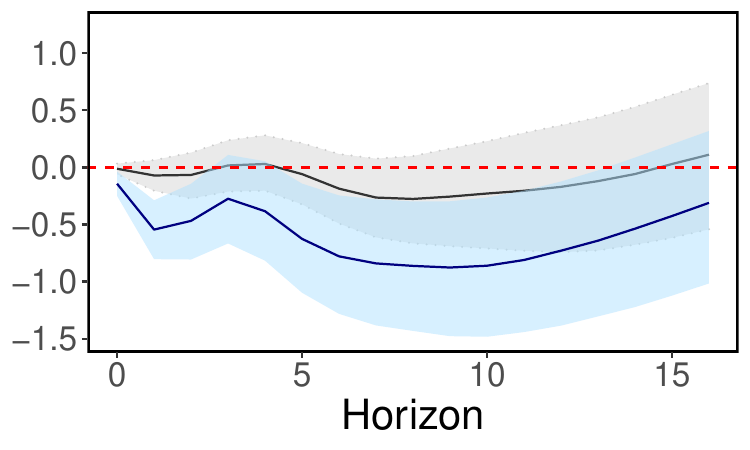}
\end{minipage}

\begin{minipage}{\textwidth}
    \footnotesize
    \emph{Note:} The plot gives the impulse responses excluding the pandemic observations in blue, i.e., the blue solid lines refer to the median response and the light blue shaded area indicate the $16$th and $84$th percentiles of the posterior distribution. The impulse responses obtained when including the COVID-19 pandemic are shown in black. The median responses are presented in the black solid lines and the confidence bands in form of the grey shaded area. The red dashed line shows the zero line. 
\end{minipage}
\end{figure}

In this section, I present the results of our second empirical application which involves simulating the effect of an uncertainty shock on key US macroeconomic and financial variables. As discussed in Section \ref{sec:ident}, I rely on the uncertainty index proposed by \cite{JLN2015uncertainty}.\footnote{Note that applying the National Financial Conditions Index (NFCI) as the uncertainty index yields very similar results.} I compare the results of the three different FAVAR approaches introduced in Section \ref{sec:favar} and show the impulse response functions for the same set of variables as in the previous section. Again, each panel of Figure \ref{fig:unc} presents the impulse response function when using the dataset until the end of 2019 in blue and the extended version with the COVID-19 pandemic included in grey. The solid lines depict the median response while the blue and grey shaded areas correspond to the $16$th and $84$th percentiles of the posterior distribution.

As shown before, the impulse responses for the scenario excluding the pandemic observations are similar in all three FAVAR approaches with the main difference being that the linear FAVAR estimates the strongest effects for most variables. When extending the observation window to the end of 2020 I find that the effects are smaller and often insignificant if the linear FAVAR is adopted. The Deep Dynamic FAVAR, on the other hand, yields significant effects for all variables.

The first variable, GDP growth, falls in response to an uncertainty shock with a certain delay. The linear and the Deep Dynamic FAVAR estimate a quite strong reaction when the pandemic is excluded. When including the pandemic observations all approaches suggest a smaller reaction, with the Deep Dynamic FAVAR suggesting the strongest effect which is also quite long-lasting. The linear FAVAR as well as the Locally Embedded FAVAR yield less pronounced dynamics. 

For the unemployment rate, all three FAVAR approaches produce very similar results. This is true for the scenario excluding as well as including the pandemic observations in the sample. The growth of unemployment rises in the periods following the uncertainty shock and peaks after seven to eight periods. 

The inflation rate shows a positive reaction to the uncertainty shock on impact when considering the scenario before the COVID-19 pandemic. Estimating the models including the pandemic observations results in no significant reaction of inflation on impact but positive reactions after a year for the linear FAVAR and the Locally Embedded FAVAR. The Deep Dynamic FAVAR suggests a significantly positive reaction on impact and a positive reaction after about six periods.

For housing starts, I observe the largest differences between the models. The linear FAVAR suggest a strong negative reaction for data through 2019Q4. The peak is reached three periods after the shock hit the economy. However, when the sample is extended to the end of 2020 I see no significant reaction of the housing variable to the uncertainty shock. Similarly, the model based on the locally linear embedding algorithm yields a negative response on impact but no reaction for the scenario including the COVID-19 pandemic. For the Deep Dynamic FAVAR I observe a negative reaction for both scenarios, with and without COVID-19 observations, but the effect is slightly larger when excluding the pandemic.

Investigating the response of the S\&P 500 stock market index reveals that on impact all models yield a significantly negative reaction. This effect is again strongest for the linear FAVAR when modelled without the COVID-19 periods. Including the pandemic observations also suggests a negative reaction on impact but to a far lesser extent. A similar pattern can be observed for the Locally Embedded FAVAR. The Deep Dynamic FAVAR suggests a similar negative response of the stock market index to the uncertainty shock in both scenarios.

For the short-term interest rate, I only observe a significant reaction to the uncertainty shock when leaving the pandemic observations aside. This pattern holds for all three modeling approaches.

\section{Closing remarks}\label{sec:conclusions}

In this paper, I propose a set of high-dimensional non-linear factor models. By applying non-linear dimension reduction techniques to a high-dimensional dataset and assuming that the resulting latent factors evolve according to a vector autoregression, two novel approaches are developed. The first is the Locally Embedded FAVAR, which is based on the linear locally embedding algorithm and the second is the Deep Dynamic FAVAR, which employs a deep learning algorithm for constructing the lower-dimensional representation of a dataset. 

When I apply the proposed techniques to synthetic data, allowing for an analysis of each model's behavior in a controlled environment, I find that the non-linear approaches yield competitive forecasting performance across all hold-outs and outperform the linear model for highly volatile observations. Depending on the specific dimension reduction technique used, the factors differ in how they extract signals from macroeconomic and financial variables and cover various sectors of the economy. As I have shown in two different empirical applications, the proposed non-linear FAVAR approaches yield tight estimates and responses in line with economic theory. This is true for tranquil times as well as for times characterized by high uncertainty. Analyzing model performances in times of crises is of particular interest as the current COVID-19 pandemic has caused unprecedented fluctuations in various economic and financial variables.

The proposed model can be seen as a very general framework, which nests several functional forms to generate the series of latent factors. This is of interest for dealing with high-dimensional datasets as well as for analyzing dynamics in uncertain and highly volatile times.

\clearpage
\small{\setstretch{0.85}
\addcontentsline{toc}{section}{References}

\ifx\undefined\BySame
\newcommand{\BySame}{\leavevmode\rule[.5ex]{3em}{.5pt}\ }
\fi
\ifx\undefined\textsc
\newcommand{\textsc}[1]{{\sc #1}}
\newcommand{\emph}[1]{{\em #1\/}}
\let\tmpsmall\small
\renewcommand{\small}{\tmpsmall\sc}
\fi
}

\newpage

\begin{appendices}
\begin{center}
\LARGE\textbf{Appendices}
\end{center}

\section{Additional results}\label{sec:App Res}

\begin{figure}[htb!]
\caption{First latent factor arising from linear and non-linear dimension reduction techniques and corresponding variable importance for 2020Q4. \label{fig:vi_factor1_2020}}

\begin{minipage}{\textwidth}
\centering
\includegraphics[scale=.42]{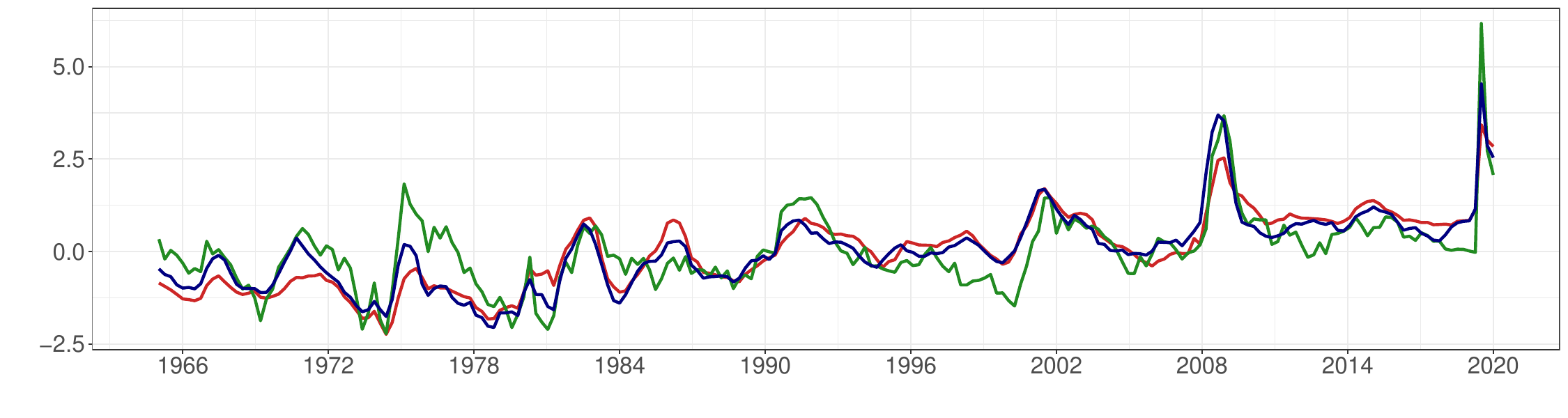}
\end{minipage}
\begin{minipage}{\textwidth}
\vspace{-1.6cm}
\centering
\hspace{1.5cm} \includegraphics[scale=.55]{legend_col.pdf}
\vspace{-0.7cm}
\end{minipage}

\begin{minipage}{0.33\textwidth}
\centering
\includegraphics[scale=0.42]{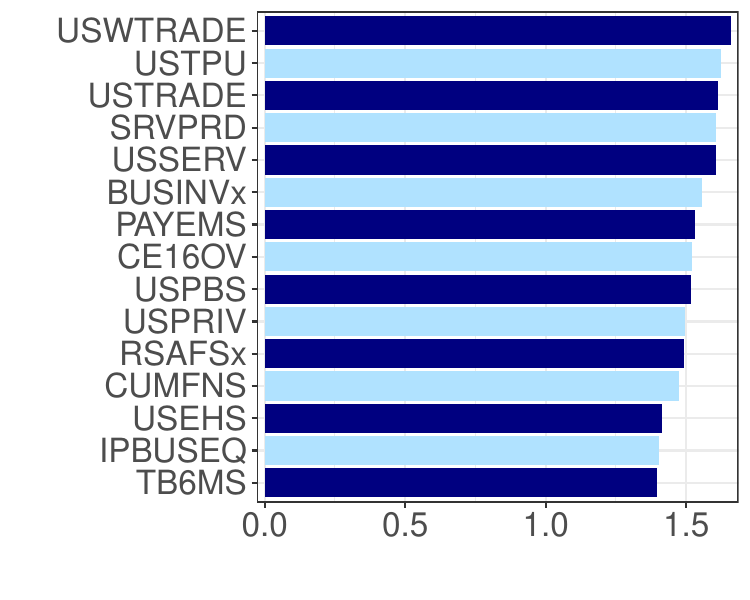}
\end{minipage}
\begin{minipage}{0.33\textwidth}
\centering
\includegraphics[scale=0.42]{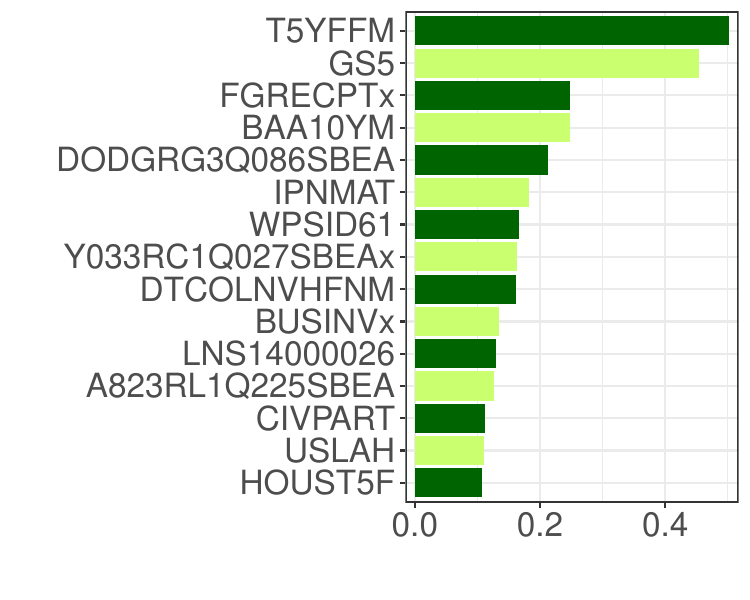}
\end{minipage}
\begin{minipage}{0.33\textwidth}
\centering
\includegraphics[scale=0.42]{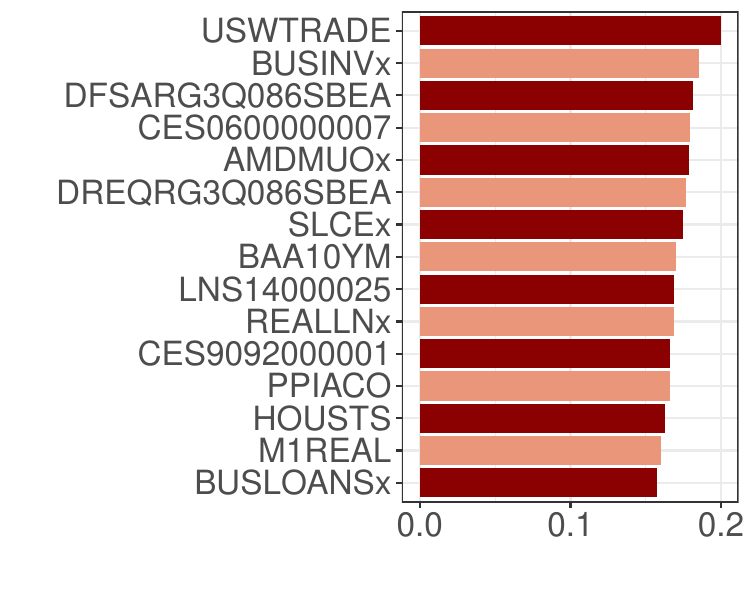}
\end{minipage}

\begin{minipage}{\textwidth}
\footnotesize
    \emph{Note:} The upper panel depicts normalized factors of the three dimension reduction techniques (mapped according to the highest correlation) with mean zero and variance one obtained from the main dataset ($N=162$) ranging from $1965$Q1 to $2020$Q4. The barplots show the 15 most important variables for each factor measured by the factor loadings for the linear FAVAR and the Locally Embedded FAVAR and Shapley values for the Deep Dynamic FAVAR. Mnemonics are those of \cite{mccracken2020fred} and can be found in Appendix \ref{sec:App Data}.
\end{minipage}

\end{figure}

\begin{figure}[htb!]
\caption{Second latent factor arising from linear and non-linear dimension reduction techniques and corresponding variable importance for 2020Q4. \label{fig:vi_factor2_2020}}

\begin{minipage}{\textwidth}
\centering
\includegraphics[scale=.42]{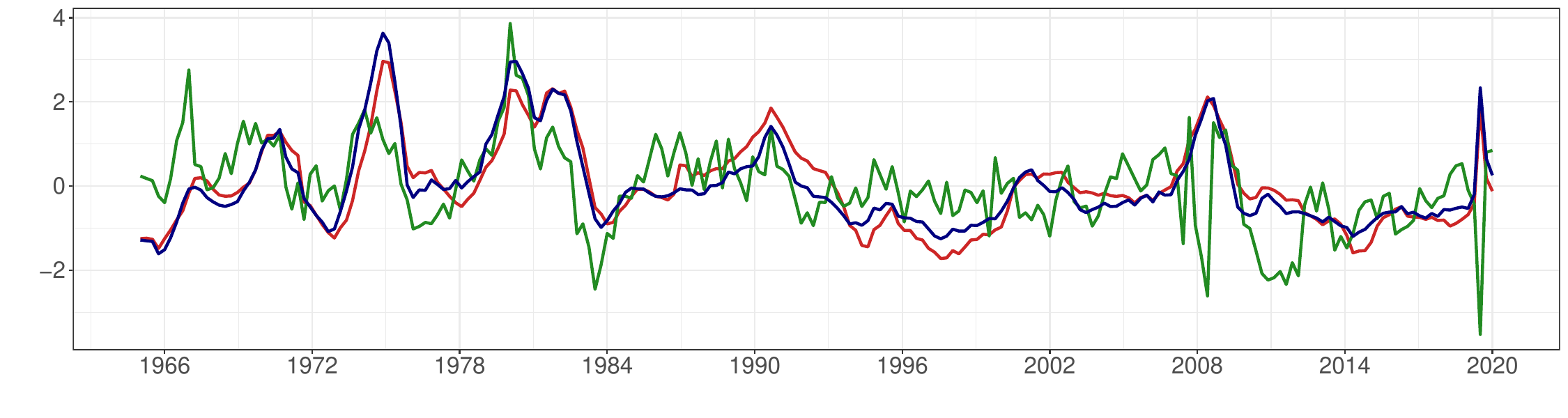}
\end{minipage}
\begin{minipage}{\textwidth}
\vspace{-1.6cm}
\centering
\hspace{1.5cm} \includegraphics[scale=.55]{legend_col.pdf}
\vspace{-0.7cm}
\end{minipage}

\begin{minipage}{0.33\textwidth}
\centering
\includegraphics[scale=0.42]{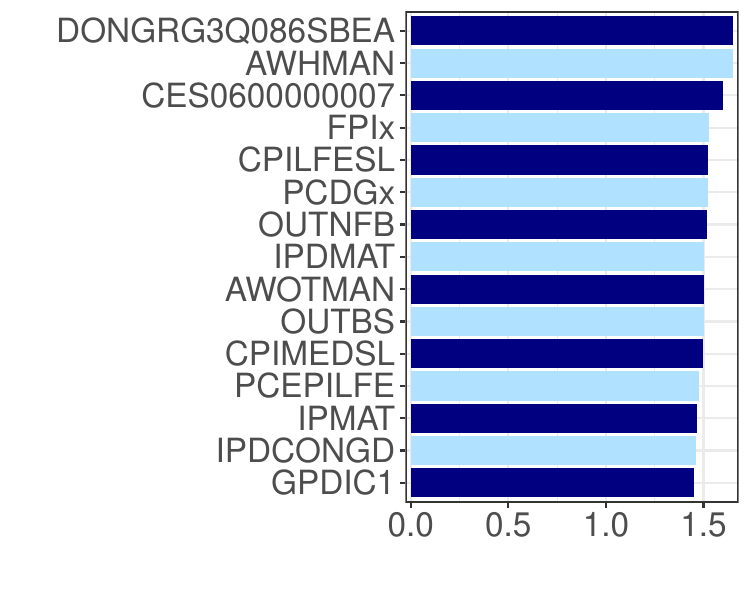}
\end{minipage}
\begin{minipage}{0.33\textwidth}
\centering
\includegraphics[scale=0.42]{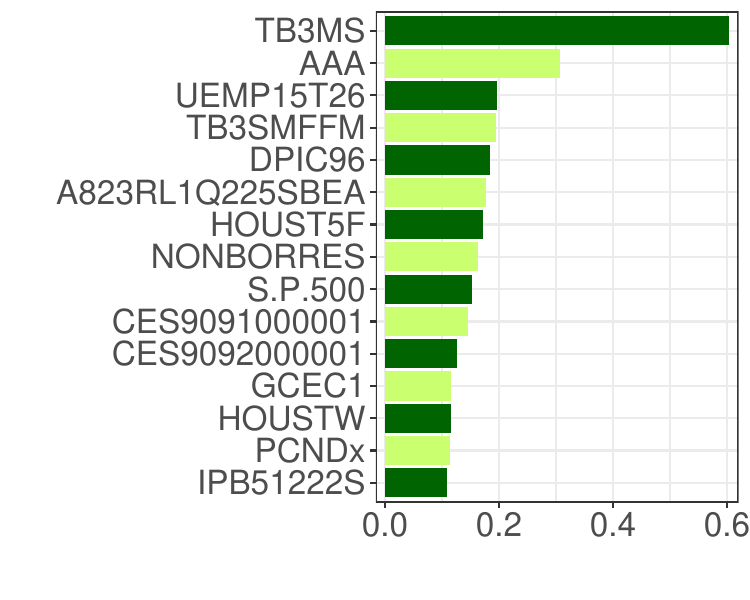}
\end{minipage}
\begin{minipage}{0.33\textwidth}
\centering
\includegraphics[scale=0.42]{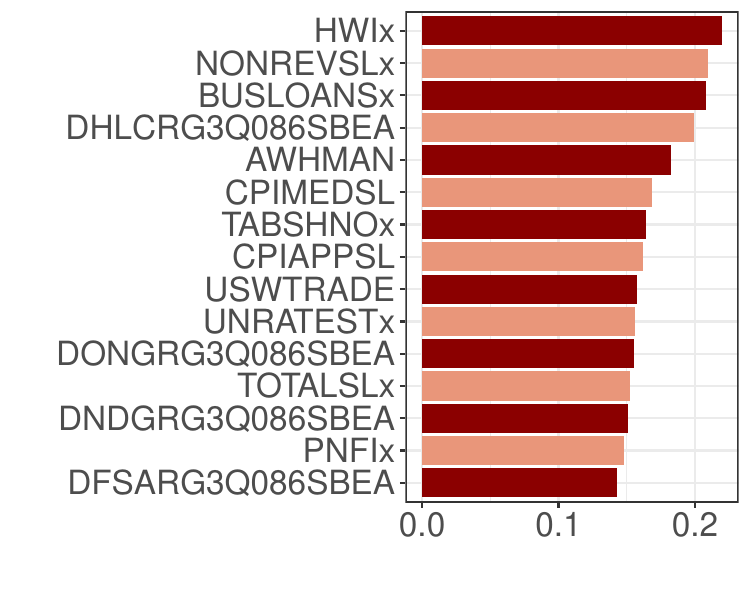}
\end{minipage}

\begin{minipage}{\textwidth}
\footnotesize
    \emph{Note:} For more details I refer to Figure \ref{fig:vi_factor1_2020}.
\end{minipage}

\end{figure}

\begin{figure}[htb!]
\caption{Third latent factor arising from linear and non-linear dimension reduction techniques and corresponding variable importance for 2020Q4. \label{fig:vi_factor3_2020}}

\begin{minipage}{\textwidth}
\centering
\includegraphics[scale=.42]{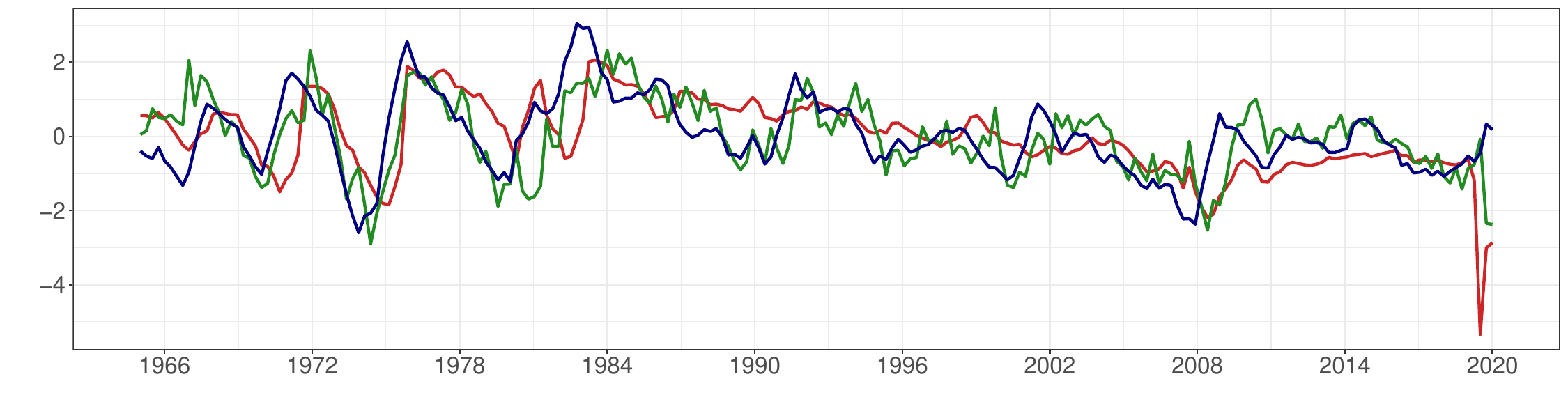}
\end{minipage}
\begin{minipage}{\textwidth}
\vspace{-1.6cm}
\centering
\hspace{1.5cm} \includegraphics[scale=.55]{legend_col.pdf}
\vspace{-0.7cm}
\end{minipage}

\begin{minipage}{0.33\textwidth}
\centering
\includegraphics[scale=0.42]{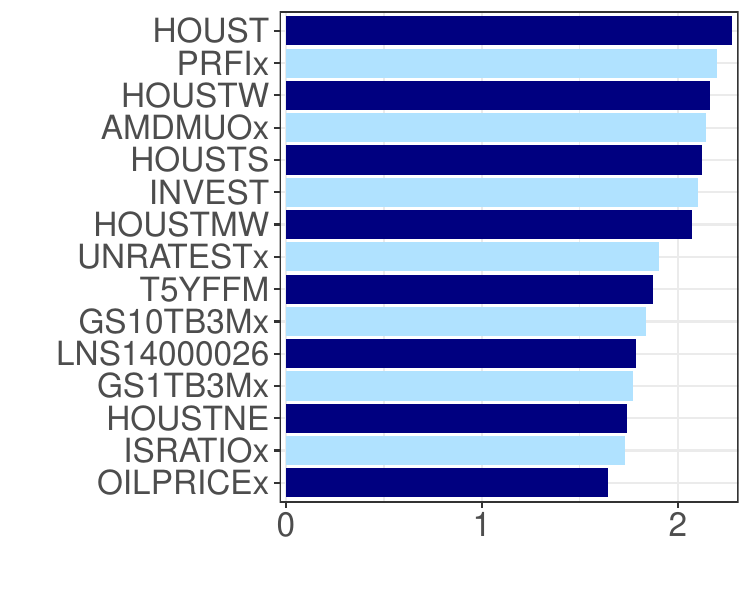}
\end{minipage}
\begin{minipage}{0.33\textwidth}
\centering
\includegraphics[scale=0.42]{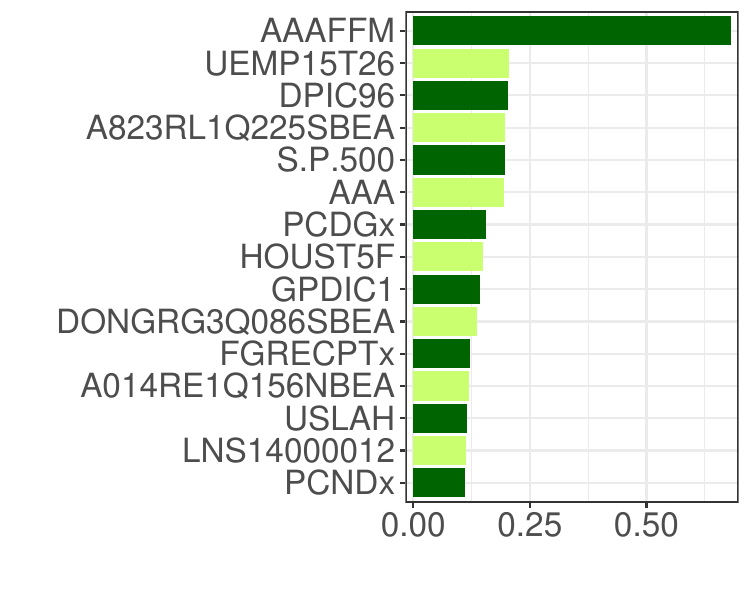}
\end{minipage}
\begin{minipage}{0.33\textwidth}
\centering
\includegraphics[scale=0.42]{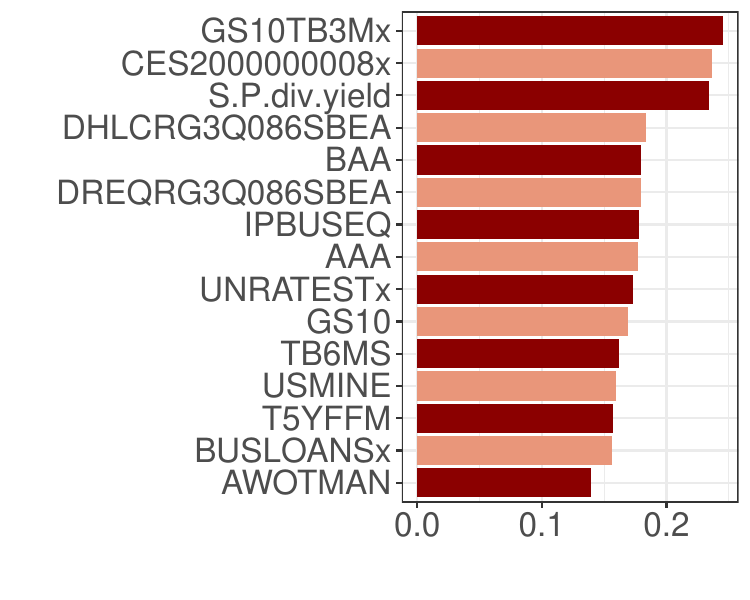}
\end{minipage}

\begin{minipage}{\textwidth}
\footnotesize
    \emph{Note:} For more details I refer to Figure \ref{fig:vi_factor1_2020}.
\end{minipage}

\end{figure}

\begin{figure}[!h]
\caption{Fourth latent factor arising from linear and non-linear dimension reduction techniques and corresponding variable importance for 2020Q4. \label{fig:vi_factor4_2020}}

\begin{minipage}{\textwidth}
\centering
\includegraphics[scale=.42]{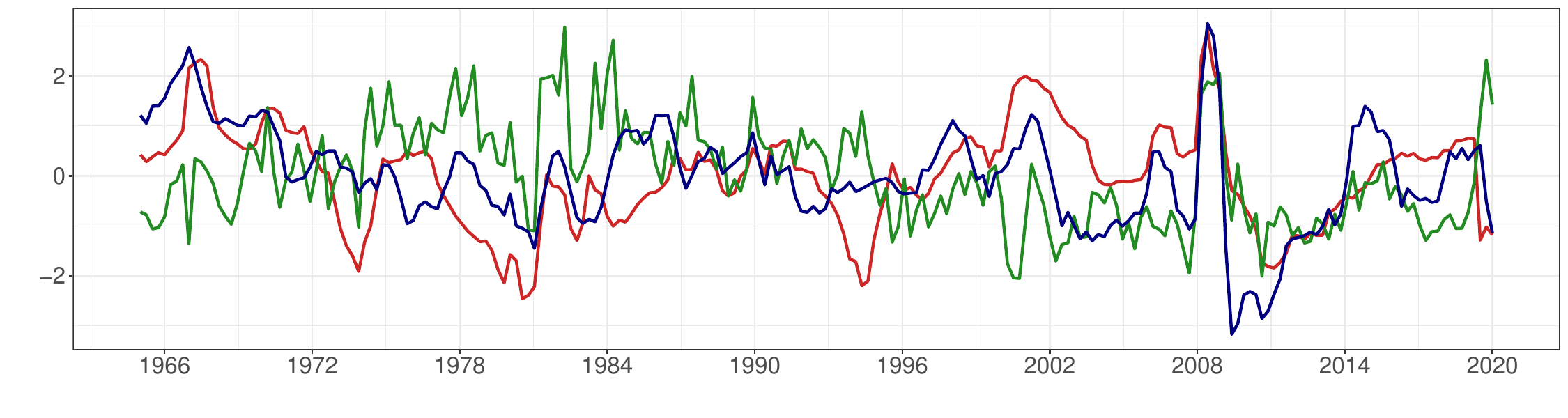}
\end{minipage}
\begin{minipage}{\textwidth}
\vspace{-1.6cm}
\centering
\hspace{1.5cm} \includegraphics[scale=.55]{legend_col.pdf}
\vspace{-0.7cm}
\end{minipage}

\begin{minipage}{0.33\textwidth}
\centering
\includegraphics[scale=0.42]{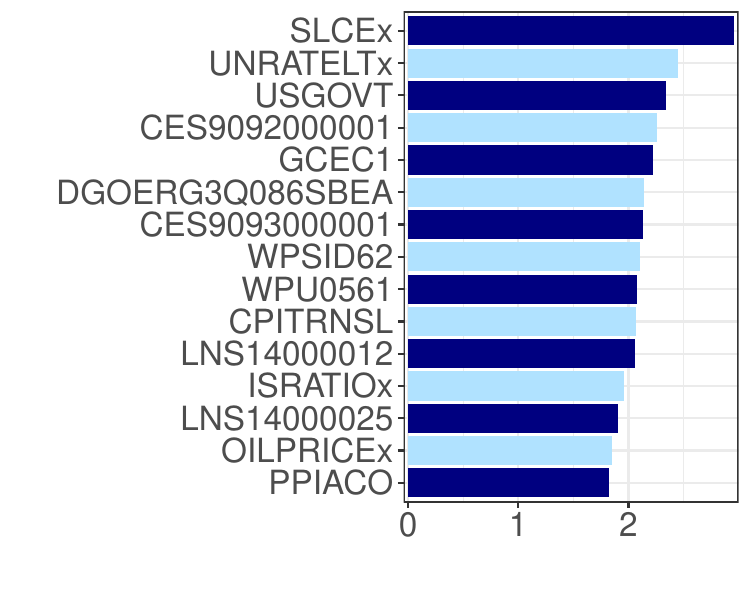}
\end{minipage}
\begin{minipage}{0.33\textwidth}
\centering
\includegraphics[scale=0.42]{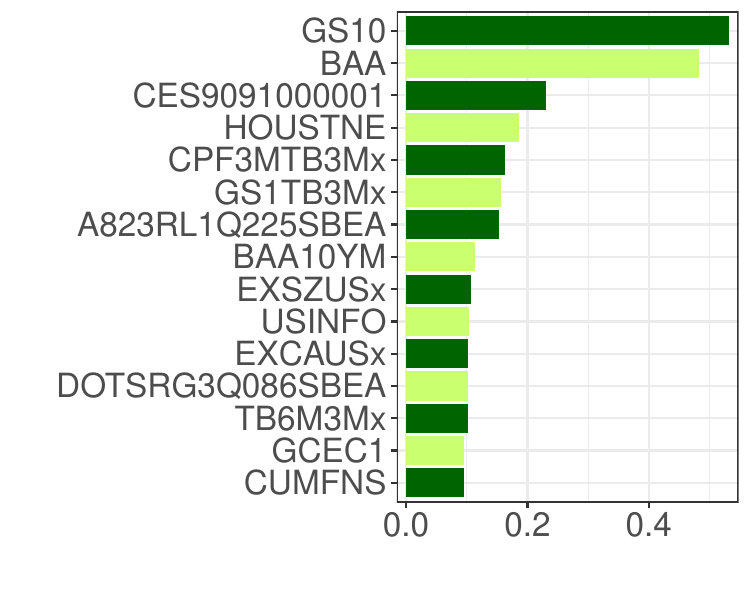}
\end{minipage}
\begin{minipage}{0.33\textwidth}
\centering
\includegraphics[scale=0.42]{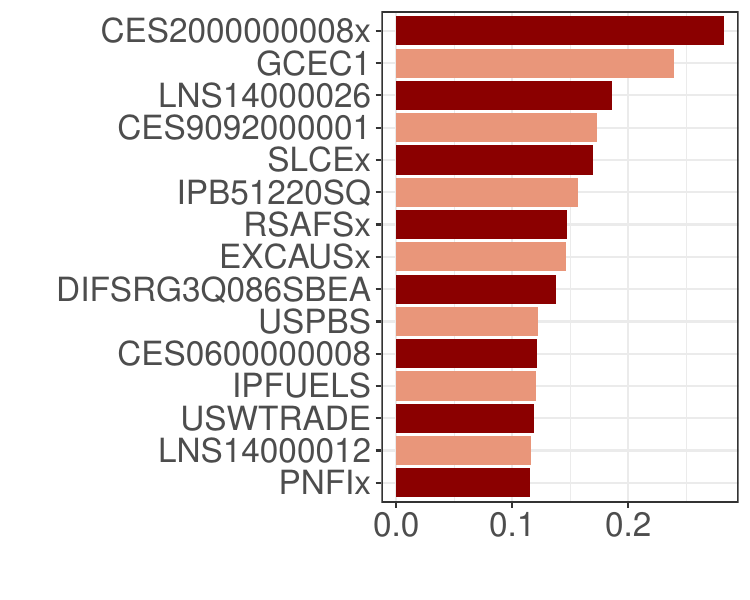}
\end{minipage}

\begin{minipage}{\textwidth}
\footnotesize
    \emph{Note:} For more details I refer to Figure \ref{fig:vi_factor1_2020}.
\end{minipage}

\end{figure}

\begin{figure}[!h]
\caption{Fifth latent factor arising from linear and non-linear dimension reduction techniques and corresponding variable importance for 2020Q4. \label{fig:vi_factor5_2020}}

\begin{minipage}{\textwidth}
\centering
\includegraphics[scale=.42]{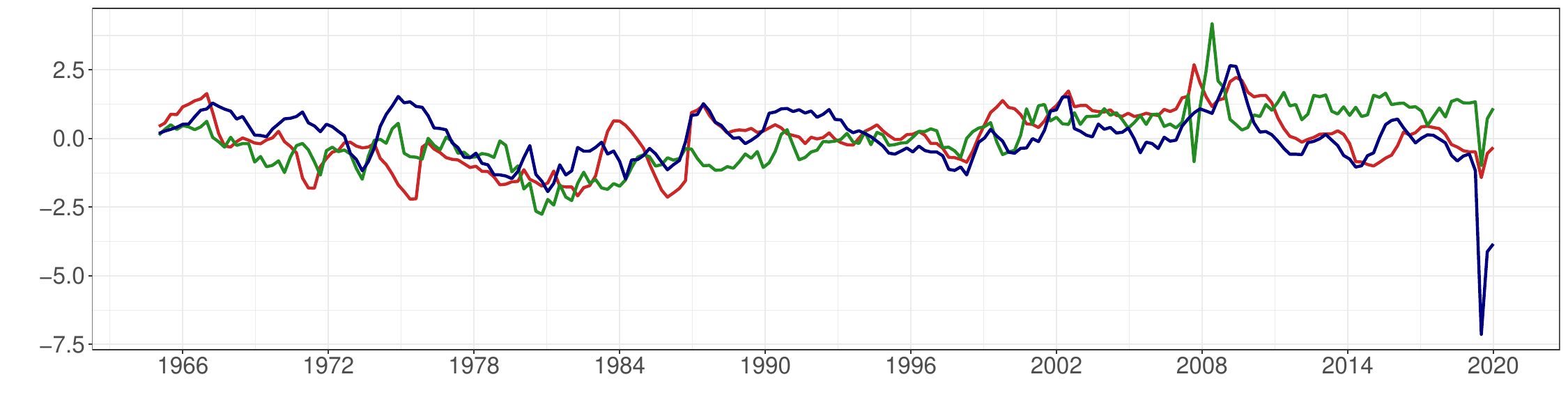}
\end{minipage}
\begin{minipage}{\textwidth}
\vspace{-1.6cm}
\centering
\hspace{1.5cm} \includegraphics[scale=.55]{legend_col.pdf}
\vspace{-0.7cm}
\end{minipage}

\begin{minipage}{0.33\textwidth}
\centering
\includegraphics[scale=0.42]{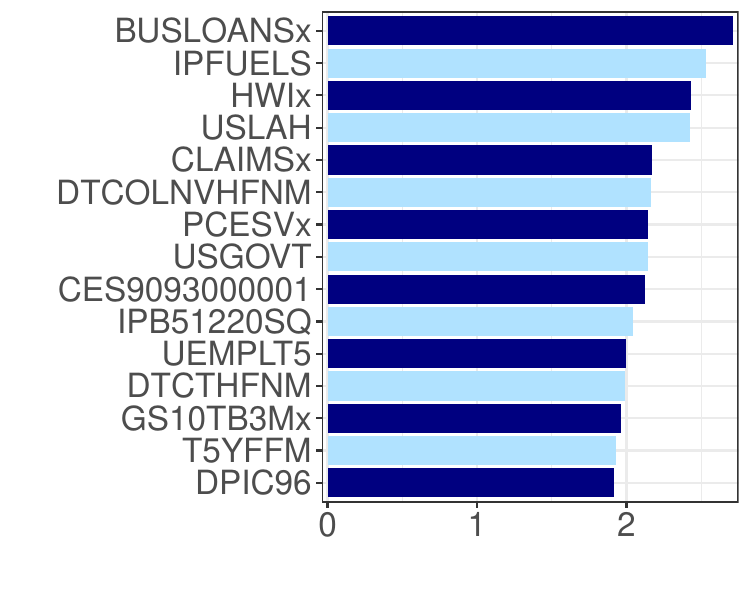}
\end{minipage}
\begin{minipage}{0.33\textwidth}
\centering
\includegraphics[scale=0.42]{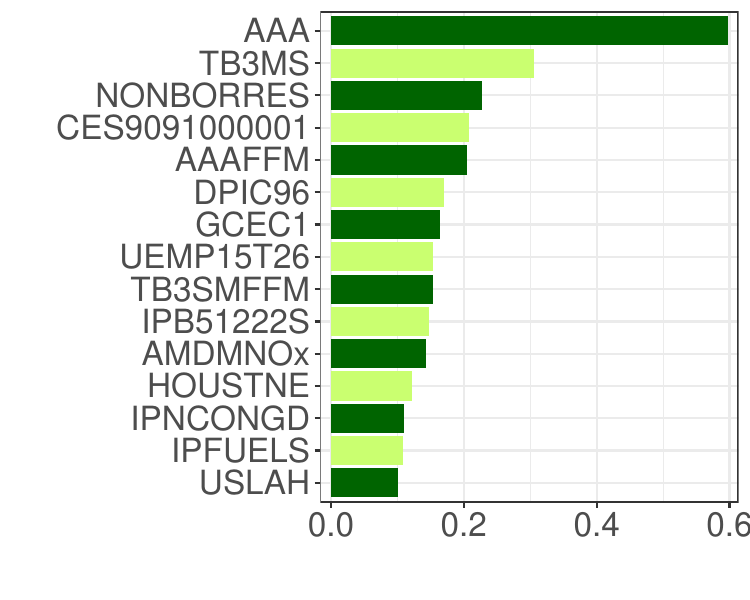}
\end{minipage}
\begin{minipage}{0.33\textwidth}
\centering
\includegraphics[scale=0.42]{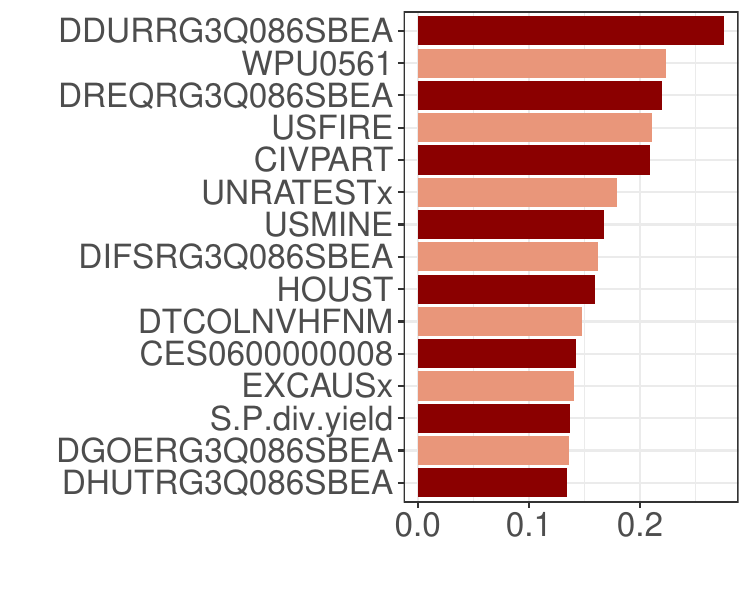}
\end{minipage}

\begin{minipage}{\textwidth}
\footnotesize
    \emph{Note:} For more details I refer to Figure \ref{fig:vi_factor1_2020}.
\end{minipage}

\end{figure}

\begin{table}[!h]
{\small
\caption{Relative point and density forecasting performance for simulated data \label{tab:sim_performance_details}} 
\begin{center}
\begin{tabular}{lcccccc}
\toprule
\multicolumn{1}{l}{\textbf{Variables}}&\multicolumn{1}{c}{}&\multicolumn{2}{c}{\textbf{Crisis Times}}&\multicolumn{1}{c}{}&\multicolumn{2}{c}{\textbf{Tranquil Times}}\tabularnewline
\addlinespace[2pt] 
 \cline{3-4} \cline{6-7}
\addlinespace[5pt] 
\multicolumn{1}{c}{}&\multicolumn{1}{c}{}&\multicolumn{1}{c}{\textbf{RMSE}}&\multicolumn{1}{c}{\textbf{CRPS}}&\multicolumn{1}{c}{}&\multicolumn{1}{c}{\textbf{RMSE}}&\multicolumn{1}{c}{\textbf{CRPS}}\tabularnewline
\midrule
\addlinespace[5pt] 
\rowcolor{gray!15} 
\multicolumn{7}{l}{\textbf{Deep Dynamic FAVAR}} \tabularnewline
\addlinespace[2pt] 
   Variable 1&   &   \textbf{0.94}&   \textbf{0.92}& &  1.00&   1.00\tabularnewline
   Variable 2&   &   \textbf{0.89}&   \textbf{0.87}& &  1.03&   1.01\tabularnewline
   Variable 3&   &   \textbf{0.83}&   \textbf{0.78}& &  1.03&   1.02\tabularnewline
   Variable 4&   &   \textbf{0.84}&   \textbf{0.84}& &  \textbf{0.97}&   \textbf{0.99}\tabularnewline
   Variable 5&   &   \textbf{0.93}&  \textbf{ 0.92}& &  \textbf{0.98}&   \textbf{0.99}\tabularnewline
   Variable 6&   &   \textbf{0.95}&   \textbf{0.96}& &  \textbf{0.99}&   \textbf{0.99}\tabularnewline
   Variable 7&   &   \textbf{0.79}&   \textbf{0.75}& &  1.00&   \textbf{0.99}\tabularnewline
   Variable 8&   &   \textbf{0.72}&   \textbf{0.67}& &  \textbf{0.98}&   \textbf{0.99}\tabularnewline
   Variable 9&   &   \textbf{0.86}&   \textbf{0.84}& &  1.00&   \textbf{0.99}\tabularnewline
   Variable 10&   &   1.02&   1.00&&   1.09&   1.04\tabularnewline
   Variable 11&   &   \textbf{0.72}&   \textbf{0.66}&&   \textbf{0.96}&   \textbf{0.97}\tabularnewline
   Variable 12&   &   1.05&   1.08&&   1.03&   1.01\tabularnewline
   Variable 13&   &   1.12&   1.16&&   1.05&   1.01\tabularnewline
   Variable 14&   &   1.08&   1.17&&   1.04&   1.01\tabularnewline
   Variable 15&   &   1.11&   1.15&&   1.02&   1.00\tabularnewline
   Variable 16&   &   \textbf{0.90}&   \textbf{0.89}&&   1.03&   1.01\tabularnewline
   Variable 17&   &   \textbf{0.85}&   \textbf{0.87}&&   \textbf{0.98}&   \textbf{0.99}\tabularnewline
   Variable 18&   &   \textbf{0.78}&   \textbf{0.77}&&   \textbf{0.97}&   \textbf{0.99}\tabularnewline
   Variable 19&   &   \textbf{0.98}&   1.04&&   \textbf{0.98}&   \textbf{0.99}\tabularnewline
   Variable 20&   &   1.05&   1.02&&   1.16&   1.11\tabularnewline
   \midrule
   \addlinespace[5pt] 
\rowcolor{gray!15} 
\multicolumn{7}{l}{\textbf{Locally Embedded FAVAR}} \tabularnewline
\addlinespace[2pt] 
   Variable 1& &  1.02&     1.03& &  1.00&   1.00\tabularnewline
   Variable 2& &  \textbf{0.95}&     \textbf{0.94}& &  1.00&   1.00\tabularnewline
   Variable 3& & \textbf{ 0.97}&     \textbf{0.97}& &  1.00&   1.01\tabularnewline
   Variable 4& &  \textbf{0.96}&     \textbf{0.95}& &  \textbf{0.99}&   1.00\tabularnewline
   Variable 5& &  1.03&     1.04& &  1.00&   1.00\tabularnewline
   Variable 6& &  \textbf{0.97}&     \textbf{0.96}& &  \textbf{0.99}&   1.00\tabularnewline
   Variable 7& &  \textbf{0.99}&     \textbf{0.98}& &  1.01&   1.00\tabularnewline
   Variable 8& &  \textbf{0.94}&     \textbf{0.94}& &  \textbf{0.99}&   1.00\tabularnewline
   Variable 9& &  1.00&     1.01& &  1.00&   1.00\tabularnewline
   Variable 10& &  \textbf{0.99}&     \textbf{0.99}& &  1.01&   1.01\tabularnewline
   Variable 11&  & 1.01&     1.01& &  \textbf{0.99}&   1.00\tabularnewline
   Variable 12&  & 1.01&     1.01& &  \textbf{0.99}&   1.00\tabularnewline
   Variable 13&  & 1.00&     \textbf{0.99}& &  1.01&   1.01\tabularnewline
   Variable 14&  & \textbf{0.97}&     \textbf{0.98}& &  \textbf{0.99}&   1.00\tabularnewline
   Variable 15&  & 1.01&     1.01& &  1.00&   1.00\tabularnewline
   Variable 16&  & \textbf{0.99}&     \textbf{0.98}& &  1.00&   1.00\tabularnewline
   Variable 17&  & \textbf{0.98}&     \textbf{0.98}& &  \textbf{0.99}&   1.00\tabularnewline
   Variable 18&  & \textbf{0.99}&     \textbf{0.99}& &  1.00&   1.00\tabularnewline
   Variable 19&  & \textbf{0.97}&     \textbf{0.98}& &  \textbf{0.99}&   1.00\tabularnewline
   Variable 20&  & \textbf{0.99}&     \textbf{0.98}& &  1.02&   1.03\tabularnewline
\bottomrule
\end{tabular}
\begin{minipage}{0.58\textwidth}
\scriptsize 
    \textit{Note}: The table shows point forecasting performance in terms of root mean squared errors (RMSE) as well as density forecasting performance in terms of continuous ranked probability scores (CRPS). All metrics are relative to the linear FAVAR. Values below one (bold numbers) show that the non-linear model outperforms the linear one. 
\end{minipage}
\end{center}}
\end{table}

\clearpage
\newpage

\section{Technical Appendix}\label{sec:App Tech}

\subsection{Equation-by-equation estimation}\label{subsec:eqbyeq}

Before I sketch the implemented prior in more detail, I briefly discuss the equation-by-equation estimation approach. As shown by \cite{carriero2019large,carriero2022corrigendum}, one can exploit the fact that $\bm \Sigma_{\bm \epsilon} = \tilde{\bm A}_0^{-1} \bm \Sigma_{\tilde{\bm \epsilon}} \tilde{\bm A}_0^{-1'}$ with $\tilde{\bm A}_0^{-1}$ being a lower triangular matrix and $\bm \Sigma_{\tilde{\bm \epsilon}}$ being diagonal and write the VAR model as $K$ independent regressions. I collect the coefficient matrices in an $K \times 2(Kp+1)$-dimensional matrix $\bm A = (\bm A_1, \dots, \bm A_p, \bm c)$ and the regressors in an $K \times 1$-dimensional vector $\bm x_t = (\bm y'_{t-1}, \dots, \bm y'_{t-p}, 1)'$. Moreover, I define $\tilde{\bm y}_t = \tilde{\bm A}_0 \bm y_t$. This allows to rewrite Eq. \ref{eq:redVAR} as
\begin{equation*}
\tilde{\bm y}_t = \tilde{\bm A}_0 \bm y_t = \tilde{\bm A}_0 \bm A \bm x_t  + \tilde{\bm \epsilon}_t, \quad \tilde{\bm \epsilon}_t \sim \mathcal{N} (\bm 0,\bm \Sigma_{\bm \tilde{\epsilon}})
\end{equation*}
The fact that the variance-covariance matrix $\bm \Sigma_{\bm \tilde{\epsilon}}$ is diagonal allows to write the VAR model as $K$ independent regressions where the first equation in the system is given by
\begin{equation*}
\tilde{y}_{1t} = \bm x'_t \bm A_{\bullet 1} + \tilde{\epsilon}_{1t}, \quad \tilde{\epsilon}_{1t} \sim \mathcal{N} (0, \tilde{\sigma}_{1})
\end{equation*}
with $\bm A_{\bullet 1}$ depicting the first column in $\bm A$ and $\tilde{\epsilon}_{1t}$ the first element in $\tilde{\bm \epsilon}_t$. $\tilde{\sigma}_{1}$ refers to the first diagonal element in $\bm \Sigma_{\tilde{\epsilon}}$. The following $k$ equations for $k=2, \dots, K$ are given by
\begin{align*}
\tilde{y}_{2t} &= \tilde{\bm A}_{0;2,1} \bm x'_t \bm A_{\bullet 1} + \bm x'_t \bm A_{\bullet 2} + \tilde{\epsilon}_{2t}, \quad \tilde{\epsilon}_{2t} \sim \mathcal{N} (0, \tilde{\sigma}_{2}) \\
 &\vdots\\
\tilde{y}_{Kt} &= \tilde{\bm A}_{0;K,1} \bm x'_t \bm A_{\bullet 1} +  \dots + \tilde{\bm A}_{0;K,K-1} \bm x'_t \bm A_{\bullet K-1} + \bm x'_t \bm A_{\bullet K} + \tilde{\epsilon}_{kt}, \quad \tilde{\epsilon}_{Kt} \sim \mathcal{N} (0, \tilde{\sigma}_{K})
\end{align*}
where $\bm A_{\bullet k}$ denotes the $k$th column in $\bm A$ and $\tilde{A}_{0,ki}$ is the ($k,i$)th element in $\tilde{\bm A}_0$. $\tilde{\sigma}_{k}$ is the $k$th diagonal element in $\bm \Sigma_{\tilde{\epsilon}}$ (for $k = 1,\dots, K$).


\subsection{Prior structure}\label{subsec:prior}
For the estimation of the models proposed in this paper I implement the well-known Minnesota prior \citep{doan1984forecasting, sims1998bayesian}. 
The main idea is to center the system on a multivariate random walk, which is assumed to reflect the notion of the variables a priori. Furthermore, own lags are assumed to provide more information on the variation of a certain variable than lags of other variables and that the variables becomes less important with increasing lag length.

I assume that each $\bm \alpha_k$ follows a normal distribution
\begin{equation*}
\bm \alpha_k \sim \mathcal{N} (\underline{\bm \alpha}_k, \underline{\bm V}_k)
\end{equation*}
where the corresponding prior mean is denoted by $\underline{\bm \alpha}_k$ and the prior variance-covariance matrix by $\underline{\bm V}_k$. 
For stationary series, such as the latent factors and the uncertainty index, I set $\underline{\bm \alpha}_k$ to zero. For non-stationary data, I define $\underline{\bm \alpha}_k=1$.

The prior variance-covariance matrix $\underline{\bm V}_k$ is diagonal with the $i$th diagonal element chosen such that
\begin{equation*}
\underline{\bm V}_{k,ii} =
\begin{cases}
\frac{\xi_1}{r^2} & \text{ on the coefficients of own lags (for } r=1,\dots,p)\\
\frac{\xi_2 \hat{\sigma}^2_i}{r^2 \hat{\sigma}^2_j} & \text{ on the coefficients of other variables' lags (} i \neq j)\\
\xi_3 & \text{on the intercept term.}
\end{cases} 
\end{equation*} \label{eq:minnesota}
$\xi_1$ and $\xi_2$ denote scaling parameters, which make sure that coefficients are more heavily shrunk to zero with increasing lag length and with $\xi_1 > \xi_2$ that own coefficients are more likely to be important than those of other variables. $\xi_3$ controls the prior on the intercept term. We set $\xi_3 = 1000$ and estimate the parameters $\xi_1$ and $\xi_2$ within the sampler as suggested by \cite{giannone2015prior}. $\hat{\sigma}^2$ is determined by estimating the variance of an AR(4) process in $\Delta \bm y$ via OLS.

\subsection{MCMC sampling algorithm}\label{subsec:mcmc}
The sampling algorithm is comprised of the following steps:

\begin{enumerate}
\item I draw $\bm \alpha_k$ for $k=1, \dots, K$ from a multivariate Normal distribution
\begin{equation*}
\bm \alpha_k \sim \mathcal{N} (\overline{\bm \alpha}_k, \overline{\bm V}_k)
\end{equation*}
where $\overline{\bm \alpha}_k$ and $\overline{\bm V}_k$ denote the posterior quantities given by
\begin{align*}
\overline{\bm \alpha}_k &= \overline{\bm V}_k (\tilde{\bm X'}_k \bm y_k / \tilde{\sigma}_k + \underline{\bm V}_k^{-1} \underline{\bm \alpha}_k)\\
\overline{\bm V}_k &= (\tilde{\bm X'}_k \tilde{\bm X}_k / \tilde{\sigma}_k + \underline{\bm V}_k^{-1})^{-1}
\end{align*}
$\underline{\bm \alpha}_k$ refers to the prior mean $\underline{\bm \alpha}$ for equation $k$ and $\underline{\bm V}_k$ to the prior variance-covariance matrix $\underline{\bm V}$ as specified in Section \ref{subsec:prior} for equation $k$.

\item The structural error variances $\tilde{\sigma}_{k}$ for $k=1, \dots, K$ are sampled from an inverse Gamma distribution:
\begin{equation*}
\tilde{\sigma}_{k} \sim \mathcal{IG} (e_k, d_k)
\end{equation*}
with $e_k = (T/2 + 0.005)$ and $d_k = (\tilde{\bm \epsilon}'_k \tilde{\bm \epsilon}_k /2 + 0.005)$. 

\item Updating the Minnesota parameters $\xi_1$ and $\xi_2$ involves drawing candidate values from a Gaussian distribution with mean  $\xi_1^{(jj-1)}$ and $\xi_2^{(jj-1)}$ and variance $\tilde{c}_1$ and $\tilde{c}_2$. $\xi_1^{(jj-1)}$ and $\xi_2^{(jj-1)}$ denote the previous draws of $\xi_1$ and $\xi_2$ and $\tilde{c}_1$ and $\tilde{c}_1$ are defined as a scaling parameters chosen such that the acceptance rate is between 15 and 30 percent. I update the parameter values with probability $\beta^{(jj)}$ and stick with the previous values with probability $(1-\beta^{jj})$. $\beta^{(jj)}$ is determined by comparing the likelihood of the parameter proposals and the likelihood of the previous parameter values conditional on the data. According to this choice we update $\underline{\bm V}_{i,j}$ for the next draw from the posterior distribution. This procedure follows \cite{giannone2015prior}.


\end{enumerate}

\subsection{Variable importance measures for the non-linear FAVARs}\label{subsec:VI}

\noindent \textbf{Neighborhood Preserving Embedding for the Locally Embedded FAVAR.} To get interpretable factor loadings from the locally linear embedding algorithm, I implement the Neighborhood Preserving Embedding of \cite{he2005neighborhood}. After computing the weight matrix $\bm \Omega$ I make use of a linear approximation and assume that $\hat{\bm F} = \bm P' \bm D$. This changes the cost function for $\hat{\bm F}$ for being the new data point in Eq. \ref{eq:LLE newdata} to
\begin{equation*}
\Phi(\hat{\bm F}) = \sum_i | \bm P' \bm d_{\bullet i} - \sum \Omega_{ij} \bm P' \bm d_{\bullet i} |^2
\end{equation*}
The last step is then to solve the standard eigenvalue problem given by
\begin{equation*}
\bm D \bm M \bm D' \bm P = \bm D \bm D' \bm P \bm \Lambda
\end{equation*}
with $\bm M = (\bm I_t + \bm \Omega)' (\bm I_t + \bm \Omega)$ and $\bm \Lambda$ denoting the eigenvalue matrix with diagonal elements being the eigenvalues of ($\bm D \bm D')^{-1} \bm D \bm M \bm D'$. Similar to PCA this can be solved by applying Singular Value Decomposition (SVD) and the resulting factors enjoy interpretability via the factor loadings.

\noindent \textbf{Shapley values for the Deep Dynamic FAVAR.} Machine learning techniques, such as the autoencoder, are very flexible approaches but often face the criticism of being hard to interpret. To tackle the black box critique the variable importance measure for the Deep Dynamic FAVAR is based on the Shapley additive explanations framework \citep{strumbelj2010efficient,lundberg2017unified}, which is itself inspired from the concept of Shapley values \citep{shapley1953value}. The recent macro and finance literature shows growing interest in this type of measures to gain interpretable results of highly non-linear techniques \citep[see, ][]{joseph2021forecasting,borup2022anatomy,bluwstein2023credit,coulombe2023maximally}. In the proposed framework, it allows to identify the key drivers of the deep dynamic factors and give them some basic economic meaning. 

Let $\phi_{t,k}$ be the Shapley value at time $t$ (for $t = 1, \dots, T$) and variable $k$ (for $k = 1, \dots, K$). Each predicted factor $\hat{F}_{t,q}$ (for $q=1,\dots,Q$) can be decomposed into 
\begin{equation*}
\hat{F}_{t,q} = \sum_{k=1}^K \phi_{t,k} + \phi_0,
\end{equation*}
where $\phi_0$ is the mean predicted value in the training set and can be interpreted as an intercept. Computing the Shapley value of each variable $k$ involves deriving its marginal contribution to the prediction of the factor. This is done by comparing the payoffs of all possible coalitions of regressors, formally given by
\begin{equation*}
\phi_{t,k} = \sum_{\mathcal{S} \subseteq K/k} \frac{S!(K-S-1)!}{K!} (f(\mathcal{S} \cup \{ k\}) - f(\mathcal{S})),
\end{equation*} 
with the payoff of a coalition $\mathcal{S} \subseteq K$ denoted by $f(\mathcal{S})$ and the payoff of the same coalition combined with regressor $k$ being $f(\mathcal{S} \cup \{ k\})$. The difference between these two measures gives the marginal contribution of variable $k$ to the coalition. Since the computational burden of Shapley values for large datasets is considerable, an almost exact hybrid algorithm of \cite{covert2021improving} is used.

\setcounter{equation}{0}
\setcounter{table}{0}
\setcounter{figure}{0}
\renewcommand\theequation{B.\arabic{equation}}
\renewcommand\thetable{B.\arabic{table}}
\renewcommand\thefigure{B.\arabic{figure}}

\clearpage

\section{Data Appendix}\label{sec:App Data}
The Federal Reserve Economic Data (FRED) contains quarterly observations of macroeconomic variables for the US and is available for download at \url{https://research.stlouisfed.org}. Details on the dataset can be found in \cite{mccracken2020fred}. The time series start from 1959 Q1 and encompasses 248 quarterly series in total. Due to missing values, I preselect 166 variables starting from 1965 Q1 covering all sectors of the economy and transform them according to \autoref{tab:data-descr1}. The last column in \autoref{tab:data-descr1} classifies the variables into slow- and fast-moving quantities for the identification by zero restrictions on contemporaneous effects.

For the determination of the average duration of the US business cycle I use the data provided by the National Bureau of Economic Research downloadable at \url{https://www.nber.org/research/data/us-business-cycle-expansions-and-contractions}. I consider 12 cycles between 1945 and 2020 with an average duration of 75 months from peak to peak, i.e. six years.

\begin{table}[!tbp]
\caption{Data description\label{tab:data-descr1}}
{\tiny
\begin{center}
\begin{tabular}{llccc}
\toprule
\multicolumn{1}{l}{FRED.Mnemonic}&\multicolumn{1}{l}{Description}&\multicolumn{1}{l}{Trans.}&\multicolumn{1}{l}{Obs. var.}&\multicolumn{1}{l}{Type}\tabularnewline
\midrule
GDPC1&Real Gross Domestic Product&$50$&&slow\tabularnewline
PCECC96&Real Personal Consumption Expenditures&$50$&&slow\tabularnewline
PCDGx&Real personal consumption expenditures:  Durable goods &$50$&&slow\tabularnewline
PCESVx&Real Personal Consumption Expenditures:  Services &$50$&&slow\tabularnewline
PCNDx&Real Personal Consumption Expenditures:  Nondurable Goods &$50$&&slow\tabularnewline
GPDIC1&Real Gross Private Domestic Investment&$50$&&slow\tabularnewline
FPIx&Real private fixed investment &$50$&&slow\tabularnewline
Y033RC1Q027SBEAx&Real Gross Private Domestic Investment:  Fixed Investment:  Nonresidential Equip&$50$&&slow\tabularnewline
PNFIx&Real private fixed investment:  Nonresidential &$50$&&slow\tabularnewline
PRFIx&Real private fixed investment:  Residential &$50$&&slow\tabularnewline
A014RE1Q156NBEA&Shares of gross domestic product: Change
in private inventories&$ 1$&&slow\tabularnewline
GCEC1&Real Government Consumption Expenditures and Gross Investment&$50$&&slow\tabularnewline
A823RL1Q225SBEA&Real Government Consumption Expenditures and Gross Investment:  Federal&$ 1$&&slow\tabularnewline
FGRECPTx&Real Federal Government Current Receipts &$50$&&slow\tabularnewline
SLCEx&Real government state and local consumption expenditures &$50$&&slow\tabularnewline
EXPGSC1&Real Exports of Goods and Services&$50$&&slow\tabularnewline
IMPGSC1&Real Imports of Goods and Services&$50$&&slow\tabularnewline
DPIC96&Real Disposable Personal Income&$50$&&slow\tabularnewline
OUTNFB&Nonfarm Business Sector:  Real Output&$50$&&slow\tabularnewline
OUTBS&Business Sector:  Real Output&$50$&&slow\tabularnewline
INDPRO&IP:Total index Industrial Production Index (Index 2012=100)&$50$&x&slow\tabularnewline
IPFINAL&IP:Final products Industrial Production (Market Group) (Index 2012=100)&$50$&&slow\tabularnewline
IPCONGD&IP:Consumer goods Industrial Production: Consumer Goods (Index 2012=100)&$50$&&slow\tabularnewline
IPMAT&Materials (Index 2012=100)&$50$&&slow\tabularnewline
IPDMAT&Durable Materials (Index 2012=100)&$50$&&slow\tabularnewline
IPNMAT&Nondurable Materials (Index 2012=100)&$50$&&slow\tabularnewline
IPDCONGD&Durable Consumer Goods (Index 2012=100)&$50$&&slow\tabularnewline
IPB51110SQ&Durable Goods:  Automotive products (Index 2012=100)&$50$&&slow\tabularnewline
IPNCONGD&Nondurable Consumer Goods (Index 2012=100)&$50$&&slow\tabularnewline
IPBUSEQ&Business Equipment (Index 2012=100)&$50$&&slow\tabularnewline
IPB51220SQ&Consumer energy products (Index 2012=100)&$50$&&slow\tabularnewline
CUMFNS&Capacity Utilization:  Manufacturing (SIC) (Percent of Capacity)&$ 1$&&slow\tabularnewline
IPMANSICS&Industrial Production:  Manufacturing (SIC) (Index 2012=100)&$50$&&slow\tabularnewline
IPB51222S&Industrial Production:  Residential Utilities (Index 2012=100)&$50$&&slow\tabularnewline
IPFUELS&Industrial Production:  Fuels (Index 2012=100)&$50$&&slow\tabularnewline
PAYEMS& Emp:Nonfarm All Employees: Total nonfarm (Thousands of Persons)&$50$&&slow\tabularnewline
USPRIV& All Employees: Total Private Industries (Thousands of Persons)&$50$&&slow\tabularnewline
MANEMP& All Employees: Manufacturing (Thousands of Persons)&$50$&&slow\tabularnewline
SRVPRD&All Employees:  Service-Providing Industries (Thousands of Persons)&$50$&&slow\tabularnewline
USGOOD&All Employees:  Goods-Producing Industries (Thousands of Persons)&$50$&&slow\tabularnewline
DMANEMP&All Employees:  Durable goods (Thousands of Persons)&$50$&&slow\tabularnewline
NDMANEMP&All Employees:  Nondurable goods (Thousands of Persons)&$50$&&slow\tabularnewline
USCONS&All Employees:  Construction (Thousands of Persons)&$50$&&slow\tabularnewline
USEHS&All Employees:  Education \& Health Services (Thousands of Persons)&$50$&&slow\tabularnewline
USFIRE&All Employees:  Financial Activities (Thousands of Persons)&$50$&&slow\tabularnewline
USINFO&All Employees:  Information Services (Thousands of Persons)&$50$&&slow\tabularnewline
USPBS&All Employees:  Professional \& Business Services (Thousands of Persons)&$50$&&slow\tabularnewline
USLAH&All Employees:  Leisure \& Hospitality (Thousands of Persons)&$50$&&slow\tabularnewline
USSERV&All Employees:  Other Services (Thousands of Persons)&$50$&&slow\tabularnewline
USMINE&All Employees:  Mining and logging (Thousands of Persons)&$50$&&slow\tabularnewline
USTPU&All Employees:  Trade, Transportation \& Utilities (Thousands of Persons)&$50$&&slow\tabularnewline
USGOVT&All Employees:  Government (Thousands of Persons)&$50$&&slow\tabularnewline
USTRADE&All Employees:  Retail Trade (Thousands of Persons)&$50$&&slow\tabularnewline
USWTRADE&All Employees:  Wholesale Trade (Thousands of Persons)&$50$&&slow\tabularnewline
CES9091000001&All Employees:  Government:  Federal (Thousands of Persons)&$50$&&slow\tabularnewline
CES9092000001&All Employees:  Government:  State Government (Thousands of Persons)&$50$&&slow\tabularnewline
CES9093000001&All Employees:  Government:  Local Government (Thousands of Persons)&$50$&&slow\tabularnewline
CE16OV&Civilian Employment (Thousands of Persons)&$50$&&slow\tabularnewline
CIVPART&Civilian Labor Force Participation Rate (Percent)&$ 1$&&slow\tabularnewline
UNRATE&Civilian Unemployment Rate (Percent)&$ 1$&x&slow\tabularnewline
UNRATESTx&Unemployment Rate less than 27 weeks (Percent)&$ 1$&&slow\tabularnewline
UNRATELTx&Unemployment Rate for more than 27 weeks (Percent)&$ 1$&&slow\tabularnewline
LNS14000012&Unemployment Rate - 16 to 19 years (Percent)&$ 1$&&slow\tabularnewline
LNS14000025&Unemployment Rate - 20 years and over, Men (Percent)&$ 1$&&slow\tabularnewline
LNS14000026&Unemployment Rate - 20 years and over, Women (Percent)&$ 1$&&slow\tabularnewline
UEMPLT5&Number of Civilians Unemployed - Less Than 5 Weeks (Thousands of Persons)&$50$&&slow\tabularnewline
UEMP5TO14&Number of Civilians Unemployed for 5 to 14 Weeks (Thousands of Persons)&$50$&&slow\tabularnewline
UEMP15T26&Number of Civilians Unemployed for 15 to 26 Weeks (Thousands of Persons)&$50$&&slow\tabularnewline
UEMP27OV&Number of Civilians Unemployed for 27 Weeks and Over (Thousands of Persons)&$50$&&slow\tabularnewline
AWHMAN&Average Weekly Hours of Prod and Nonsuperv Employees:  Manufacturing &$ 1$&&slow\tabularnewline
AWOTMAN&Avg Weekly Overtime Hours of Prod and NonsupervEmployees: Manufacturing &$ 1$&&slow\tabularnewline
HWIx&Help-Wanted Index&$ 1$&&slow\tabularnewline
CES0600000007&Average Weekly Hours of Prod and NonsupervEmployees:  Goods-Producing&$ 1$&&slow\tabularnewline
CLAIMSx&Initial Claims&$50$&&slow\tabularnewline
HOUST&Housing Starts: Total: New Privately Owned Housing Units Started&$50$&&slow\tabularnewline
HOUST5F&Privately Owned Housing Starts: 5-Unit Structures or More&$50$&&slow\tabularnewline
PERMIT&New Private Housing Units Authorized by Building Permits&$50$&&slow\tabularnewline
HOUSTMW&Housing Starts in Midwest Census Region (Thousands of Units)&$50$&&slow\tabularnewline
HOUSTNE&Housing Starts in Northeast Census Region (Thousands of Units)&$50$&&slow\tabularnewline
HOUSTS&Housing Starts in South Census Region (Thousands of Units)&$50$&&slow\tabularnewline
HOUSTW&Housing Starts in West Census Region (Thousands of Units)&$50$&&slow\tabularnewline
RSAFSx&Real Retail and Food Services Sales (Millions of Chained 2012 Dollars)&$50$&&slow\tabularnewline
AMDMNOx&Real Manufacturers New Orders:  Durable Goods (Millions of 2012 Dollars)&$50$&&slow\tabularnewline
AMDMUOx&Real Value of Manufacturers Unfilled Orders for Durable Goods Industries&$50$&&slow\tabularnewline
\bottomrule
\end{tabular}
\end{center}}
\end{table}

\begin{table}[!tbp]
\captionsetup{labelformat=empty}
\caption{Data description (cont.)\label{tab:data-descr1}}
{\tiny
\begin{center}
\begin{tabular}{llccc}
\toprule
\multicolumn{1}{l}{FRED.Mnemonic}&\multicolumn{1}{l}{Description}&\multicolumn{1}{l}{Trans.}&\multicolumn{1}{l}{Obs. var.}&\multicolumn{1}{l}{Type}\tabularnewline
\midrule
BUSINVx&Total Business Inventories (Millions of Dollars)&$50$&&slow\tabularnewline
ISRATIOx&Total Business:  Inventories to Sales Ratio&$ 1$&&slow\tabularnewline
PCECTPI&Pers Cons Exp: Chain-type Price Index &$50$&&slow\tabularnewline
PCEPILFE&Personal Consumption Expenditures Excluding Food and Energy&$50$&&slow\tabularnewline
GDPCTPI&Gross Domestic Product: Chain-type Price Index&$50$&&slow\tabularnewline
GPDICTPI&Gross Private Domestic Investment: Chain-type Price Index &$50$&&slow\tabularnewline
IPDBS&Business Sector:  Implicit Price Deflator (Index 2012=100)&$50$&&slow\tabularnewline
DGDSRG3Q086SBEA&Pers Cons Exp:  Goods &$50$&&slow\tabularnewline
DDURRG3Q086SBEA&Pers Cons Exp:  Durable goods &$50$&&slow\tabularnewline
DSERRG3Q086SBEA&Pers Cons Exp:  Services &$50$&&slow\tabularnewline
DNDGRG3Q086SBEA&Pers Cons Exp:  Nondurable goods&$50$&&slow\tabularnewline
DHCERG3Q086SBEA&Pers Cons Exp:  Services:  Household consumption expenditures&$50$&&slow\tabularnewline
DMOTRG3Q086SBEA&Pers Cons Exp:  Durable goods:  Motor vehicles and parts&$50$&&slow\tabularnewline
DFDHRG3Q086SBEA&Pers Cons Exp:  Durable goods:  Furnishings and durable household equipment&$50$&&slow\tabularnewline
DREQRG3Q086SBEA&Pers Cons Exp:  Durable goods:  Recreational goods and vehicles&$50$&&slow\tabularnewline
DODGRG3Q086SBEA&Pers Cons Exp:  Durable goods:  Other durable goods&$50$&&slow\tabularnewline
DFXARG3Q086SBEA&Pers Cons Exp:  Nondurable goods:  Food and beverages for off-premises cons&$50$&&slow\tabularnewline
DCLORG3Q086SBEA&Pers Cons Exp:  Nondurable goods:  Clothing and footwear&$50$&&slow\tabularnewline
DGOERG3Q086SBEA&Pers Cons Exp:  Nondurable goods:  Gasoline and other energy goods&$50$&&slow\tabularnewline
DONGRG3Q086SBEA&Pers Cons Exp:  Nondurable goods:  Other nondurable goods&$50$&&slow\tabularnewline
DHUTRG3Q086SBEA&Pers Cons Exp:  Services:  Housing and utilities&$50$&&slow\tabularnewline
DHLCRG3Q086SBEA&Pers Cons Exp:  Services:  Health care&$50$&&slow\tabularnewline
DTRSRG3Q086SBEA&Pers Cons Exp:  Transportation services&$50$&&slow\tabularnewline
DRCARG3Q086SBEA&Pers Cons Exp: Recreation services&$50$&&slow\tabularnewline
DFSARG3Q086SBEA&Pers Cons Exp:  Services:  Food services and accomodations&$50$&&slow\tabularnewline
DIFSRG3Q086SBEA&Pers Cons Exp:  Financial services and insurance&$50$&&slow\tabularnewline
DOTSRG3Q086SBEA&Pers Cons Exp:  Other services &$50$&&slow\tabularnewline
CPIAUCSL&Consumer Price Index for All Urban Consumers:  All Items&$50$&x&slow\tabularnewline
CPILFESL&Consumer Price Index for All Urban Consumers:  All Items Less Food \& Energy&$50$&&slow\tabularnewline
WPSFD49207&Producer Price Index by Commodity for Finished Goods &$50$&&slow\tabularnewline
PPIACO&Producer Price Index for All Commodities &$50$&&slow\tabularnewline
WPSFD49502&Producer Price Index by Commodity for Finished Consumer Goods &$50$&&slow\tabularnewline
WPSFD4111&Producer Price Index by Commodity for Finished Consumer Foods&$50$&&slow\tabularnewline
PPIIDC&Producer Price Index by Commodity Industrial Commodities &$50$&&slow\tabularnewline
WPSID61&PPI by Commodity Intermediate Materials:  Supplies \& Components&$50$&&slow\tabularnewline
WPU0561&Producer Price Index by Commodity for Fuels and Related Products and Power&$50$&&slow\tabularnewline
OILPRICEx&Real Crude Oil Prices:  West Texas Intermediate (WTI) - Cushing, Oklahoma&$50$&&slow\tabularnewline
WPSID62&Producer Price Index:  Crude Materials for Further Processing &$50$&&slow\tabularnewline
PPICMM&PPI:  Commodities:  Metals and metal products:  Primary nonferrous metals&$50$&&slow\tabularnewline
CPIAPPSL&Consumer Price Index for All Urban Consumers:  Apparel&$50$&&slow\tabularnewline
CPITRNSL&Consumer Price Index for All Urban Consumers:  Transportation&$50$&&slow\tabularnewline
CPIMEDSL&Consumer Price Index for All Urban Consumers:  Medical Care&$50$&&slow\tabularnewline
CUSR0000SAC&Consumer Price Index for All Urban Consumers:  Commodities&$50$&&slow\tabularnewline
CES2000000008x&Real Average Hourly Earnings of Prod and Nonsuperv Employees: Construction&$50$&&slow\tabularnewline
CES3000000008x&Real Average Hourly Earnings of Prod and Nonsuperv Employees: Manufacturing&$50$&&slow\tabularnewline
COMPRNFB&Nonfarm Business Sector:  Real Compensation Per Hour (Index 2012=100)&$50$&&slow\tabularnewline
CES0600000008&Average Hourly Earnings of Production and Nonsupervisory Employees:&$50$&&slow\tabularnewline
\midrule
Shadow Rate& Shadow Federal Funds Rate (Percent)&$ 1$&x&policy\tabularnewline
\midrule
TB3MS&3-Month Treasury Bill: Secondary Market Rate (Percent)&$ 1$&&fast\tabularnewline
TB6MS&6-Month Treasury Bill: Secondary Market Rate (Percent)&$ 1$&&fast\tabularnewline
GS1&1-Year Treasury Constant Maturity Rate (Percent)&$ 1$&&fast\tabularnewline
GS10&10-Year Treasury Constant Maturity Rate (Percent)&$ 1$&&fast\tabularnewline
AAA&Moodys Seasoned Aaa Corporate Bond Yield (Percent)&$ 1$&&fast\tabularnewline
BAA&Moodys Seasoned Baa Corporate Bond Yield (Percent)&$ 1$&&fast\tabularnewline
BAA10YM&Moodys Seasoned Baa Corporate Bond Yield Rel. to Yield on 10-Year Treasury&$ 1$&&fast\tabularnewline
TB6M3Mx&6-Month Treasury Bill Minus 3-Month Treasury Bill, secondary market (Percent)&$ 1$&&fast\tabularnewline
GS1TB3Mx&1-Year Treasury Constant Maturity Minus 3-Month Treasury Bill, second market&$ 1$&&fast\tabularnewline
GS10TB3Mx&10-Year Treasury Constant Maturity Minus 3-Month Treasury Bill, second market&$ 1$&&fast\tabularnewline
CPF3MTB3Mx&3-Month Commercial Paper Minus 3-Month Treasury Bill, second market&$ 1$&&fast\tabularnewline
GS5&5-Year Treasury Constant Maturity Rate&$ 1$&&fast\tabularnewline
TB3SMFFM&3-Month Treasury Constant Maturity Minus Federal Funds Rate&$ 1$&&fast\tabularnewline
T5YFFM&5-Year Treasury Constant Maturity Minus Federal Funds Rate&$ 1$&&fast\tabularnewline
AAAFFM&Moodys Seasoned Aaa Corporate Bond Minus Federal Funds Rate&$ 1$&&fast\tabularnewline
M1REAL& Real M1 Money Stock&$50$&&fast\tabularnewline
M2REAL&Real M2 Money Stock&$50$&&fast\tabularnewline
BUSLOANSx&Real Commercial and Industrial Loans, All Commercial Banks&$50$&&fast\tabularnewline
CONSUMERx&Real Consumer Loans at All Commercial Banks &$50$&&fast\tabularnewline
NONREVSLx&Total Real Nonrevolving Credit Owned and Securitized, Outstanding&$50$&&fast\tabularnewline
REALLNx&Real Real Estate Loans, All Commercial Banks&$50$&&fast\tabularnewline
TOTALSLx&Total Consumer Credit Outstanding&$50$&&fast\tabularnewline
TOTRESNS&Total Reserves of Depository Institutions &$50$&&fast\tabularnewline
NONBORRES&Reserves Of Depository Institutions, Nonborrowed&$ 7$&&fast\tabularnewline
DTCOLNVHFNM&Consumer Motor Vehicle Loans Outstanding Owned by Finance Companies&$50$&&fast\tabularnewline
DTCTHFNM&Total Consumer Loans and Leases Outstanding Owned and Sec by Finance Comp. &$50$&&fast\tabularnewline
INVEST&Securities in Bank Credit at All Commercial Banks &$50$&&fast\tabularnewline
TABSHNOx&Real Total Assets of Households and Nonprofit Organizations&$50$&&fast\tabularnewline
EXSZUSx&Switzerland / U.S. Foreign Exchange Rate&$50$&&fast\tabularnewline
EXJPUSx&Japan / U.S. Foreign Exchange Rate&$50$&&fast\tabularnewline
EXUSUKx&U.S. / U.K. Foreign Exchange Rate&$50$&&fast\tabularnewline
EXCAUSx&Canada / U.S. Foreign Exchange Rate&$50$&&fast\tabularnewline
S.P.500&S\&Ps Common Stock Price Index:  Composite&$ 5$&&fast\tabularnewline
S.P..indust&S\&Ps Common Stock Price Index:  Industrials&$50$&&fast\tabularnewline
S.P.div.yield&S\&Ps Composite Common Stock:  Dividend Yield&$ 1$&&fast\tabularnewline
\bottomrule
\end{tabular}

\begin{minipage}{\textwidth}
     \tiny
    \emph{Note:} Transformation codes (in Column \texttt{Trans.}): (1) no transformation, (5) $\Delta \log(x_t)$, (50) $\Delta \log(x_t)$ yoy-transformation, (7) $\Delta (x_t/x_{t-1}-1.0)$
    \end{minipage}

\end{center}}
\end{table}

\setcounter{equation}{0}
\setcounter{table}{0}
\setcounter{figure}{0}
\renewcommand\theequation{C.\arabic{equation}}
\renewcommand\thetable{C.\arabic{table}}
\renewcommand\thefigure{C.\arabic{figure}}


\end{appendices}
\end{document}